\documentclass{aa}
\usepackage[varg]{txfonts}
\usepackage{graphicx}
\usepackage{aecompl,color, amsmath}
\usepackage{xcolor}
\usepackage{siunitx}
\DeclareSIUnit\erg{erg}
\newcommand{\kms}{km~s$^{-1}$}

\graphicspath{ {figures/} }

\begin{document}

\title{Stellar feedback in M83 as observed with MUSE}
\subtitle{I. Overview, an unprecedented view of the stellar and gas kinematics and evidence of outflowing gas}
\author{
Lorenza Della Bruna\inst{\ref{SU}} \and
Angela Adamo\inst{\ref{SU}} \and
Philippe Amram\inst{\ref{marseille}} \and
Erik Rosolowsky\inst{\ref{Alberta}} \and
Christopher Usher\inst{\ref{SU}} \and
Mattia Sirressi\inst{\ref{SU}} \and
Andreas Schruba\inst{\ref{MPE}} \and
Eric Emsellem\inst{\ref{ESO-garching}, \ref{CNRS}} \and
Adam Leroy\inst{\ref{ohio}} \and
Arjan Bik\inst{\ref{SU}} \and
William~P. Blair\inst{\ref{johns-hopkins}} \and
Anna~F. McLeod\inst{\ref{durham1}, \ref{durham2}} \and
Göran Östlin\inst{\ref{SU}}  
Florent Renaud\inst{\ref{lund}}\and
Carmelle Robert\inst{\ref{UL}} \and
Laurie Rousseau-Nepton\inst{\ref{CFHT}} \and
Linda~J. Smith\inst{\ref{space-telescope}}
}

\institute{
The Oskar Klein Center, Department of Astronomy, Stockholm University, AlbaNova, SE-10691 Stockholm, Sweden\label{SU}
\and
Aix-Marseille Université, CNRS, CNES, LAM, Marseille, France\label{marseille}
\and
Department of Physics, University of Alberta, Edmonton, AB T6G 2E1, Canada\label{Alberta}
\and
Max-Planck-Institut für extraterrestrische Physik, Giessenbachstraße 1, D-85748 Garching, Germany\label{MPE}
\and
European Southern Observatory, Karl-Schwarzschild-Str. 2, 85748, Garching, Germany\label{ESO-garching}
\and
Univ. Lyon, Univ. Lyon1, ENS de Lyon, CNRS, Centre de Recherche Astrophysique de Lyon, UMR5574, 69230, Saint-Genis-Laval, France\label{CNRS}
\and
Department of Astronomy, The Ohio State University, 140 West 18th Avenue, Columbus, Ohio 43210, USA\label{ohio}
\and
The William H. Miller III Department of Physics and Astronomy, Johns Hopkins University, 3400 N. Charles Street, Baltimore, MD 21218, USA\label{johns-hopkins}
\and
Centre for Extragalactic Astronomy, Department of Physics, Durham University, South Road,  Durham DH1 3LE, UK\label{durham1}
\and
Institute for Computational Cosmology, Department of Physics, University of Durham, South Road, Durham DH1 3LE, UK\label{durham2}
\and
Department of Astronomy and Theoretical Physics, Lund Observatory, Box 43, 221 00 Lund, Sweden\label{lund}
\and
Département de physique, de génie physique et d'optique, Université Laval\label{UL}, 
\and
Canada-France-Hawaii Telescope, Kamuela, HI, 96743, USA; Department of Physics and Astronomy, University of Hawaii at Hilo, Hilo, HI, 96720-4091, USA\label{CFHT}
\and
European Space Agency (ESA), ESA Office, Space Telescope Science Institute, 3700 San Martin Drive, Baltimore, MD 21218, USA\label{space-telescope}
}

\authorrunning{Della Bruna et al.}
\titlerunning{Stellar feedback in M83}

\date{Received XXX / Accepted YYY}

 \abstract
{
Young massive stars inject energy and momentum into the surrounding gas, creating a multi-phase interstellar medium (ISM) and regulating further star formation. The main challenge of studying stellar feedback proves to be the variety of scales spanned by this phenomenon, ranging from the immediate surrounding of the stars (\ion{H}{ii} regions, 10s pc scales) to galactic-wide kiloparsec scales. 
}
{
We present a large mosaic (3.8 $\times$ 3.8 kpc) of the nearby spiral galaxy M83, obtained with the MUSE instrument at ESO Very Large Telescope (VLT).
The integral field spectroscopy data cover a large portion of the optical disk at a resolution of $\sim$ 20 pc, allowing the characterisation of single \ion{H}{ii} regions while sampling diverse dynamical regions in the galaxy.
}
{
We obtained the kinematics of the stars and ionised gas, and compared them with molecular gas kinematics observed in CO(2-1) with the ALMA telescope array.
We separated the ionised gas into \ion{H}{ii} regions and diffuse ionised gas (DIG) and investigated how the fraction of H$\alpha$ luminosity originating from the DIG ($f_{\rm DIG}$) varies with galactic radius.
}
{
We observe that both stars and gas trace the galactic disk rotation, as well as a fast-rotating nuclear component (30\arcsec $\simeq$ 700 pc in diameter), likely connected to secular processes driven by the galactic bar.
In the gas kinematics, we observe a stream east of the nucleus (50\arcsec $\simeq$ 1250 pc in size), redshifted with respect to the disk.
The stream is surrounded by an extended ionised gas region (1000 $\times$ 1600 pc) with enhanced velocity dispersion and a high ionisation state, which is largely consistent with being ionised by slow shocks. 
We interpret this feature as either the superposition of the disk and an extraplanar layer of DIG, or as a bar-driven inflow of shocked gas.
A double Gaussian component fit to the H$\alpha$ line also reveals the presence of a nuclear biconic structure whose axis of symmetry is perpendicular to the bar. The two cones (20\arcsec $\simeq$ 500 pc in size) appear blue- and redshifted along the line of sight.
The cones stand out for having an H$\alpha$ emission separated by up to 200~\kms\ from that of the disk, and a high velocity dispersion $\sim$ 80 -- 200~\kms. At the far end of the cones, we observe that the gas is consistent with being ionised by shocks. These features had never been observed before in M83; we postulate that they are tracing a starburst-driven outflow shocking into the surrounding ISM.
Finally, we obtain $f_{\rm DIG} \sim$ 13\% in our field of view, and observe that the DIG contribution varies radially between 0.8 and 46\%, peaking in the interarm region. We inspect the emission of the \ion{H}{ii} regions and DIG in `BPT' diagrams, finding that in \ion{H}{ii} regions photoionisation accounts for 99.8\% of the H$\alpha$ flux, whereas the DIG has a mixed contribution from photoionisation (94.9\%) and shocks (5.1\%).
}{}
\keywords{Galaxies: general - Galaxies: individual: NGC 5236 - Galaxies: ISM - Galaxies: kinematics and dynamics - ISM: structure -  \ion{H}{ii} regions}
\maketitle 

\section{Introduction}
\label{section:introduction}

\begin{figure*}
\center
\sidecaption
\includegraphics[width=12.5cm]{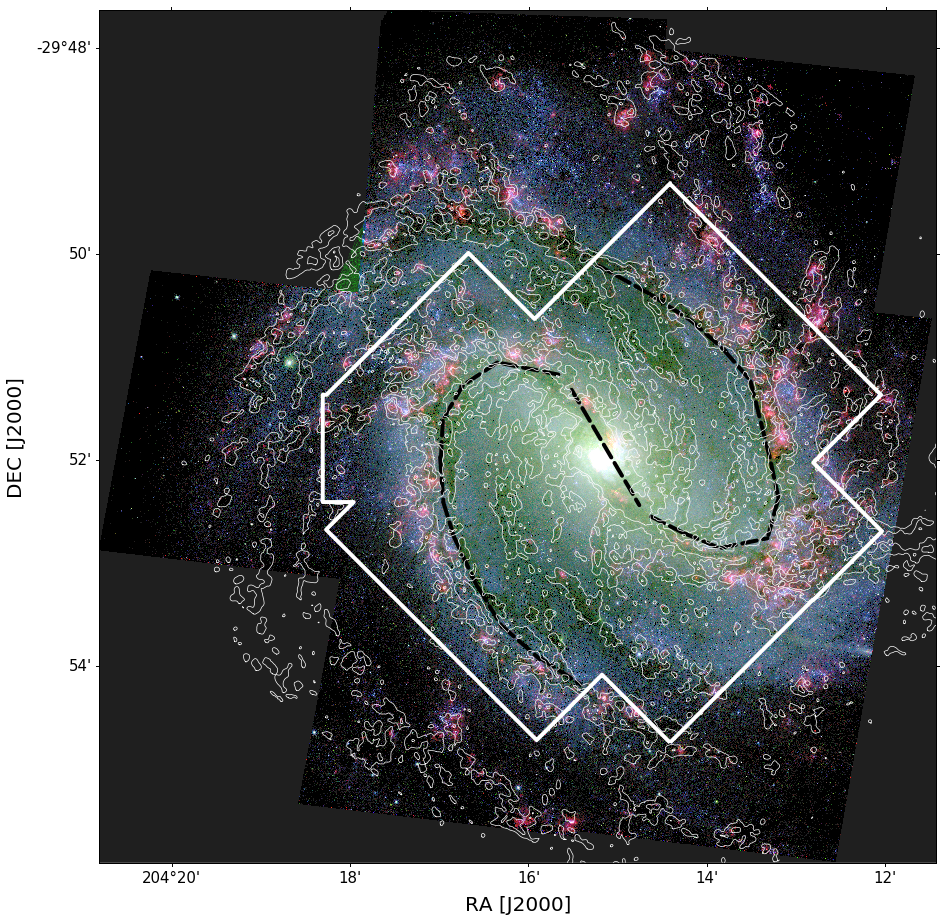}
\caption{HST colour-composite image of M83 (Red: F657N, Blue: F814W, Green: F438W). The footprint of the MUSE mosaic is overlaid in white. The position of the bar and the two spiral arms is sketched in black (see Section~\ref{section:introduction}). The white contours trace molecular gas mass surface density from ALMA CO(2-1) (contours of 5 and 25 K~\kms, corresponding to a surface gas density $\simeq 30$ and 170 $M_\odot$ pc$^{-2}$).
}
\label{fig:rgb_hst}
\end{figure*}

Young massive stars originate from the gravitational collapse of giant molecular clouds (GMCs), and they inject energy and momentum into the surrounding interstellar medium (ISM) via different feedback processes such as thermal feedback from protostars, photoionisation, and mechanical feedback from stellar winds and supernovae (for a review, see \citealp{krumholz14_review} and \citealp{dale15}).
These combined effects can disrupt the parent GMCs \citep{dale15, howard17}, resulting in a self-regulating mechanism which inhibits future star formation. On the other hand, positive feedback can occasionally facilitate further collapse in neighbouring regions, boosting the formation of new stars. Stellar feedback directly affects star-forming regions by carving channels and `bubbles' of ionised gas into the surrounding cool gas and dust, and it also impacts the ISM on wider galactic scales, as it clears the path for winds and outflows to escape the star-forming regions. Overall, stellar feedback can shape the global properties of a galaxy \citep[e.g.][]{scannapieco12, hopkins13}, and it plays a key role in the recycling of gas, the regulation of star formation, and the chemical enrichement and mixing of star-forming galaxies \citep[e.g.][]{leroy15a, maiolino19}. Modelling feedback has therefore proven to be essential in simulations of GMCs \citep[e.g.][]{dale14} as well as galaxy formation and evolution \citep{schaye15,hopkins18} in order to correctly reproduce key observables and relations between them.
\par A more detailed understanding of the process of stellar feedback is also key to explain the origin of a warm and diffuse component of the ISM (diffuse ionised gas or DIG, see \citealp{haffner09} for a review), which has been observed to make up a considerable fraction of the H$\alpha$ emission (up to 50\%) in local spiral galaxies \citep{ferguson96,zurita00, thilker02,hoopes03, oey07}.
The origin of this ISM component is still under study, and has been linked to radiation leaking from the star-forming regions \citep{zurita02, weilbacher18,paperII, belfiore21}, field stars \citep{hoopes00, zhang17, belfiore21}, shocks \citep{collins01}, cosmic rays \citep{vandenbroucke18} or scattering by dust \citep{seon12}. Studying stellar feedback in nearby galaxies will allow to probe the role of each of these processes in ionising the DIG.
\par The advent of integral ﬁeld spectroscopy (IFS) has been a leap forward in the study of stellar feedback, as it allows to acquire spectral information of multiple lines simultaneously over a relatively wide field of view (FoV) at good angular resolution. This makes it possible for example to obtain at the same time the kinematics of gas and stars, and to disentangle different gas ionisation mechanisms \citep[see][for a review]{kewley19}.
\par IFS instruments have led the way to multi-scale studies of kinematics and physical properties of ionised gas in nearby galaxies. At the largest (kpc) scales, surveys such as CALIFA \citep{sanchez12}, MANGA \citep{bundy15} and SAMI \citep{croom12, bryant15} provided a census of the $z \sim 0$ galaxy population. The large sample of \ion{H}{ii} regions provided by these surveys allowed for a better understanding of the general properties of star forming regions. For example, it was possible to investigate the link between the physical conditions of the ISM and the stellar population in the regions \citep{sanchez15}, and between the stellar population and the leakage of ionising photons \citep{morisset16}. Large scale surveys also allowed to quantify and characterise the presence of extraplanar DIG (eDIG) in edge-on spiral galaxies \citep{jones17, bizyaev17, levy19}.

\par At intermediate scales, two instruments whose wide ﬁeld coverage and high sensitivity have been particularly important for the study of feedback are: the MUSE integral field unit \citep[IFU,][]{bacon10}, mounted on the ESO Very Large Telescope (VLT) at Cerro Paranal observatory and the SITELLE imaging fourier transform spectrograph at Canada-France-Hawaii Telescope \citep[CFHT,][]{drissen19}.
The relatively large FoV (1 arcmin$^2$ and 11 $\times$ 11 arcmin$^2$, respectively) and high spatial sampling ($0\farcs2$ and $0\farcs32$/pixel) of these instruments are ideal to study global properties of nearby galaxies, while at the same time achieving a high level of detail at small scales. This was exploited by galaxy surveys such as MAD \citep{erroz-ferrer19, denbrok20}, SIGNALS \citep{rousseau19} and PHANGS-MUSE \citep{emsellem21}, that are targeting each $\sim$ 20 - 50 galaxies, at distances where 1\arcsec $\sim$ 50 -- 200 pc, enabling the identification and characterisation of individual \ion{H}{ii} regions. This makes it possible to study the dependence of region properties on their location in the galaxy (e.g. in arm vs interarm regions) and on the local ISM conditions \citep{kreckel16, kreckel19, erroz-ferrer19}.

\par Finally, a few specific studies of local galaxies with MUSE (\citealp[LMC,][]{mcleod19_lmc}, \citealp[NGC 300,][]{mcleod20, mcleod21}; \citealp[NGC~7793,][]{paperI, paperII}) are starting to probe scales of tens of parsec, resolving individual \ion{H}{ii} regions in detail and detecting the most massive individual stars, which gives a way to bridge small and galaxy-scale stellar feedback.
For individual regions, this allows one to investigate, for example, the contribution of various types of feedback \citep{mcleod19_lmc, mcleod20, mcleod21}, the ionisation structure of the region and to derive an escape fraction by modelling the expected ionising photon flux based on the observed stellar population \citep{mcleod19_lmc, mcleod20,paperI, paperII}.

\par In this study, we exploit MUSE IFS data to study the ISM in the nearby galaxy M83 at a scale of $\sim$ 20 pc. Our work complements the very high resolution (10 pc) studies and the large scale (100 pc -- kpc) studies, by covering a wide area ($\sim$ 20 arcmin$^2$) at a resolution that still allows the characterisation of individual \ion{H}{ii} regions while at the same time sampling a large number of regions ($\sim 4700$, Della Bruna et al., in prep) and a wide range of ISM conditions across the galactic disk.

\par
\begin{figure*}
\center
\includegraphics[width=9.1cm]{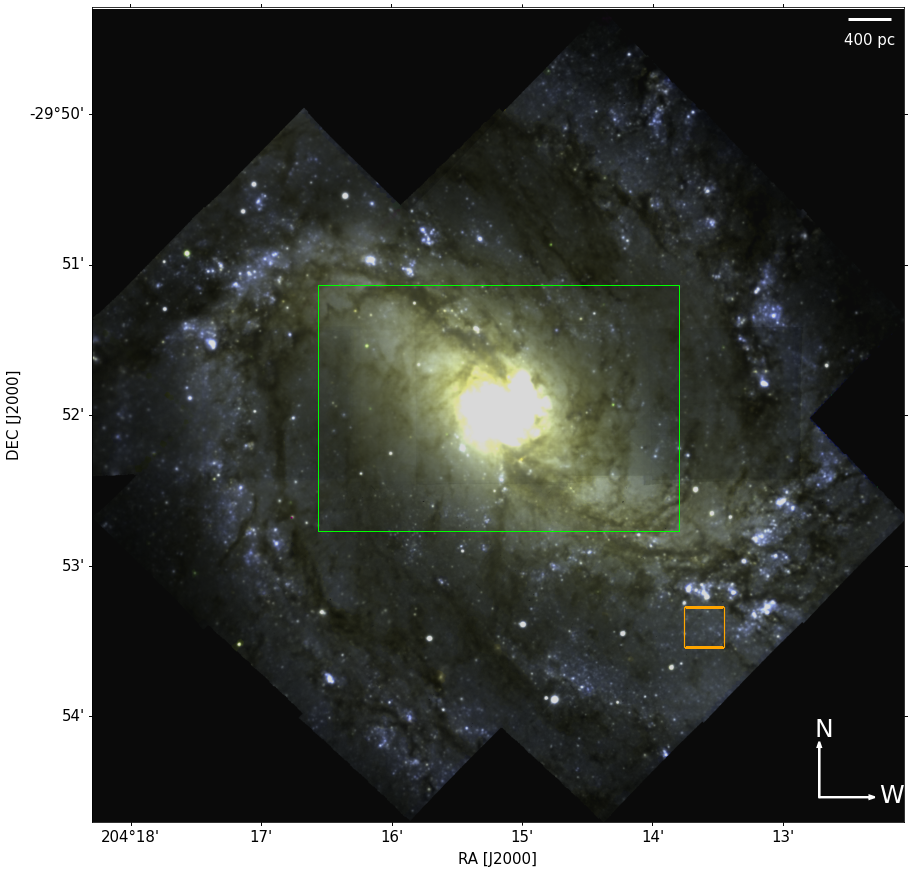}
\includegraphics[width=9.1cm]{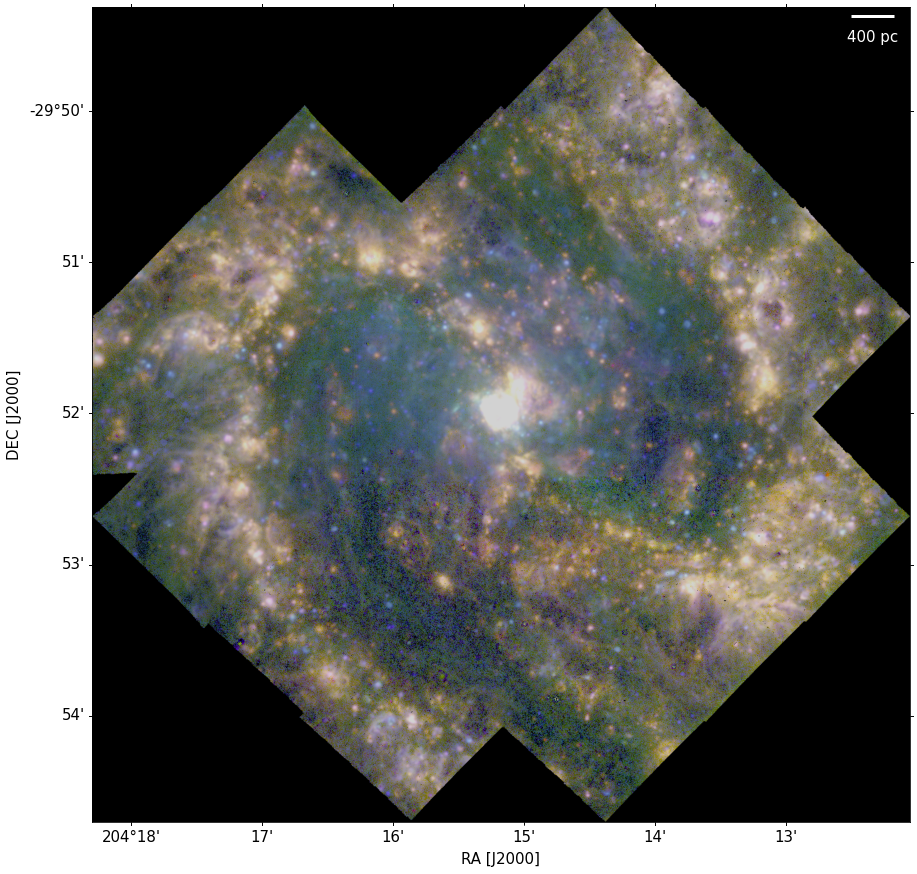}
\caption{Three-colour composites of the MUSE data. \textit{Left}: composite of stellar bands (Blue: 4875 -- 4950~\AA{}, Green: 6520 -- 6528~\AA{}, Red: 6750 -- 6810~\AA{}). The green rectangle indicates the central starburst region analysed in Sect.~\ref{section:starburst_region}. The orange square indicates the position of the H$\alpha$ cloud highlighted in Fig.~\ref{fig:kinematics} (central panels). \textit{Right}: composite of gas emission (Blue: [\ion{O}{iii}], Green: [\ion{S}{ii}] Red: H$\alpha$).}
\label{fig:rgb_muse}
\end{figure*}

 M83 (also known in the literature as NGC~5236) is a nearby spiral galaxy, at a distance of 4.89 Mpc \citep{jacobs09}. Fig.~\ref{fig:rgb_hst} and ~\ref{fig:rgb_muse} show M83 observed with the Hubble Space Telescope (HST) and the Atacama Large Millimeter/submillimeter Array (ALMA), and with MUSE (this work). M83 has a grand design barred spiral morphology, and its full optical disk is $\sim$ 20 kpc across; the position of the stellar bar and the two spiral arms\footnote{We determine the outline of spiral arms as the inner radius of the star-forming regions identified in Sect.~\ref{section:hii_dig}, and the bar position angle based on archival Spitzer S$^4$G data (\citealp{sheth10, salo15}, \url{https://irsa.ipac.caltech.edu/data/SPITZER/S4G/galaxies/NGC5236.html}).} are sketched in Fig.~\ref{fig:rgb_hst}.
The galaxy has a stellar mass $\log_{10} M_\star = 10.53 M_\odot$ \citep{leroy21} and hosts a nuclear starburst ring \citep{sersic65,buta93,calzetti04,knapen10,comeron10}, resulting in a high star formation rate (SFR = 4.2 $M_\odot$ yr$^{-1}$, \citealp{leroy21}). Its offset from the SFR-$M_\star$ main sequence for star forming galaxies ($\Delta_{\rm MS}$ = 0.44 dex) is consistent with the typical scatter at this stellar mass \citep{popesso19}.
M83 has been extensively studied over all wavelengths, and its stellar population has been mapped across the galactic disk. Catalogues of  young star clusters \citep[YSCs,][]{silva-villa14,adamo15},
Wolf-Rayet stars \citep{hadfield05} and supernova remnants \citep{blair14,winkler17,williams19, russell20} are publicly available. Overall, this makes M83 an ideal candidate for the detailed study of stellar feedback at scales of 10s of parsecs.

\par The MUSE data presented in this work consist of a large mosaic covering the central 3.8 kpc in radial extent at $\sim$ 20 pc resolution. This is the first extensive IFS study of stars and ionised gas across the disk of M83. Previous studies have either targeted the ionised gas using a wide field spectrograph \citep[][$\sim$ 40 pc resolution]{poetrodjojo19} or Fabry-Pérot observations \citep{fathi08}, or have focused on spectroscopy of the nuclear region \citep[e.g.][]{knapen10, piqueras_lopez12, gadotti20, callanan21}. We complement our observations with high-resolution archival imaging from HST ($<$ 2 pc resolution) and CO(2-1) observations from ALMA at $\sim$ 50 pc resolution. An overview of the dataset is shown in Fig.~\ref{fig:rgb_hst}.
The MUSE data allow us to map for the first time the large-scale stellar kinematics, as well as unprecedentedly detailed H$\alpha$ kinematics. The wealth of information provided by IFU spectroscopy enables us to investigate spatially resolved physical properties of the gas.
At the same time, the HST coverage provides us with a detailed catalogue of massive stars and clusters, and the ALMA data trace the distribution of the molecular gas in the galaxy. The goal of this project is the study of star formation from galactic scales (this paper) to small scale ($\sim$ 20 pc, Della Bruna et al., in prep.). The data allow also for a census of Wolf-Rayet stars (Smith et al., in prep), planetary nebulae (Della Bruna et al., in prep.) and supernova remnants (Long et al., in prep.). Ultimately, we aim to provide a comprehensive picture of the star formation cycle in M83 at tens of parsec scales.

\par This paper is organised as follows: in Section~\ref{section:data_description} we give an overview of the dataset and of the MUSE data reduction. In Sect.~\ref{section:kinematics} we present the kinematics of stars and ionised gas. In Sect.~\ref{section:extinction} we estimate the extinction from the ionised gas and compare it with the distribution of the molecular gas. In Sect.~\ref{section:hii_dig} we identify \ion{H}{ii} regions complexes, determine the fraction of DIG and inspect the properties of the ionised gas in emission line diagrams. In Sect.~\ref{section:starburst_region} we take a closer look at the kinematics and emission of the central starburst region. Finally, in Sections~\ref{section:discussion} and \ref{section:conclusions} we discuss the results and draw our conclusions.

\section{Dataset overview and MUSE data reduction}
\begin{table*}
\caption{Physical properties of M83.}
\begin{tabular}{llcc}
\hline \hline
 & Parameter & Value & Ref. \\ \hline
Distance and morphology & Distance & 4.89 Mpc ($1\arcsec = 24$ pc) & (1) \\
& Effective radius & 3.5 kpc & (2)\\
& Morphological Type & SAB(s)c & (3) \\ 
\hline
Physical properties & $\log_{10} M_\star$ &  10.53 $M_\odot$ & (2) \\ 
& SFR & 4.2 $M_\odot$ yr$^{-1}$ & (2) \\ 
& $\Delta_{\rm MS}$\tablefootmark{(a)} & 0.44 dex & (2)\\
& $\log_{10} \tau_{\rm dep}^{\rm mol}$\tablefootmark{(b)} & 8.9 yr & (2) \\
& Central abundance & 12 + log(O/H) = 8.99 & (4) \\
\hline
\end{tabular}
\tablebib{(1) \citet{jacobs09}; (2) \citet{leroy21}; (3) RC3 catalogue \citep{RC3_cat}, \citet{buta15}; (4) \citet{bresolin16}.}
\tablefoot{
\tablefoottext{a}{offset from the SFR-$M_\star$ main sequence}
\tablefoottext{b}{derived assuming $\log_{10} L_{\rm CO}$ = 8.84 K~\kms~pc$^{-2}$ \citep{leroy21}, a standard Galactic
$\alpha_{\rm CO} = 4.35$ $M_\odot$ pc$^{-2}$ (K~\kms)$^{-1}$ and $R_{21} = 0.65$ \citep{denbrok21_alma, leroy21_subm}.}
}
\label{table:m83_param}
\end{table*}

\label{section:data_description}
We observed M83 with the MUSE instrument at VLT (observing programmes 096.B-0057(A) and 0101.B-0727(A), PI Adamo). The main properties of the target are summarised in Table~\ref{table:m83_param}. The MUSE data cover the inner $\sim$ 3.8 kpc in galactocentric radial extent ($1.1 \times R_e$ in effective radius, see Table~\ref{table:m83_param}), and a total physical area of 40.5 kpc$^2$.
The dataset consists of 20 pointings (indicated in green in the top panel of Fig.~\ref{fig:texp_coverage}) in Wide Field Mode (WFM) and extended wavelength configuration (4650 -- 9300~\AA{}), for a total of 46 exposures. The pointings were observed with science exposures of 550 sec (4 exposures for pointings 1 to 8, a single exposure for remaining pointings). Sky frames of 180 sec were obtained such that every science exposure is preceded or followed by a sky frame acquisition. Because of the large extent of the target in the sky, sky frame offsets were larger than 4 arcmin.
We complemented our observations with MUSE data available in the ESO archive: observing programmes 097.B-0899(B) (PI Ibar) and 097.B-0640(A) (PI Gadotti). Both programmes were observed in WFM with nominal wavelength range (4800 -- 9300~\AA{}). From these programmes, we excluded some frames that had calibration issues, poor seeing or that were not fully overlapping with the rest of our dataset. From archival dataset 097.B-0899(B) we included five pointings (indicated in blue in Fig.~\ref{fig:texp_coverage}, top panel), each observed during three science exposures of 600 sec with two external sky frames with a 180 sec exposure. Archival dataset 097.B-0640(A) consists of one pointing (purple square in Fig.~\ref{fig:texp_coverage}, top panel), of which we included four 480 s exposures, and an additional two sky exposures of 300 s.

\par The final mosaic consists of 26 pointings for a total of 65 MUSE exposures. The central coordinates and exposure time of all the included pointings can be found in appendix~\ref{section:appendix_muse_data_coverage} (Table~\ref{table:pointings_info} and Fig.~\ref{fig:texp_coverage}). The pointings were imaged over a wide range of observing conditions. We derived a PSF for each pointing by fitting a Moffat profile as a function of wavelength to bright, isolated point sources with \textsc{PampelMUSE} \citep{kamann18}, and report the values in Table~\ref{table:pointings_info}. Across the full mosaic we measure a median PSF of $0\farcs7$ at 7000~\AA{} (17 pc at the distance of our target). The full width at half maximum of the Moffat profile declines with increasing wavelength, with a median difference of $0\farcs16$ between the blue and red end.

\par M83 was also observed with HST during the WFC3 Early release science programme (GO11360, PI O'Connell), using narrow and broad band imaging ranging from the UV to the NIR. The coverage of the inner 4.5 kpc ($1.3 \times R_e$) in galactocentric radius was later completed with the programme GO12513 (PI Blair). In total, the HST mosaic consists of seven contiguous pointings \citep{blair14}, with a FWHM $\sim 0\farcs08$ (1.9 pc).
\\
\\
Finally, archival ALMA CO (J = 2–1) data\footnote{Project codes 2013.1.01161.S, 2015.1.00121.S and 2016.1.00386.S, PI Sakamoto.} covering the inner 7.2 kpc ($2.1 \times R_e$) of the galaxy in galactocentric radius have been processed as part of the PHANGS-ALMA survey~\citep{leroy21} using the PHANGS-ALMA pipeline \citep{leroy21_pipeline}. The data have an angular resolution of $2\farcs14$ (50 pc), a spectral resolution of 2.5~\kms\ channel$^{-1}$ and a rms brightness temperature sensitivity of 0.17 K. The data are described in detail in \citet{leroy21}, and the data reduction in \citet{leroy21_pipeline}.

\par In Fig.~\ref{fig:rgb_hst} we show a 3-colour composite of the HST data.
The footprint of the MUSE mosaic is shown in white.
The white contours indicate the ALMA CO emission at levels of 5 and 25 K~\kms. Using Eq. (1) in \citet{sun20} and assuming a CO(2–1)-to-CO(1–0) line ratio $R_{21} = 0.65$ \citep{denbrok21_alma, leroy21_subm} and a standard Galactic CO-to-H2 conversion factor $\alpha_{\rm CO} = 4.35$ $M_\odot$ pc$^{-2}$ (K~\kms)$^{-1}$, these values corresponds to a cold gas surface density of $\simeq 30$ and 170 $M_\odot$ pc$^{-2}$.
The position of the bar and the spiral arms is shown with black dashed lines (see Sect.~\ref{section:introduction}).
The HST and ALMA data cover the inner $\sim$ 4.5 -- 7 kpc of the disk, imaging the spiral arms in their entirety, whereas the MUSE mosaic is limited to the central 3.8 kpc. 

\par We reduced the MUSE data with the ESO pipeline \citep[v2.8.3]{weilbacher14, weilbacher20_pipeline}, following the standard reduction procedure.
In a first phase, the instrumental signature was removed using \texttt{muse\_scibasic}. For this purpose, we combined bias frames to account for the readout noise and lamp frames to correct for flat fielding. Wavelength calibration was performed by combining arc-lamp frames, and correction of 3D illumination and a refined flat-fielding was achieved by combining twilight frames. Moreover, we used the line spread function (LSF) and geometry table (describing the spatial location of IFU slicers) distributed with the pipeline.
In a second phase, we performed flux calibration using a standard star, and we modelled the sky spectrum from the sky frames with \texttt{muse\_create\_sky}. We then ran \texttt{muse\_scipost} on the object and sky frames. For the object frame, we used the `subtract-model' sky subtraction method, that uses sky lines and continuum estimated from the sky frames. We saved whitelight images as well as individual pixel table, that we later used when aligning and combining the individual exposures. 
In a third step, we assembled the final mosaic. We used \texttt{muse\_exp\_align} to compute offsets between multiple exposures of the same pointing. Offsets between different pointings were instead computed by matching each MUSE whitelight image with the B-band HST mosaic obtained from the MAST archive\footnote{\url{https://archive.stsci.edu/prepds/m83mos/}}, and registered to the Gaia DR2 \citep{gaia_dr2}. We selected in each MUSE image at least ten point-like sources within the FOV that corresponded to isolated bright clusters in the HST frame. The list of offsets in WCS was then used by \texttt{muse\_exp\_combine} to combine the individual pixel tables into a large mosaic. A colour composite of the stellar continuum in different broad bands (left) and of three line emission maps (right) extracted from the final MUSE mosaic are shown in Fig.~\ref{fig:rgb_muse}.

\section{Kinematics of stars and gas}
\label{section:kinematics}

\begin{figure*}
\center
\includegraphics[width=9.1cm]{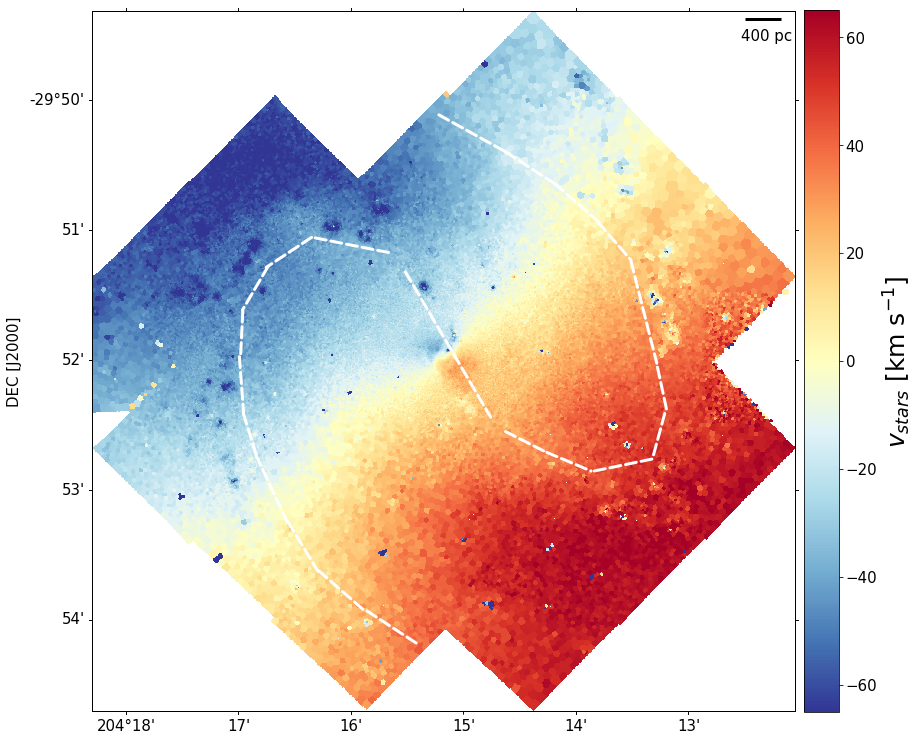}
\includegraphics[width=8.8cm]{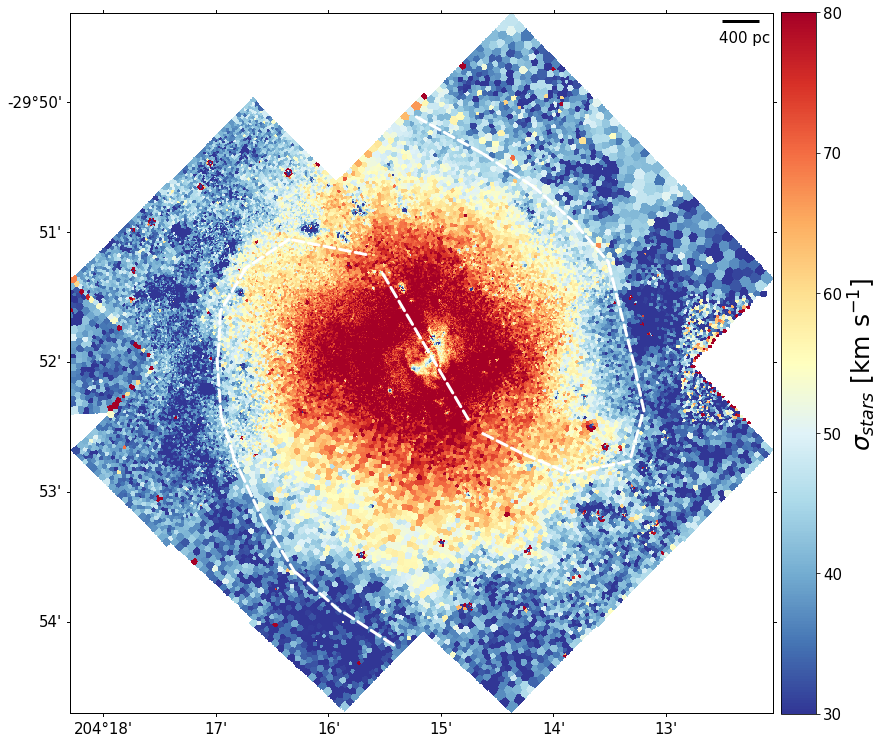}
\includegraphics[width=9.1cm]{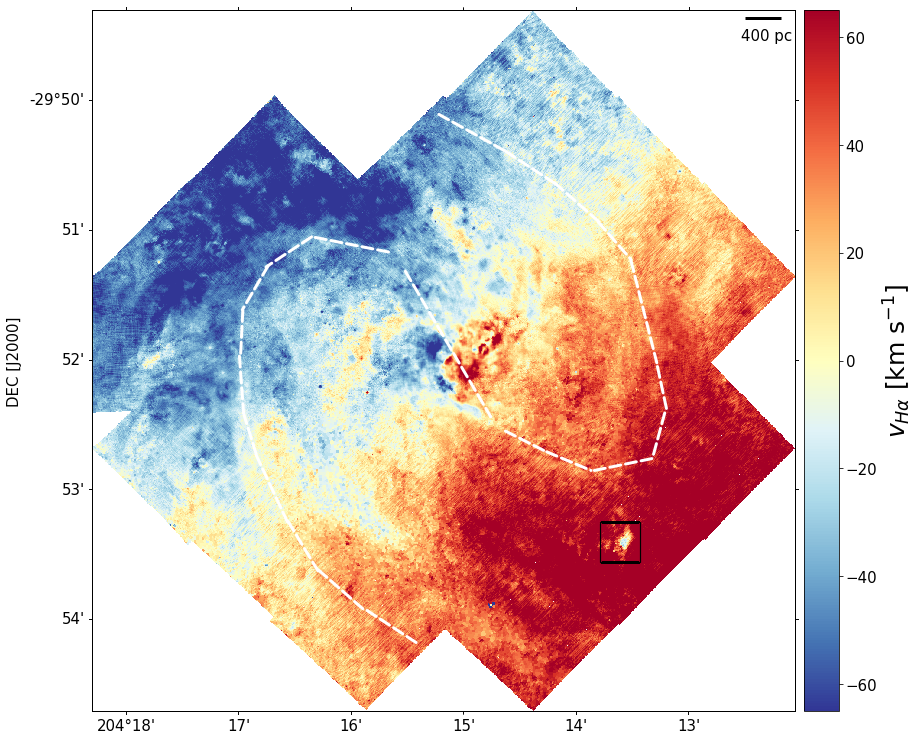}
\includegraphics[width=8.8cm]{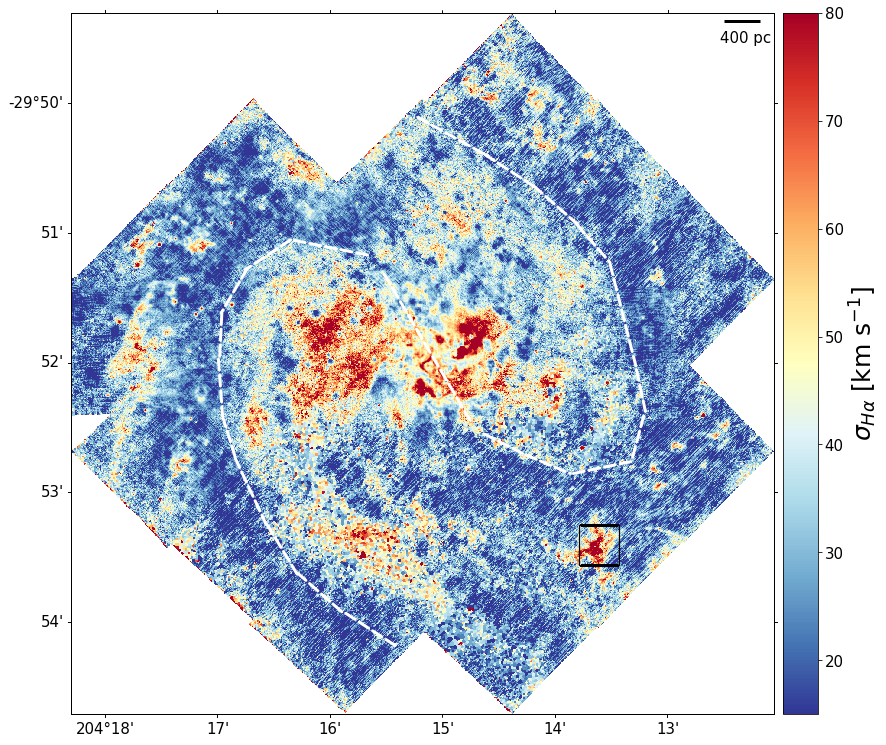}
\includegraphics[width=9.1cm]{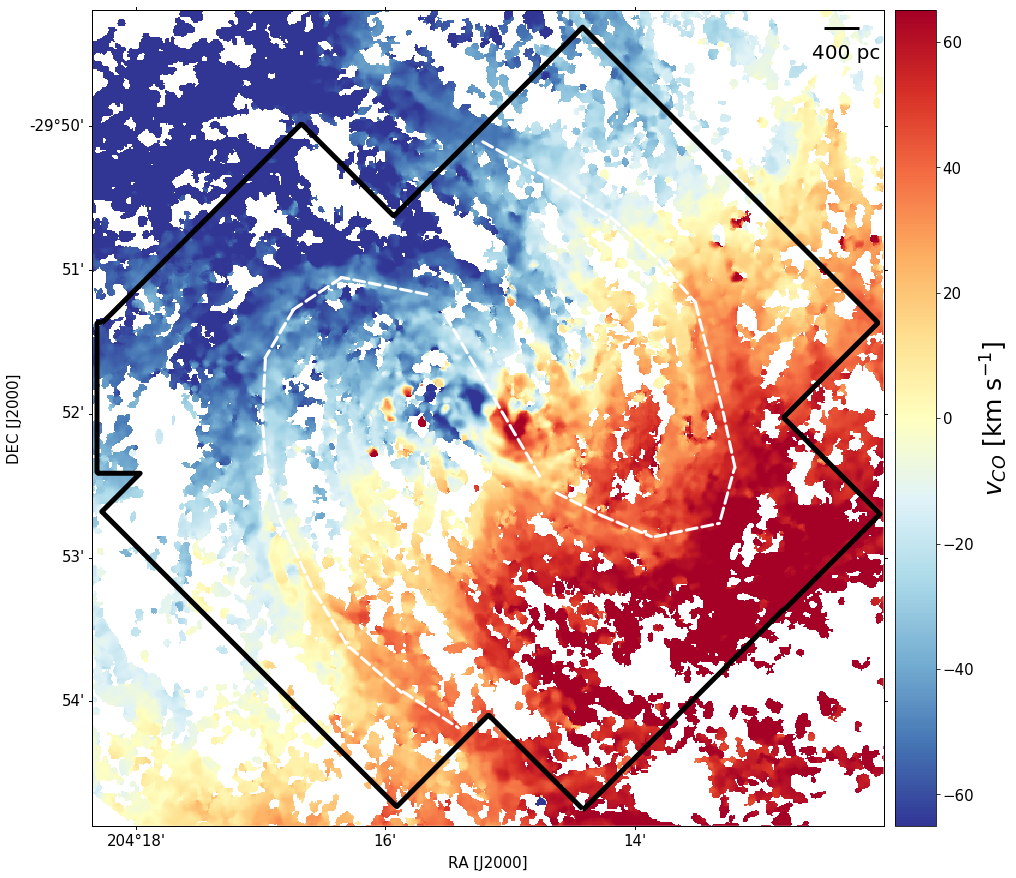}
\includegraphics[width=8.8cm]{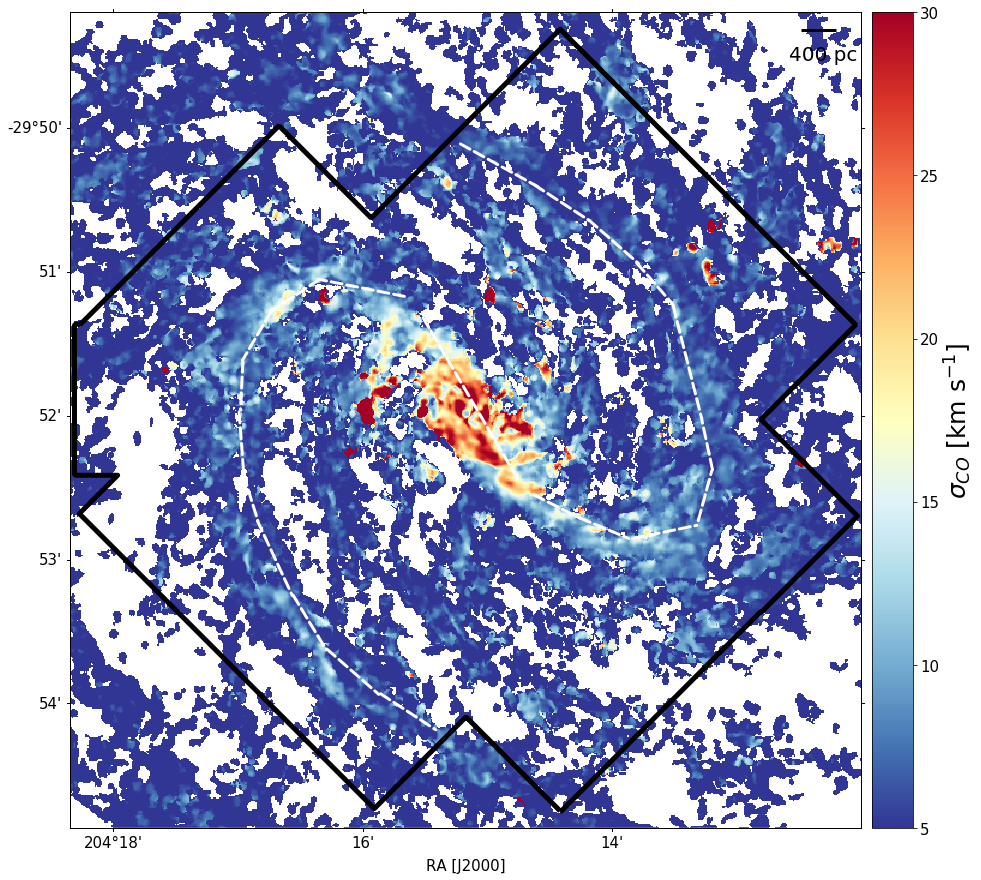}
\caption{Kinematics of the stars and gas. \textit{Top row}: MUSE stellar kinematics. 
\textit{Centre row}: MUSE H$\alpha$ kinematics. The black box in the bottom right corner indicates the position of what we interpret as an extraplanar H$\alpha$ cloud.
\textit{Bottom row}: molecular gas kinematics from ALMA CO(2-1) (mom1 and mom2 maps). The footprint of the MUSE data is overlaid in black.
All velocity maps have been corrected for systemic velocity; no inclination correction has been applied. The velocity dispersion maps refer to the intrinsic dispersion (instrumental effects have been removed). The white dashed lines sketch the position of the bar and spiral arms as shown in Fig.~\ref{fig:rgb_hst} (see Sect.~\ref{section:introduction}). A zoom-in into the central kinematics is shown in Fig.~\ref{fig:kinematics_zoomin}.
}
\label{fig:kinematics}
\end{figure*}

\subsection{Stellar kinematics}
\label{section:kinematics_stars}
We fitted the stellar continuum with \textsc{pPXF}~\citep{cappellari04,cappellari17}, using the E-MILES simple stellar populations models~\citep{vazdekis16} with an unimodal IMF and Padova 2000 isochrones~\citep{girardi00}. The data are spatially binned using the 
weighted adaptation~\citep{diehl06} of the Voronoi tessellation method~\citep{cappellari03}. The bins were targeted to a signal to noise ratio (S/N) $\simeq$ 250 in the continuum range 5025 -- 5065~\AA{}, corresponding to a S/N $\simeq$ 40 \AA{}$^{-1}$.
We fitted the range 4600 -- 8740~\AA{} (in order to include the \ion{Ca}{ii} triplet). All relevant emission lines, as well as sky residuals were masked during the fit; in particular, we excluded entirely the range 7220 -- 8507~\AA{} due to the strong residual sky emission. We used the Gaussian parametrisation of the instrumental LSF by~\citet{guerou17}\footnote{$FWHM_{\rm LSF} [\AA{}] = \num{6.266e-8} \lambda^2 - \num{9.824e-4} \lambda + 6.286$.}. We note that following this parametrisation, velocity dispersions below $\sim$ 50~\kms\ at H$\alpha$ are undersampled by the MUSE instrument.

\par The resulting stellar velocity and dispersion maps are shown in Fig.~\ref{fig:kinematics} (top row). We observe that the stars exhibit overall a regular rotation, and that the velocity dispersion increases from $\sigma \simeq 30$~\kms\ in the outskirts to $\sigma \simeq 80$~\kms\ towards the centre.
At the very centre, we observe a fast rotating component and a dip in the velocity dispersion ($\sigma \simeq 50$~\kms). We discuss this feature in Sect.~\ref{section:central_kinematics}, where we analyse the kinematics of the starburst region.

\subsection{Ionised gas kinematics}
\label{section:kinematics_ionised_gas}
We created a stellar continuum subtracted cube (`gas cube') by rescaling the best stellar population fit from \textsc{pPXF} in each Voronoi bin by the height of the continuum in each spatial pixel (spaxel). We studied the kinematics of the ionised gas from the H$\alpha$ emission line, the brightest line in the MUSE spectral range throughout our FoV. We Voronoi binned the gas cube to a S/N $\sim$ 20 in H$\alpha$\footnote{Throughout this work, we estimate the S/N of Voronoi maps using a window of 10~\AA{} centred on $\lambda_{\rm obs}$.} and fitted the line with a single Gaussian profile. We used the python scipy module \texttt{curve\_fit}, with an initial guess ($v_i, \sigma_i) = (v_{\rm syst}, 20)$ km/s for the velocity and velocity dispersion, where $v_{\rm syst}$ is the systemic velocity estimated from the stellar kinematics.
We set the initial flux $f_i$ to the integral of the binned spectrum in a window of width 22~\AA{} ($\pm$ 500~\kms) centred on the line. We then performed the fit in the same wavelength window, in order to avoid being affected by a poorly subtracted stellar continuum. We constrained $f \geq 0$ and $\sigma \geq 1$~\kms, and leave all other parameters free. We removed the instrumental signature using the parametrisation of the MUSE LSF described in Sect.~\ref{section:kinematics_stars}. The resulting velocity and dispersion maps are shown in Fig.~\ref{fig:kinematics} (middle row panels). We remark that while the spectral resolution of MUSE at H$\alpha$ is coarse ($\sigma_{\rm H\alpha} \simeq$ 120~\kms), the S/N on the H$\alpha$ line yields a median centroiding accuracy of $\simeq$ 2~\kms. We also note that the artefacts in both maps are arising from minute calibration uncertainties at the scale of a few 10s of km s$^{-1}$ and to undersampling of the MUSE LSF (see Sect.~\ref{section:kinematics_stars}).

\par The kinematics of the ionised gas are more complex than the stellar kinematics. At galactic scales, we observe an overall disk rotation, while the velocity dispersion increases in the interarm regions ($\sigma \simeq 80$~\kms) with respect to the star-forming regions along the spiral arms ($\sigma \simeq 20$~\kms). The black square in the central panels of Fig.~\ref{fig:kinematics} indicates a compact region that features a peculiar H$\alpha$ velocity which appears to be 100~\kms\ lower than the surrounding disk rotation, and a high velocity dispersion $\sim$ 80~\kms. Given that this region does not stand out in the stellar RGB image (see orange square in Fig.~\ref{fig:rgb_muse}) and that it shows a typical spectrum of a low-luminosity H$\alpha$ region, we interpret it as a cloud of extraplanar gas, possibly falling into or being ejected from the galactic disk.

\bigskip We also performed a double Gaussian component fit to the H$\alpha$ line. A double component can better capture the emission in the case of extreme line broadening on top of a single emission line peak (caused e.g. by shocks or unresolved flows) and in the case where the line has a double-peaked profile, tracing gas motions on top of the galactic rotation (e.g. resolved flows). The fit was performed as following: in a first step, we fitted a single component, as described above. This fit was then adopted as initial guess for the first component, whereas for the second component, we set an initial guess $(v_i, \sigma_i) = (v_{\rm 1comp}, \sigma_{\rm 1comp})$ and $f_0 = 0.25 \times f_{\rm 1comp}$. All boundary conditions were the same as described above. In order to prevent over-fitting by the broad component (e.g. in the case of a noisy or poorly subtracted continuum), we limited the fit to a window of width 22~\AA{} ($\pm$ 500~\kms) around the rest wavelength of H$\alpha$. We determined the optimal number of parameters based on the Bayesian information criterion (BIC) statistic \citep{schwarz78}, following the approach of \citet{koch21}. The BIC statistic consists of a likelihood term $\mathcal{L}$ plus an additional term that penalises models with more free parameters, and helps preventing overfitting:
\begin{equation}
\mbox{BIC} = \ln(m) k - 2 \ln(\mathcal{L}).  
\end{equation}
Here, $m$ is the number of fitted data points and $k$ is the number of free parameters. We selected the fit minimising the BIC statistic; we furthermore ignored double component fits in which one component contributes to less than 5\% of the total flux. We separated the resulting components into a first component, tracing the galactic rotation, and a second component having a blue or redshift with respect to the disk. In the case where the two components are closer in velocity than 57~\kms\ (corresponding to the MUSE spectral sampling of 1.25\AA{} at $\lambda_{\rm H\alpha}$), we picked as the first component the one with the largest amplitude. We found that a double component Gaussian improves the fit only in the starburst region (black rectangle in the left panel of Fig.~\ref{fig:rgb_muse}); we therefore present and discuss the resulting kinematic maps in Sect.~\ref{section:central_kinematics}, where we study in detail the central kinematics.

\subsection{Molecular gas kinematics}
\label{section:kinematics_molecular_gas}
In Fig.~\ref{fig:kinematics} (bottom row) we show the kinematics of the molecular gas observed in CO(2-1) emission with ALMA. We selected the mom1 and mom2 data obtained with the high conﬁdence `strict' mask \citep[see][]{leroy21_pipeline}. Despite the difference in resolution, we observe that the molecular gas velocity compares well with the velocity of the ionised gas (centre left panel in Fig.~\ref{fig:kinematics}).
The velocity dispersion is very low ($< 15$~\kms) throughout the disk, as expected for the cold gas component, and is only enhanced ($\sim$ 25 -- 30~\kms) around the starburst region. We discuss the kinematics of the central starburst region in Sect.~\ref{section:central_kinematics}.

\subsection{Kinematic fitting}
\label{section:kinematic_fitting}

\begin{table*}
\caption{Parameters derived from the kinematic fitting described in Sect.~\ref{section:kinematic_fitting}.}
\begin{center}
\begin{tabular}{l|cccc|ccccc|ccc}
\hline
&    \multicolumn{4}{c}{Stellar Component} &    \multicolumn{5}{c}{H$\alpha$ Component}&    \multicolumn{3}{c}{CO Component}\\ 
\hline
Model								&	$(A)$		&	$(B)$			&	$(C)$			&	$(D)$			&	$(a)$		&	$(b)$			&	$(c)$			&	$(d)$			&    $(e)$ &$(\alpha)$&$(\beta)$&$(\gamma)$\\
\hline
$\chi^2_r$$^{(1)}$					&	1.21	&	1.23		&	1.27		&	1.36		&	43.7	&	44.2		&	46.7		&	44.6		&	 46.9			&	1836	&	1870		&2007			\\
$incl^{(2)}$ [$^{\circ}$] 			&	20.3	&	20.0		&	20.1		&	19.9		&	37.9	&	\fbox{20.3}	&	\fbox{20.3}	&	\fbox{20.3}	&	 \fbox{20.3}	&	35.0	&	\fbox{20.3}	&\fbox{20.3}	\\
$v_{\rm syst}^{(3)}$ [km s$^{-1}$]			&	514.2	&	513.4		&	513.0		&	512.8		&	513.8	&	513.9		&	518.0		&	513.3		&	 516.1			&	512.7	&	512.5		&512.9			\\
$x_c^{(4)}$ [$\alpha_{2000}$/arcsec]&	13:37:00.43&\fbox{+3.7}	&	\fbox{+4.6}	&	\fbox{+9.4}	&	+0.7	&	+0.7		&	+2.0		&	\fbox{0}	&	 \fbox{0}		&	+2.8		&	-1.6		&\fbox{0.0}		\\
$y_c^{(5)}$ [$\delta_{2000}$/arcsec]&	-29:51:56.2&\fbox{-1.0}	&	\fbox{-0.8}	&	\fbox{-4.8}	&	+2.0	&	+2.0		&	-6.5		&	\fbox{0}	&	 \fbox{0}		&	+3.0		&	-1.8		&\fbox{0.0}		\\
$PA^{(6)}$ [$^{\circ}$]				&	223.1	&	223.0		&	223.0		&	222.9		&	225.8	&	225.8		&	225.9		&	225.8		&	 225.9			&	225.0	&	225.0		&224.8			\\
$\Delta V_{\rm rot}^{(7)}$ [km s$^{-1}$]&	-1.4	&	0.0			&	0.9			&	1.6			&	5.9		&	9.8			&	\fbox{0}	&	10.7		&	 \fbox{0}		&	-1.6	&	-2.1 		&-2.3			\\
\hline
\end{tabular}
\end{center}
\begin{small}
$^{(1)}$ Reduced $\chi^2$ obtained in fitting the velocity model to the observed velocity field.\\
$^{(2)}$ Galaxy inclination.\\
$^{(3)}$ Heliocentric systemic LoS velocity.\\
$^{(4)}$$^{(5)}$ For column (A), right ascension $\alpha$ and declination $\delta$ of the kinematic centre. For the next columns: shift in $\alpha$ and $\delta$ (in arcsec) from the reference centre given in column A.\\
$^{(6)}$ Position angle of the major axis of the velocity field.\\
$^{(7)}$ Mean velocity difference between both sides of the rotation curve.\\
The parameters framed in a box are those that are fixed for the fit.\\[0.1cm]
$(A)$ Best Fit Model, all nine parameters are free to vary: five parameters for the disk ($incl$, $v_{\rm syst}$, $x_c$, $y_c$ and $PA$) and four for the two 2D-Plummer functions (not shown).\\  
$(B)$ All the parameters are free, except $x_c$ and $y_c$ that are fixed to the centre of the yellow box in Fig. \ref{fig:kinem_center} (i.e. the kinematic centre from Fabry-Perot data, \citealp{fathi08}). \\
$(C)$ All the parameters are free, except $x_c$ and $y_c$ that are fixed to the centre of the grey box in Fig. \ref{fig:kinem_center} (i.e. the Pa $\beta$ kinematic centre, \citealp{knapen10}).\\
$(D)$ All the parameters are free, except $x_c$ and $y_c$ that are fixed to the grey cross in Fig. \ref{fig:kinem_center} (i.e. the corner of the errorbox given by \citealp{knapen10} that is the furthest away from the centre determination obtained in model $(A)$).\\[0.1cm]
$(a)$ Same as model $(A)$.\\ 
$(b)$ All the parameters are free, except $incl$ (fixed to the value of stellar component)\\
$(c)$ All the parameters are free, except $incl$ (fixed to the value of stellar component) and $\Delta V_{\rm rot}$ (fixed to 0).\\
$(d)$ All the parameters are free, except $incl$, $x_{\rm c}$ and $y_{\rm c}$ (fixed to the value of stellar component).\\
$(e)$ Same as $(d)$, except $\Delta(V_{\rm rot})$ that is fixed.\\[0.1cm]
$(\alpha)$ Same as model $(A)$.\\
$(\beta)$ All the parameters are free, except $incl$ (fixed to the value of stellar component).\\
$(\gamma)$ All the parameters are free, except $incl$, $x_c$ and $y_c$ (fixed to the value of stellar component).\\
\end{small}
\label{table:kinematic_models}
\end{table*}

\begin{figure}
\center
\includegraphics[width =\linewidth]{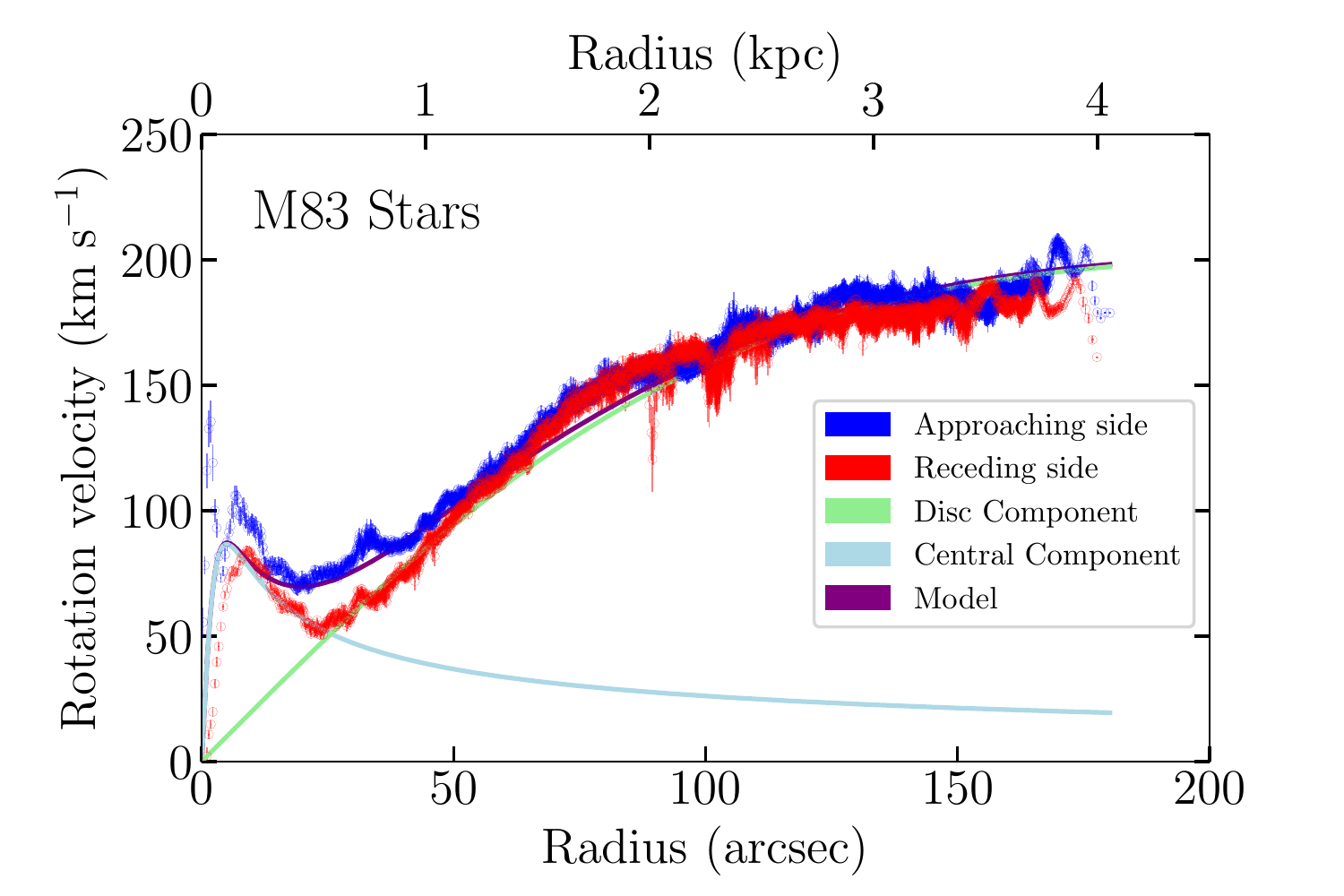}\\
\includegraphics[width =\linewidth]{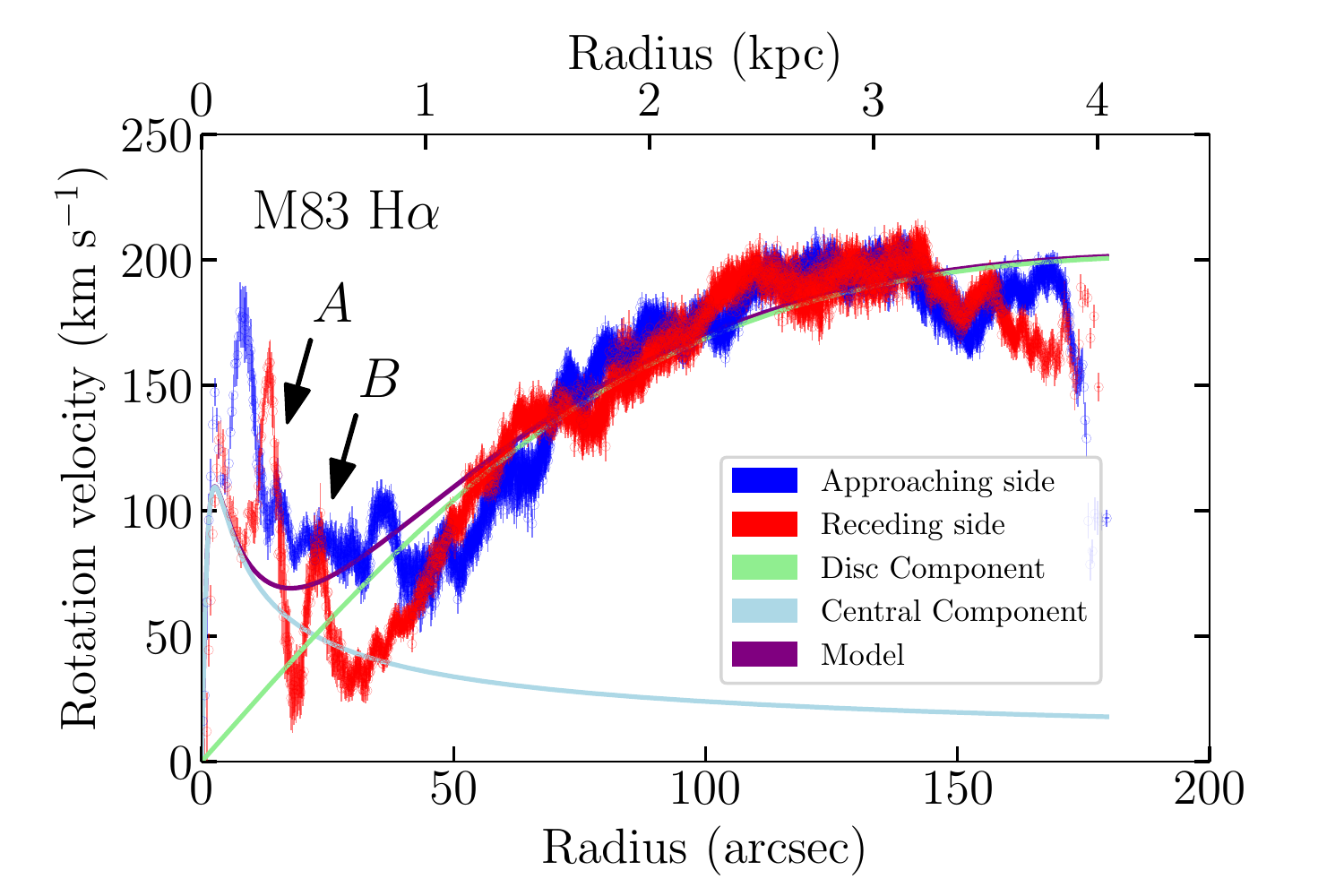}\\
\includegraphics[width =\linewidth]{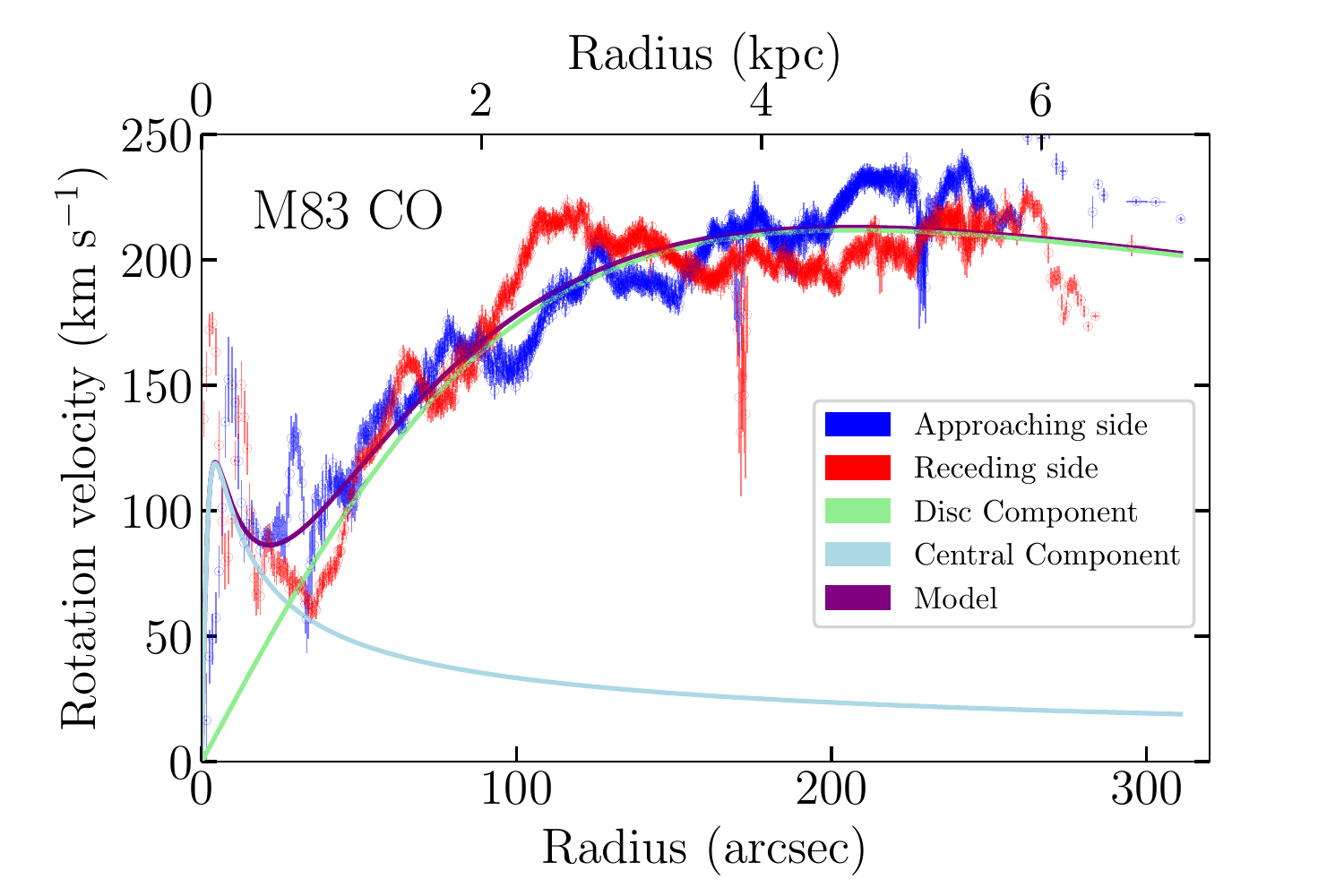}
\caption{Rotation curves for the stellar (top), H$\alpha$ (centre) and CO (bottom) velocity fields (models $(A)$, $(d)$ and $(\gamma)$ in Table~\ref{table:kinematic_models}). We remark on the larger radial extent of the CO data on the bottom plot. The contributions of the two best-fit Plummer components - modelling the galactic and circumnuclear disk - are shown in cyan and light green, and the sum of their contributions in purple. The red and blue curves correspond to the receding and approaching side of the velocity map. In Fig.~\ref{fig:rotcurve_extra}, we additionally show the H$\alpha$ rotation curve for model $(e)$, where we enforce a more symmetrical curve. In the middle panel, the radial location of features A and B discussed in Sect.~\ref{section:central_kinematics} is indicated.}
\label{fig:rotcurve}
\end{figure}

In order to derive the global kinematic parameters of M83, we fitted the velocity maps of stars and gas with the 2D fitting method described in \citet{epinat08} and based on the Levenberg-Marquardt non-linear least-square algorithm. In our work, the method has been upgraded using the MocKinG software\footnote{\url{https://gitlab.lam.fr/bepinat/MocKinG}.} to take into account uncertainties on the line of sight (LoS) velocities, the flux distribution, the linewidth of the profiles and the spatial resolution.

\subsubsection{Analysis of the stellar velocity field}
\par We fitted the raw stellar LoS velocity field with several two-dimensional theoretical velocity distributions, and found that the best fit is obtained using two Plummer components \citep{plummer1911}, one describing the galactic disk and the other tracing the central structure. The Plummer density profile has a finite density core and falls off as $r^{-5}$ at large radii; this very steep fall-off was essential to fit at best the central component.

The velocity field is described by nine free parameters (see Table \ref{table:kinematic_models}). They consist of four geometrical parameters -- the disk inclination ($incl$), the position angle of the major axis ($PA$), the location of the kinematic centre ($x_c$, $y_c$) -- plus the heliocentric systemic velocity of the galaxy ($v_{\rm syst}$). Additionally, each Plummer component has two free parameters which describe the velocity amplitude and the turnover radius, accounting for four other free parameters. We used an angular sector of inclusion of 67.5 degrees (in the galaxy plane) around the major axis\footnote{The reason for this choice is the fact that the LoS velocities along and close to the minor axis are dominated by radial motions (expansion and contraction along the radius) rather than by galactic rotation.}, and the LoS velocities were weighted according to their angular distance to the major axis by a cosine function, in order to minimise the contamination due to the predominance of radial motions around the minor axis.
Regardless of the initial values for the nine parameters, the best fit model converged towards the same solution (model $(A)$ presented in Table \ref{table:kinematic_models}). 
We assessed the robustness of the fit by masking low-S/N regions in the velocity map using Kernel filters, testing various several S/N thresholds (see also Appendix~\ref{section:appendix_kinem_fitting}). We found that even when reducing by a factor two the number of fitted spaxels, the output parameters are remarkably stable.
The best-fit inclination $incl$=20.3$\pm$0.1${\degr}$ is in excellent agreement with the one of 21${\degr}$ listed by \citet{gadotti19}, the position angle $PA$=223.1$\pm$0.1${\degr}$ is also very close to the one of 227${\degr}$ found by \citet{sheth10} and the systemic velocity $v_{\rm syst} = 514.2$ km s$^{-1}$ is only 6 km s$^{-1}$ higher than the one from LEDA \citep{paturel03, makarov14}. The best-fit centre is consistent - within uncertainties - with the Pa$\beta$ kinematic centre determined by \citet{knapen10} (see Fig.~\ref{fig:kinem_center}). Given the large number of spaxel used in the fit, the statistical uncertainties derived on the fitted parameters are very low; in order to study the robustness of the fit in Appendix~\ref{section:appendix_kinem_fitting} we test three additional models against model $(A)$, adopting different centres from literature values.
Using the parameters found with model $(A)$, we computed the rotation curve shown in the top panel of Fig. \ref{fig:rotcurve}, on top of which we overplot the contribution of the two Plummer components. The agreement between both sides of the rotation curve is very good for r $\gtrsim$ 50 arcsec (1.1 kpc). Within the first kpc, large discrepancies are observed, with the approaching side rotating faster than the receding one. Regardless of the chosen parameters, it is not possible to make the slopes on both sides of the galaxy coincide within the first 100 pc, indicating that the central stellar component is non-axisymmetric. The discrepancy at the very end of the rotation curve, on the other hand, is due to the decreasing S/N at the edge of the disk.

\subsubsection{Analysis of the ionised gas velocity field}
\par We fitted the observed H$\alpha$ velocity field with the same theoretical velocity distribution as for the stars. We tested several models, whose parameters are also given in Table \ref{table:kinematic_models}. The best-fit model is model $(a)$. However, the H$\alpha$ velocity field is so perturbed that the hypothesis of an axisymmetric disk used by the model provides a $\chi^2_r$ much higher than the one computed for the stellar disk, and does not allow to correctly determine the inclination of the galaxy, notoriously the most difficult parameter to fit \citep[e.g.][]{epinat08}. The problem persisted even when masking the central region.
To overcome this, we fitted a second model (model $(b)$) in which we fixed the inclination to the one of the stellar disk. This change only affected the $\chi^2_r$ parameter (which increases by $<$ 1\%), clearly indicating that the inclination is relatively decoupled from the other parameters.   
The bottom row of Table \ref{table:kinematic_models} shows the average difference between the two sides of the rotation curve. In the literature, the systemic velocity is often determined by fitting a one-dimensional rotation curve. In this case, the amplitude of each side of the curve is matched by adjusting the centre location. 
On the other hand, when fitting a two-dimensional velocity field as in this work, if the receding and approaching sides display an asymmetric LoS velocity distribution, a difference in rotational velocity ($\Delta V_{\rm rot}$) may appear between the two sides of the rotation curve.
We therefore fitted a third model (model $(c)$), where we additionally fixed $\Delta V_{\rm rot}=0$, in order to make the rotation curve more symmetric. This resulted in a moderate increase in $\chi^2_r$ (5\%).
Finally, we produced two more models (models (d) and (e)) in which we fixed the centre of rotation to the one determined from the stellar disk. This was motivated by the fact that the potential well of the galaxy is dominated by the mass of the stars and of the dark matter halo. In addition, this will facilitate the comparison between the stellar and the gaseous disks.
With respect to model $(d)$, in model $(e)$, we also fixed $\Delta V_{\rm rot}=0$. The only parameter marginally affected was the systemic velocity ($v_{\rm syst}$). We show the rotation curve corresponding to model $(d)$ in  Fig.~\ref{fig:rotcurve} (middle panel). In the Figure, we also indicate the radial location of the central kinematic features that will be discussed in Sect.~\ref{section:central_kinematics} (A and B, black arrows). In Fig.~\ref{fig:rotcurve_extra} of the Appendix, we include the H$\alpha$ velocity rotation curve produced with model $(e)$ as a comparison. 

\subsubsection{Analysis of the CO gas velocity field}
Finally, we fitted the CO data (`mom1 with prior' map\footnote{This map uses a high completeness, in order to cover the larger possible number of sightlines.}, see~\citealp{leroy21}) with the same two Plummer profiles, obtaining the best fit model $(\alpha)$ in Table~\ref{table:kinematic_models}. However, the best fit model exhibits a large inclination $incl$=35$\degr$, which is interestingly very close to the one derived from the best-fit H$\alpha$ model (37.9$\degr$); this indicates that the warm and cold gas kinematics are similar. For the same reasons mentioned above, in model $(\beta)$, we fixed the inclination of the CO component to the one of the stellar disk, leaving the other parameters free. None of the fitted parameters were affected by this; only the $\chi^2_r$ increases by $<$ 2\%.  For model $(\gamma)$, we also fixed the centre of rotation of the CO disk to that of the stars. This increased the $\chi^2_r$ by a further $\sim$7\% with respect to model $(\beta)$. To facilitate the comparison with the stellar and ionised gas component, in Fig.~\ref{fig:rotcurve} (bottom panel) we show the rotation curve obtained from model $(\gamma)$.
We remark that since the area spanned by the ALMA data is much more extended than that of MUSE, the CO rotation curve extends $\sim$ 50\% farther. The first kpc is as difficult to fit as for the H$\alpha$ component, but we observe a fairly good agreement between the warm and the cold gaseous component. The receding side of both components shows a larger velocity up to the end of the MUSE rotation curve,
Beyond this radius, the CO rotation curve indicates that the receding side velocities are lower than the approaching side ones.

\section{Extinction traced by the molecular and ionised gas}
\label{section:extinction}

\begin{figure*}
\center
\includegraphics[width=9.1cm]{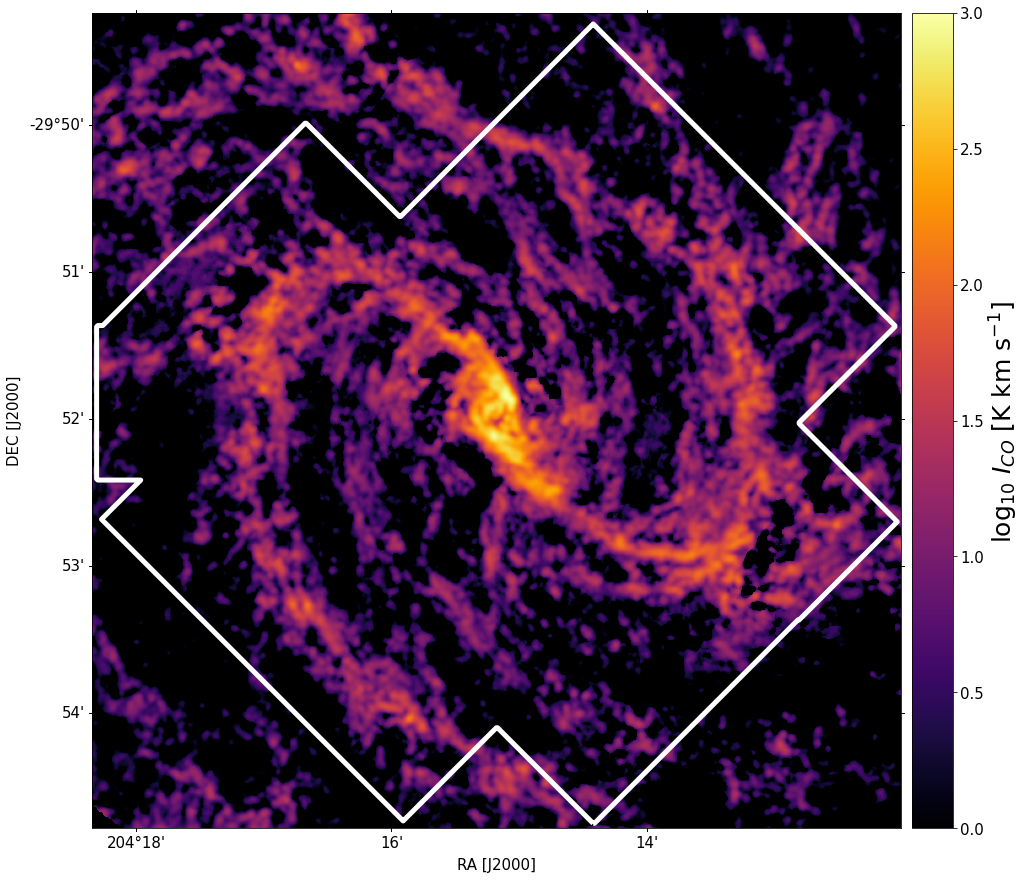}
\includegraphics[width=9cm]{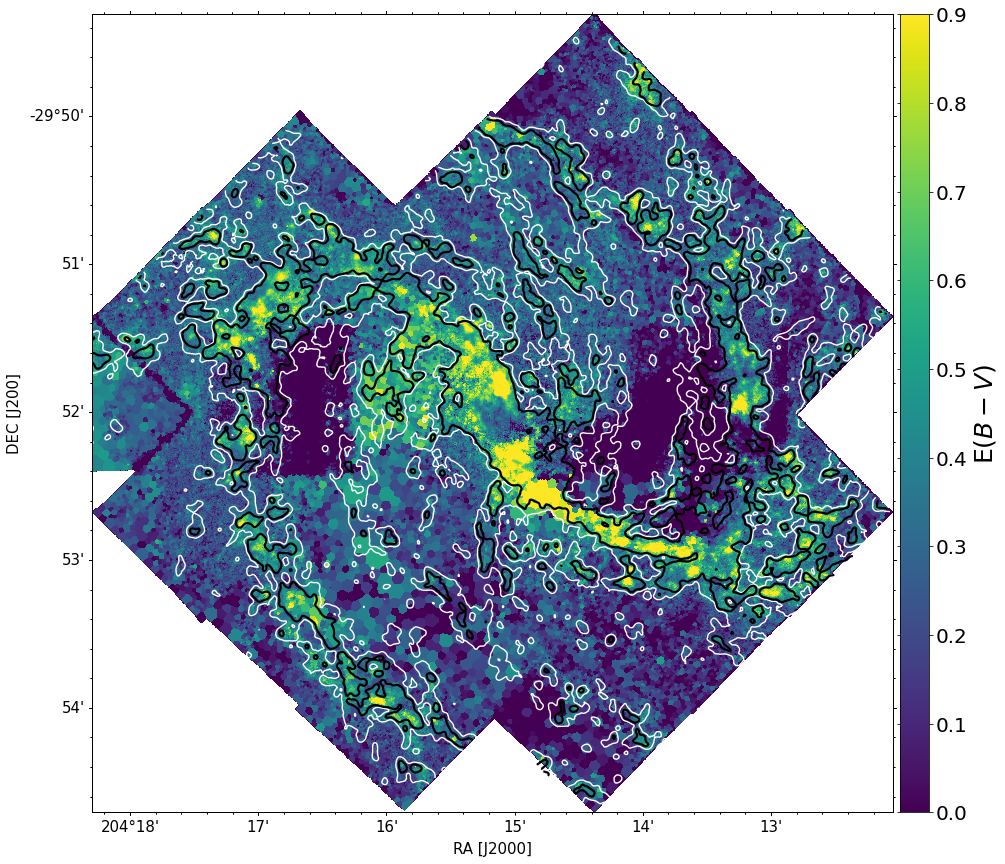}\\
\includegraphics[width=8.9cm]{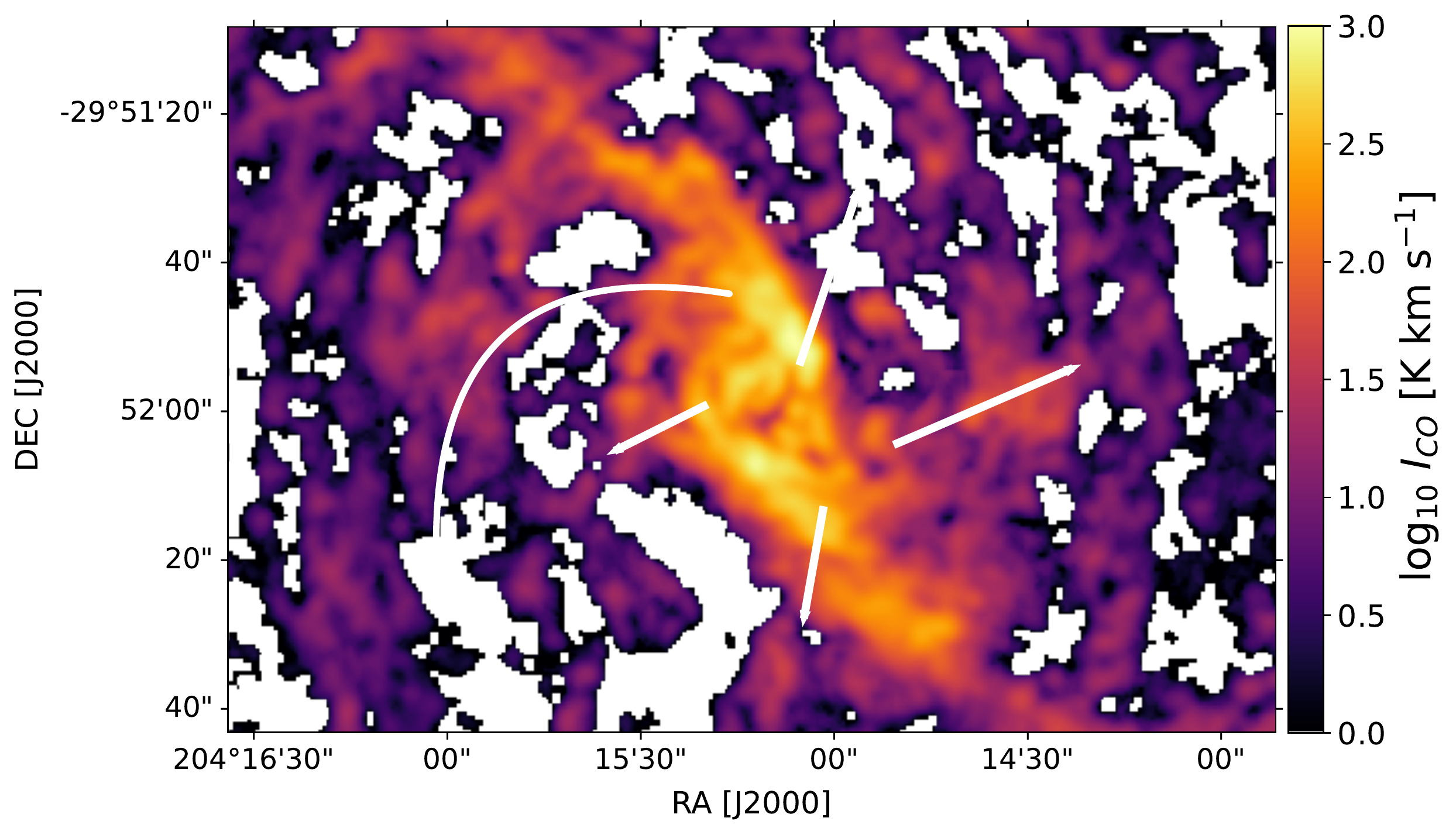}
\includegraphics[width=8.9cm]{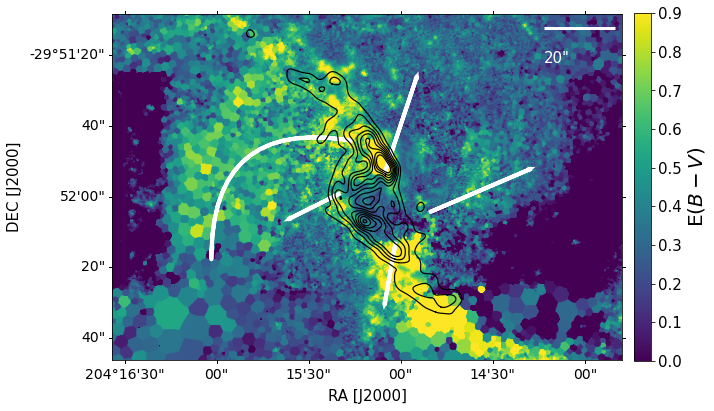}
\caption{Extinction traced by the ALMA and MUSE data. \textit{Left panels}: CO(2-1) molecular gas emission from ALMA. \textit{Right panels}: extinction map derived from the MUSE H$\beta$/H$\alpha$ ratio. The bottom panels shows a zoom-in into the central region, where the white lines mark the position of the features discussed in Sect.~\ref{section:starburst_region}. The contours in the top right panel indicate CO(2-1) emission at 5 (white) and 15 (black) K~\kms\ (corresponding to $\Sigma_{\rm mol} \sim$ 30 and 100 $M_\odot$ pc$^{-2}$); the black contours in the bottom right panel range from 90 to 800 K~\kms\ in steps of $\sim 90$ (corresponding to $\Sigma_{\rm mol}  \sim 600 - 5350~M_\odot$ pc$^{-2}$).
}
\label{fig:ebv_map}
\end{figure*}

\begin{figure*}
\center
\includegraphics[width=9.1cm]{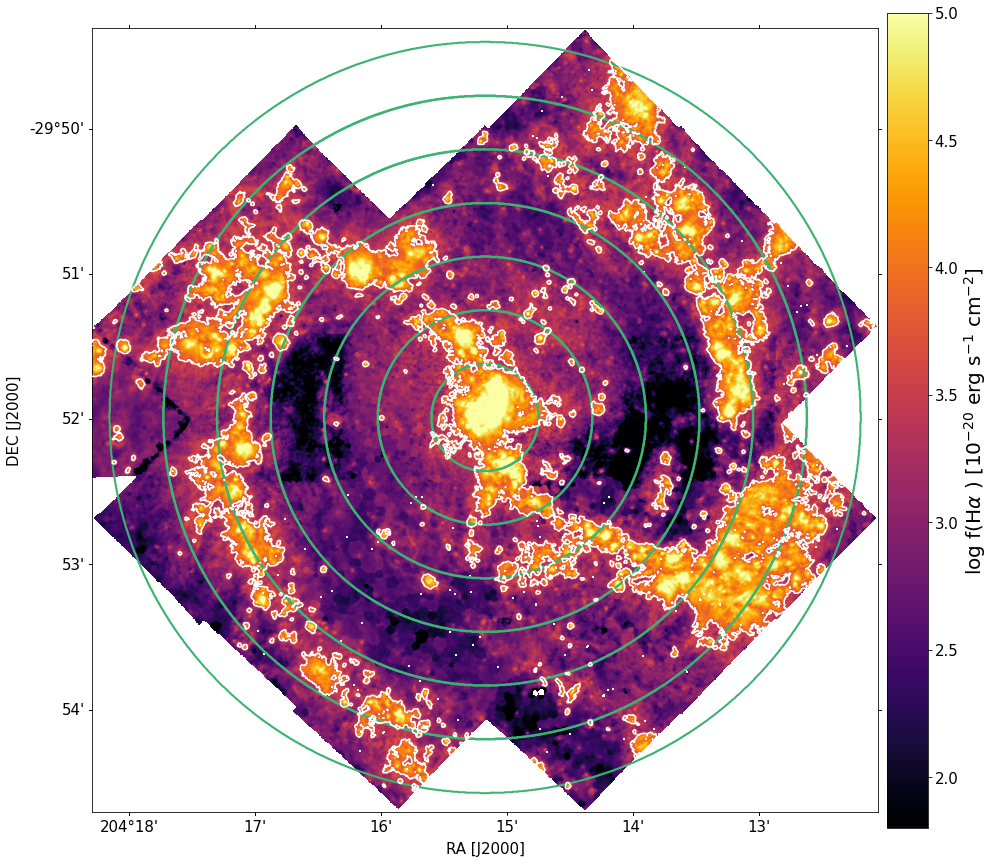}
\includegraphics[width=8.9cm]{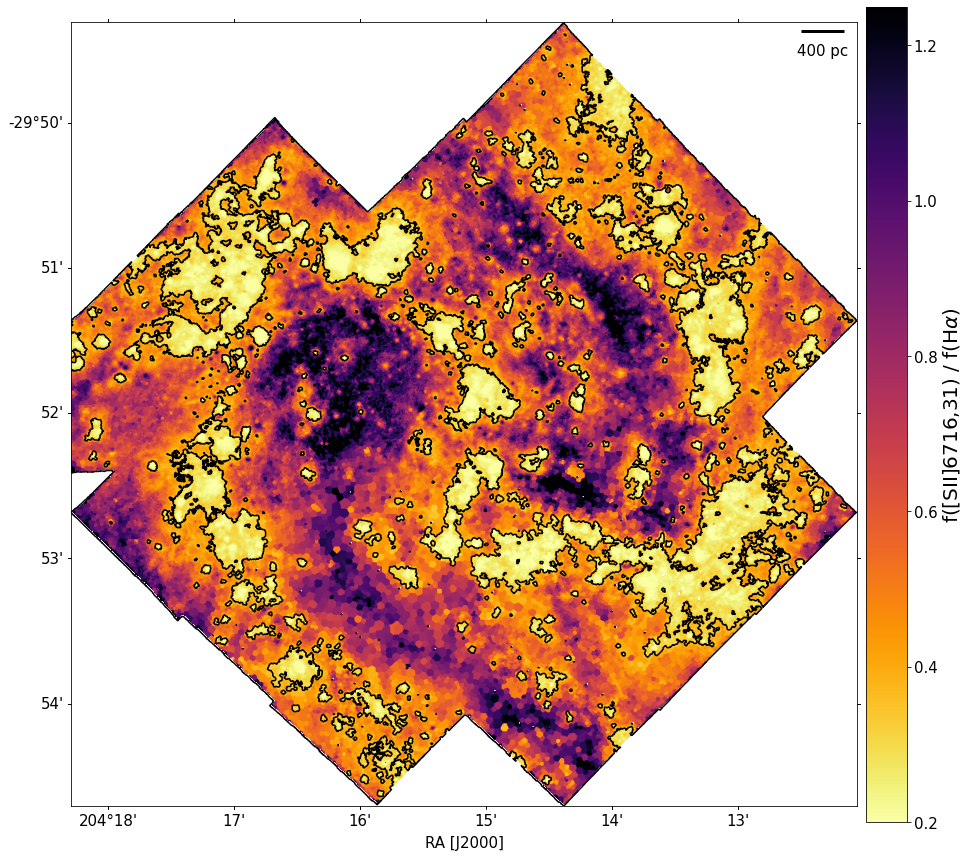}
\caption{Maps of H$\alpha$ (\textit{left panel}) and [\ion{S}{ii}]~$\lambda$6716,31/H$\alpha$ (\textit{right panel}) emission. Both maps have been corrected for extinction and deprojected. The contours indicate the outer limit of star-forming regions and correspond, respectively, to a cut in SB$_{\rm H\alpha} = 1.23 \times 10^{-15}$ erg s$^{-1}$ cm$^{-2}$ arcsec$^{-2}$ and [\ion{S}{ii}]/H$\alpha$ = 0.29. In green we indicate the annular sectors for which we compute the radial trends in Fig.~\ref{fig:fdig_vs_r}.}
\label{fig:halpha_map}
\end{figure*}

\begin{figure}
\center
\includegraphics[width=8cm]{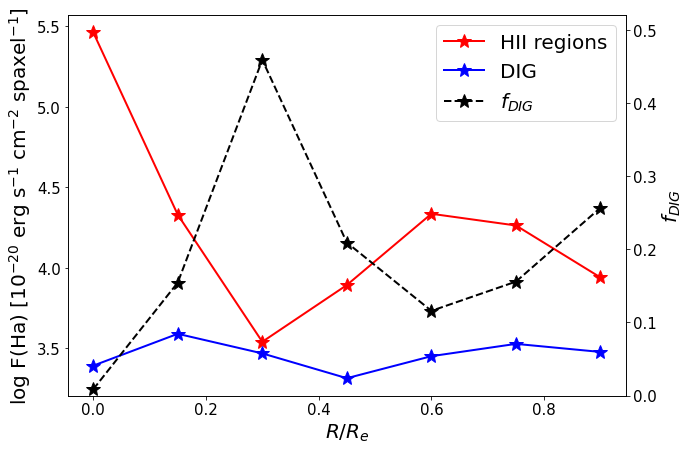}
\caption{H$\alpha$ normalised flux as function of radius for the \ion{H}{ii} regions (red) and DIG (blue), and resulting DIG fraction (black). The data are radially binned in annuli of width $dr = 0.15~R_e$ (shown in Fig.~\ref{fig:halpha_map}), and the flux has been normalised by the number of spaxels in each annulus. The \ion{H}{ii} regions flux has been corrected for diffuse emission.}
\label{fig:fdig_vs_r}
\end{figure}

We estimated the extinction from the H$\beta$/H$\alpha$ ratio with \textsc{Pyneb}~\citep{Luridiana15}. We assumed an intrinsic ratio H$\alpha$/H$\beta$ = 2.863 (corresponding to case B recombination with $T_e = 10^4$~K and $n_e$ = 100 cm$^{-3}$, \citealp{osterbrock}) and a \citet{cardelli89} extinction law. The H$\beta$/H$\alpha$ ratio map was obtained by spatially binning the gas cube to a S/N $\simeq$ 20 in H$\beta$ with the Voronoi tessellation technique. The lines were fitted with a single component Gaussian profile, as described in Sect.~\ref{section:kinematics_ionised_gas}.

\par The resulting extinction map is shown in Fig.~\ref{fig:ebv_map} (right panels).
Because dust and gas are usually well mixed \citep[e.g.][]{bohlin78},  we expect extinction and emission from gas to trace one another, modulo complications due to geometry and the presence of foreground and background sources, as observed for example with CALIFA by \citet{barrera_ballesteros20}. By comparing the extinction map with the CO intensity emission traced by ALMA (left panels in Fig.~\ref{fig:ebv_map}), we observe that indeed regions with high extinction correspond to dense molecular gas, with a surface density larger than $100~M_\sun$ pc$^{-2}$ (black contours in the top right panel Fig.~\ref{fig:ebv_map}). This is particularly evident along the spiral arms, where the extinction is clearly higher (E($B - V$) $\gtrsim$ 0.5) with respect to interarm region, due to the presence of dense gas. In the bottom right panel of Fig.~\ref{fig:ebv_map}, we show a zoom-in of the central region with CO emission overlaid in black on the extinction map. We observe two high density peaks in the molecular gas density distribution. These regions spatially coincide with the circumnuclear ring and dust inner bar reported in \citet{elmegreen98}, and also studied by \citet{callanan21} in dense gas tracers. The northern peak in CO coincides with the approaching part of the circumnuclear disk observed in Sect.~\ref{section:kinematics_stars}, and with a peak in the gas extinction map. The southern CO peak (receding part of the circumnuclear disk) shows very low extinction values. Given the inclination of the galaxy, we interpret this as a perspective effect on the vertical separation of the gas, due to the denser molecular gas residing deeper in the disk: the low extinction traced by the ionised gas would then be estimated based on line ratios from gas that is on top of the CO layer.

\section{Properties of the ionised gas}
\label{section:hii_dig}

\subsection{\ion{H}{ii} region identification and fraction of DIG}

In Fig.~\ref{fig:halpha_map} (left panel), we show a map of the intensity of the H$\alpha$ line, obtained by fitting the line in the gas cube spatially binned to a S/N $\simeq$ 20 in H$\alpha$.
We separated the H$\alpha$ emission into \ion{H}{ii} regions and DIG using the Python package \textsc{astrodendro}\footnote{\url{https://dendrograms.readthedocs.io}.}. We remark that within the scope of this work we only require the outer boundaries of star-forming complexes. The detection of individual \ion{H}{ii} regions will be presented in an upcoming work (Della Bruna et al., in prep.). The data were organised into a hierarchical tree structure of a given depth and minimum leaf size. We set the minimum size of the leaves based on the typical PSF measured at H$\alpha$ (FWHM = 4.3 pixels $\equiv 0\farcs86$) and we fixed the optional \texttt{min\_delta} parameter (minimum difference in flux between two separate structures) to zero. We set the depth of the tree to a surface brightness (SB) threshold corresponding to an \ion{H}{ii} region ionised by a single low-luminosity O star. We determined this low luminosity threshold using the models of \citet{martins05} (Table 1 in their work), which predict an ionising photon flux $\log Q(H^0) = 47.56$ photons s$^{-1}$ for a O9.5 class V star. In the case B approximation\footnote{Case B recombination assumes that photons emitted during recombination are immediately reabsorbed, so that recombinations directly to $n = 1$ are ignored.}, the H$\alpha$ luminosity is related to the ionising photon flux as
$$L(H\alpha)~[\mbox{erg s}^{-1}] = \frac{\alpha^{eff}_{\rm H\alpha}}{\alpha_B} h \nu_{\rm H\alpha} \cdot Q(H^0)~[\mbox{s}^{-1}],$$
where $\alpha^{eff}_{\rm H\alpha}$ is the effective recombination coefficient at H$\alpha$ and $\alpha_B$ is the case B recombination coefficient. Assuming an electron temperature and density $T_e \sim \num{10 000}$ K and $n_e = 10^3~\mbox{cm}^{-3}$, this gives \citep{draine}:
$$L(H\alpha) = 1.37 \times 10^{-12}~Q(H^0).$$
We obtained thus $L(H\alpha) = 4.97 \times 10^{35} \mbox{erg s}^{-1}$, corresponding to a SB threshold of 1.23 $\times 10^{-15}$ erg s$^{-1}$ cm$^{-2}$ arcsec$^{-2}$, where we assumed an \ion{H}{ii} region diameter of 10 pc (lower limit estimate). We note that our SB cut is slightly deeper than the one applied in the recent work by \citet{poetrodjojo19}, which used a cut-oﬀ of $1.86 \times 10^{-15}$ erg s$^{-1}$ cm$^{-2}$ arcsec$^{-2}$ for their \ion{H}{ii} regions sample. The outer contours of the resulting tree are shown in white in Fig.~\ref{fig:halpha_map} (left panel). By inverting the \ion{H}{ii} mask obtained from the dendrogram, we found an (extinction corrected) DIG fraction $f_{\rm DIG} = F(H\alpha)_{\rm DIG}/F(H\alpha)_{\rm TOT} \sim$ 13\%. In order to get a first order correction for DIG contamination, we obtained the distribution of the H$\alpha$ emission over the entire FoV, and estimated its median value via sigma clipping. We then subtracted this value from the \ion{H}{ii} regions emission.
Correcting for the diffuse emission coincident with \ion{H}{ii} regions results in a negligible increase of $\lesssim 0.1 \%$.

\par We also estimated the fraction of DIG based on a cut in [\ion{S}{ii}]/H$\alpha$, as recently done by \citet{kreckel16} and \citet{poetrodjojo19}. The advantage of this ratio over a simple cut in H$\alpha$ SB is that it is sensitive to the ionisation state of the gas, as shown in Fig.~\ref{fig:halpha_map} (right panel): a high ratio (in black) traces regions where the ionising photons have $I_{\rm S^+}$ = 10.4 eV $< h\nu < $ 13.6 eV = $I_{\rm H^+}$, where $I$ denotes the ionisation potential. An intermediate ratio (orange to purple) indicates gas with $I_{\rm H^+}$ = 13.6 eV $< h\nu < $ 23.3 eV = $I_{\rm S^{++}}$ and a low ratio (yellow to orange) marks regions where sulfur is largely doubly ionised ($h\nu > $ 23.3 eV). As remarked by \citet{kreckel16}, a cut in [\ion{S}{ii}]/H$\alpha$ is better to detect fainter regions, but the generally lower S/N of the [\ion{S}{ii}] lines can result in irregular boundaries. Fig.~\ref{fig:halpha_map} (right panel) shows a map of the [\ion{S}{ii}]~$\lambda$6716,31/H$\alpha$ ratio, obtained by fitting the emission lines in the gas cube tessellated to S/N $\simeq$ 20 in the [\ion{S}{ii}]~$\lambda$6731 line.
We observe that the ratio is enhanced in the interarm regions, consistent with DIG observations in the Milky Way and in nearby galaxies \citep{madsen06, haffner09}. If the DIG is solely ionised by radiation leaking from \ion{H}{ii} regions, the enhanced ratios can be explained with the fact that photons escaping from density bounded regions have a harder spectrum between the \ion{H}{i} and \ion{He}{i} ionisation energies, due to partial absorption, and a softer spectrum at shorter wavelengths, as has been shown for instance by the simulations of a stratified ISM from \citet{wood04}. In their work, \citet{poetrodjojo19} applied a cut [\ion{S}{ii}]~$\lambda$6716,31/H$\alpha$ = 0.29, based on typical ratios observed in \ion{H}{ii} regions and DIG in the MW \citep{madsen06}.
We adopted the same limit, which results in the contours shown in Fig.~\ref{fig:halpha_map} (right panel). We observe that the regions contours are similar but somewhat more conservative than the one selected from the H$\alpha$ map. We recovered a (reddening corrected) $f_{\rm DIG} \sim$ 20\%; also in this case correcting for the diffuse emission coincident with \ion{H}{ii} regions has a negligible impact ($\lesssim 0.01 \%$). The recovered value is somewhat lower than the value of 30\% estimated by \citet{poetrodjojo19} in M83 in the radial range $R \leq 2 R_e$ (using the distance and $R_e$ adopted in this work).
The discrepancy could be due to the more accurate fitting of the stellar continuum in the MUSE data or to the difference in spatial coverage, depth and spatial and spectral resolution of the two datasets. For the remainder of this work, we adopt the \ion{H}{ii} region contours selected on the H$\alpha$ map as a reference .

\par In Fig.~\ref{fig:fdig_vs_r}, we show the radial flux profile for the \ion{H}{ii} regions and the DIG. The profile was obtained by radially binning the reddening corrected and de-projected H$\alpha$ map in annular sectors of width $dr = 0.15~R_e$ (indicated in green in Fig.~\ref{fig:halpha_map}, left panel), and matching it with the \ion{H}{ii} regions contours. The DIG flux was corrected for the diffuse background emission as described above. The first point refers to the emission in the starburst region, $ 0 \leq R \leq 0.15~R_e$. We see that the luminosity of the \ion{H}{ii} regions is highest in the centre, due to the starburst activity, and then reflects the configuration of the spiral arms. The SB of the DIG follows a similar trend, although the minimum is offset by 0.15 $R_e$. The resulting $f_{\rm DIG}$ ratio is shown in black in Fig.~\ref{fig:fdig_vs_r}; we see that the DIG contribution to the total luminosity varies between 0.8\% and 46\%, peaking in the interarm region. The radial trends are in good agreement with what observed by \citet[][Fig. 6 in their work]{poetrodjojo19}\footnote{We note that using the distance and $R_e$ adopted in \citet{poetrodjojo19}, the MUSE data presented in this work only span the range $R \leq 0.8~R_e$ in their Fig.~6.}.

\begin{figure}
\center
\includegraphics[width=8cm]{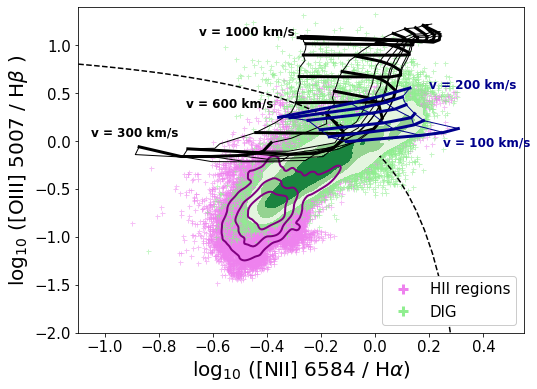}\\
\includegraphics[width=8cm]{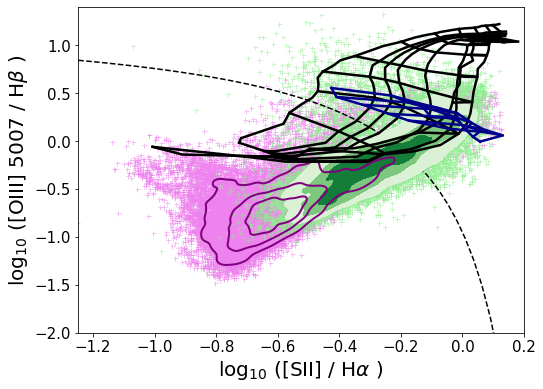}\\
\includegraphics[width=8cm]{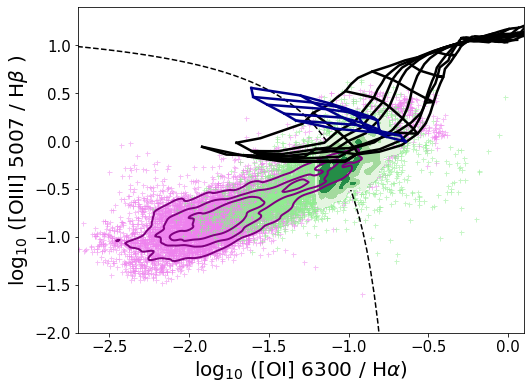}
\caption{BPT emission line diagrams. Each point corresponds to a Voronoi bin, and is colour-coded as located in an \ion{H}{ii} region (pink) or in the DIG (green). The contours overplotted on the data correspond to iso-proportions of the density, with a probability mass $\leq$ 0.25, 0.5, 0.75 and 1. The data are compared to the extreme starburst line from \citet[][black dashed line]{kewley01}, the fast shock models from \citet[][black grid]{allen08}, and slow shock models from \citet[][blue grid]{rich11}.}
\label{fig:BPT_diagrams}
\end{figure}

\begin{figure*}
\center
\includegraphics[width=8cm]{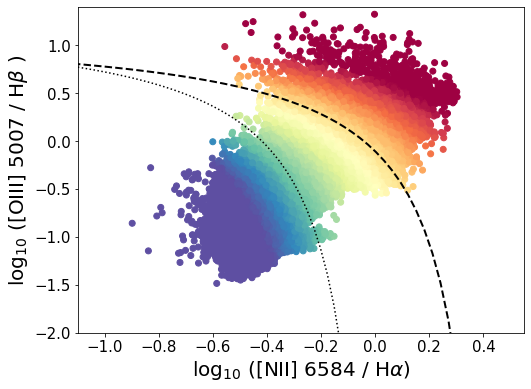}\\
\includegraphics[width=13cm]{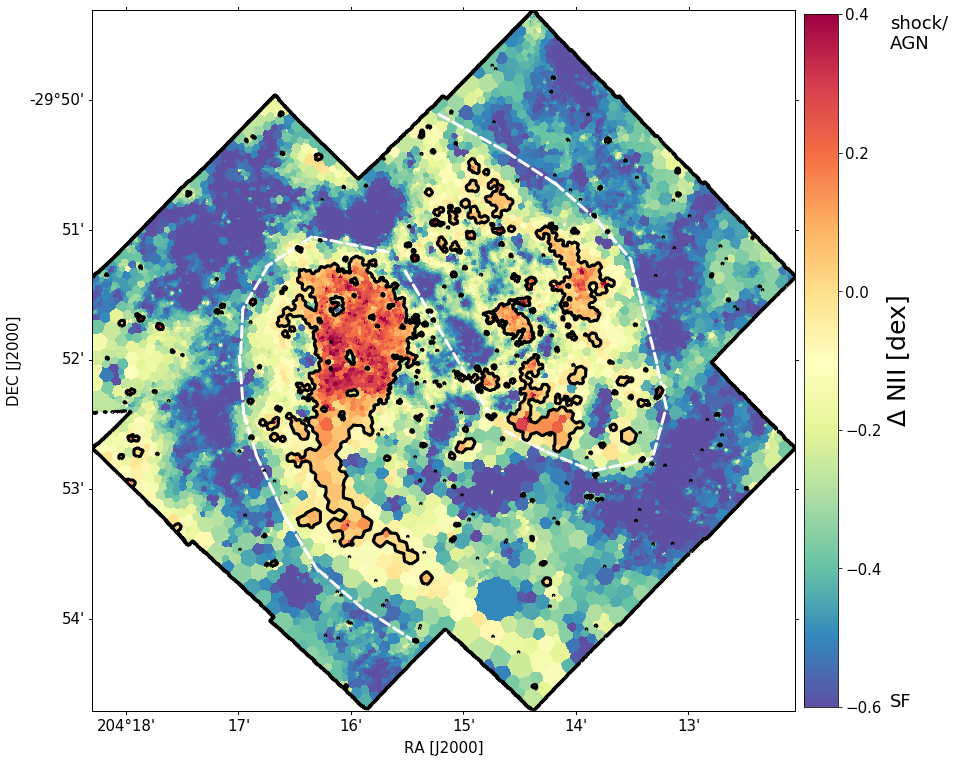}
\caption{1D (top panel) and 2D (bottom panel) \ion{N}{ii}-BPT diagram, with each point colour coded according to its distance from the extreme starburst line of \citet[][black dashed line in the top panel]{kewley01}. The distance ($\Delta$~\ion{N}{ii}), ranges from dark blue for pure SF to red for pure shock or AGN emission. In the top panel, we also show the empirical demarcation line from \citet[][black dotted line]{kauffmann03}, indicating a more stringent limit for photoionised gas.
In the bottom panel, the black contours correspond to $\Delta$~\ion{N}{ii} = 0 and white dashed lines indicate the position of the bar and spiral arms.}
\label{fig:2D_bpt}
\end{figure*}

\subsection{BPT diagram analysis}
\label{section:BPT}

We studied the physical conditions of the ionised gas in `BPT' emission line diagrams \citep{baldwin81, veilleux87}. This set of diagrams offers a powerful tool to interpret ionised gas emission, as the location in the diagram is sensitive to parameters such as the electron density, metallicity, strength and hardness of the radiation field \citep[see e.g. the review by][]{kewley19}.
We inspected the full set of diagrams, showcasing [\ion{N}{ii}]/H$\alpha$, [\ion{S}{ii}]/H$\alpha$ and [\ion{O}{i}]/H$\alpha$ as function of [\ion{O}{iii}]/H$\beta$. In each diagram the fluxes were obtained by single component Gaussian fitting of the emission lines in the gas cube tessellated to a S/N = 20 in the weakest line of interest ([\ion{O}{iii}]~$\lambda$5007 for the N2- and S2-BPT diagrams, and [\ion{O}{i}]~$\lambda$6300 for the O1 diagram).
Typical Voronoi bin sizes in the \ion{H}{ii} regions vs the DIG are: $0\farcs3$ vs $0\farcs9$ in the \ion{O}{iii} binning and $0\farcs7$ vs 2\arcsec{} in the \ion{O}{i} binning. We caution that bins located in \ion{H}{ii} regions are typically below the seeing limit, especially in the blue end of the spectrum (where PSF sizes range between 0.7 -- 0.9 arcsec). However, this should not have a relevant impact within the scope of this work.  

\par The resulting diagrams are shown in Fig.~\ref{fig:BPT_diagrams}: each point corresponds to a Voronoi bin, and is colour-coded as located in an \ion{H}{ii} region (purple) or in the DIG (green). The black dashed line indicates the location of the `extreme starburst' line by \citet{kewley01}, corresponding to the upper limit for gas excited purely by SF. Emission above this limit likely originates from shocks or active galactic nuclei (AGN) activity. In Fig.~\ref{fig:2D_bpt} (top panel) we show again the N2-BPT diagram, with each point colour-coded according to its orthogonal distance from the extreme starburst line ($\Delta$~\ion{N}{ii}). $\Delta$~\ion{N}{ii} increases from the bottom left (dark blue, purely SF gas) towards the top right corner (orange-red, purely ionised by shocks or AGN) of the diagram. In the Figure, we additionally show the empirical line of \citet[][black dotted line]{kauffmann03}, denoting a more stringent limit for gas excited by pure photoionisation. Points located between this line and the extreme starburst line (yellow-green) are likely excited by a mix of SF and shocks or AGN.
In the bottom panel of Fig.~\ref{fig:2D_bpt} we show the corresponding `2D-BPT' diagram, where each spatial bin is colour-coded according to $\Delta$~\ion{N}{ii}. We observe that the spiral arms regions stand out as purely SF, the diffuse gas immediately surrounding the regions shows a composite emission, and some of the interarm regions - especially at 0.2 -- 0.4~$R_e$ - show a clear signature of shocks.

\par Finally, we assessed the overall fraction of H$\alpha$ luminosity originating from SF (regions with $\Delta$~\ion{N}{ii} $\leq$ 0) and shocks ($\Delta$~\ion{N}{ii} > 0). We observe that, as expected, in \ion{H}{ii} regions SF accounts for most of the H$\alpha$ flux (99.8\% of the flux originates from regions where photoionisation is the dominant mechanism), whereas bins classified as DIG have a mixed contribution from both photoionisation-dominated regions (accounting for 94.9\% of the flux) and shock-dominated regions (accounting for the remaining 5.1\%). 

\par We compared our observations with models of fast and slow shocks (black and blue grid in Fig.~\ref{fig:BPT_diagrams} and ~\ref{fig:2D_bpt}). The black grid corresponds to the fast shock models of \citet{allen08}. We use the models that include the photoionising shock precursor, with metallicity $Z = 2~Z_\odot$ and electron density $n_e = 1$ cm$^{-3}$. We show the model grid spanning $b$ = (0.001 -- 100) $\mu$G in magnetic field strength and $v_s$ = 290 -- 1000 km/s in shock velocity. The blue grid displays the slow shock models described in \citet{farage10} and \citet{rich11}. These models describe shocks driven into the galactic disk by ram pressure originating as a cloud of cool gas (possibly a filament from a merger remnant) falls into the hot ISM halo of a galaxy.
The full model grid covers the range $12 + \log(\mbox{O/H})$ = 7.39 -- 9.39 ($\sim$ 0.05 -- 5 $Z_\odot$) in metallicity and $v_s$ = 100 -- 200 km/s in shock velocity; we show however only the super-solar range, $Z \geq 8.69$. We observe that the points located beyond the SF limit overlap with shock model grids. In particular in the [\ion{N}{ii}] diagram, we observe a cloud of points with log[\ion{N}{ii}]/H$\alpha > 0$ that only overlaps with the slow shock models: we investigate more closely these regions in Sect.~\ref{section:starburst_region}, where we study the kinematics and ionisation state of the starburst region.

\section{Analysis of the starburst region}
\label{section:starburst_region}

\begin{figure}
\center
\includegraphics[width=7.6cm]{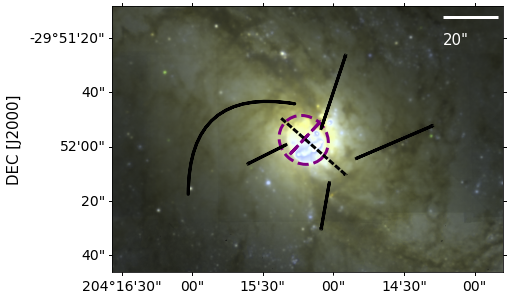}
\caption{Morphology of the M83 starburst region. The background image shows a zoom-in into the three colour stellar composite from MUSE (Fig.~\ref{fig:rgb_muse}). The purple dashed lines indicate the schematic location of the outer circumnuclear ring and dust inner bar from \citep{elmegreen98}. The brightest region of star formation (`starburst arc') is visible inside the ring. Black lines indicate the location of the kinematic features discussed in Sect.~\ref{section:starburst_region}. The scalebar of 20\arcsec{} corresponds to $\simeq$ 500 pc at the distance of our target.}
\label{fig:central_region}
\end{figure}

We now analyse in more detail the kinematics and ionisation state of the starburst region, indicated by a black rectangle in Fig.~\ref{fig:rgb_muse} (left panel) and shown in more detail in Fig.~\ref{fig:central_region}. In the Figure, we outline the approximate location of the outer dust ring (of radius $\simeq$ 9\arcsec) and inner bar as determined by \citet{elmegreen98} (purple lines) and mark the kinematic features that will be discussed in the rest of this Section (black lines).

\subsection{Kinematics of the central starburst region}
\label{section:central_kinematics}

\begin{figure*}
\center
\includegraphics[width=16cm]{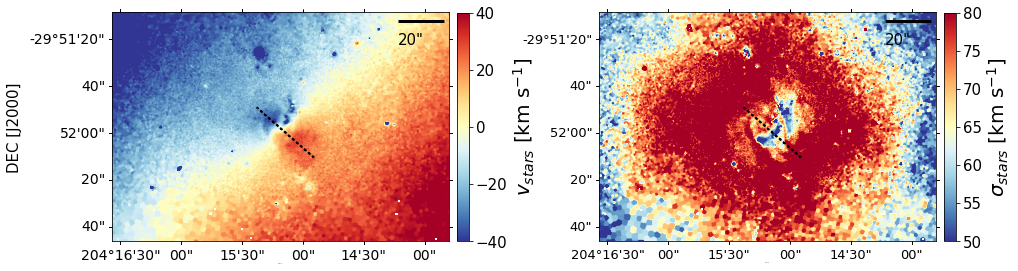}
\includegraphics[width=16cm]{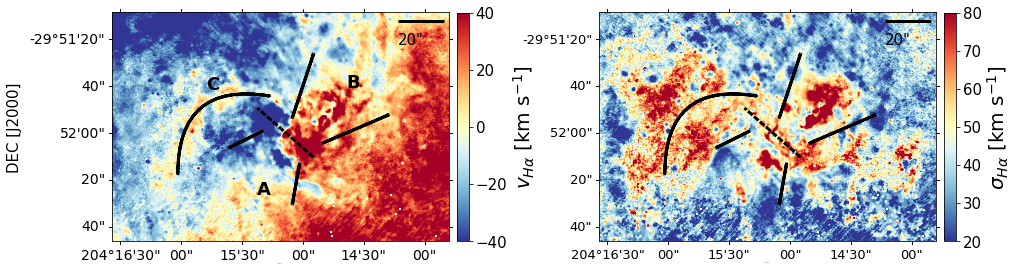}
\includegraphics[width=16cm]{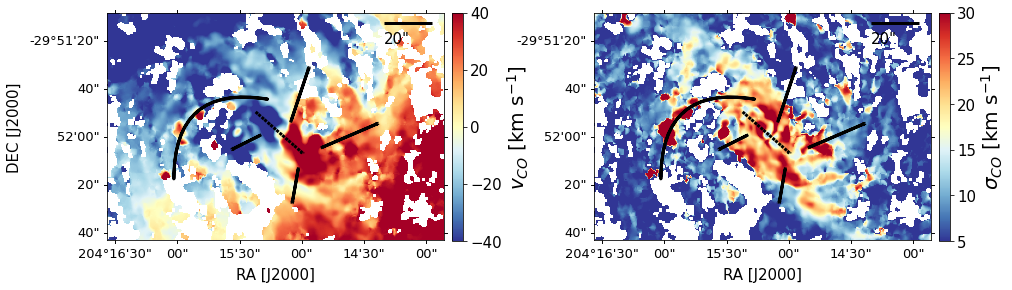}
\caption{Close up view of Fig.~\ref{fig:kinematics} showcasing the kinematics of the starburst region. All velocity maps (left panels) have been corrected for the systemic velocity; no inclination correction has been applied. Black lines mark the kinematic features discussed in Sect.~\ref{section:starburst_region}.}
\label{fig:kinematics_zoomin}
\end{figure*}

\begin{figure*}
\center
\includegraphics[width=9.2cm]{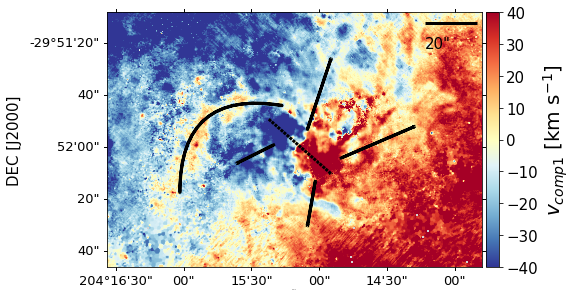}
\includegraphics[width=8.8cm]{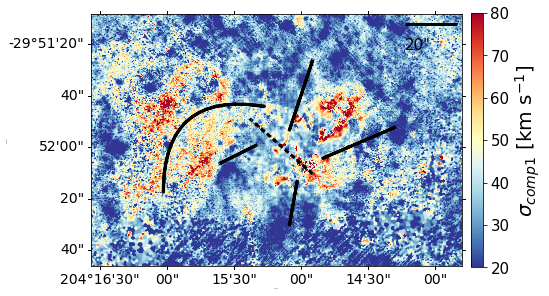}\\
\includegraphics[width=9.2cm]{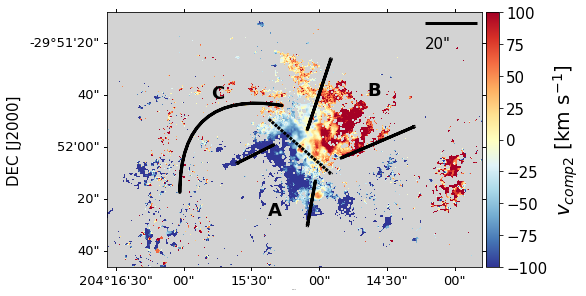}
\includegraphics[width=8.8cm]{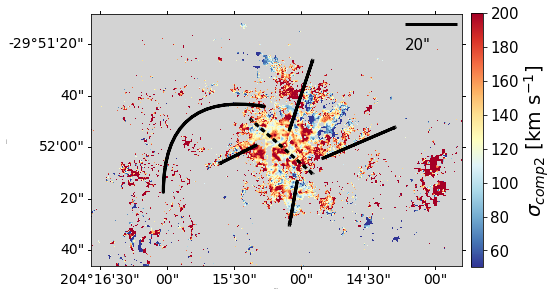}
\includegraphics[width=9.1cm]{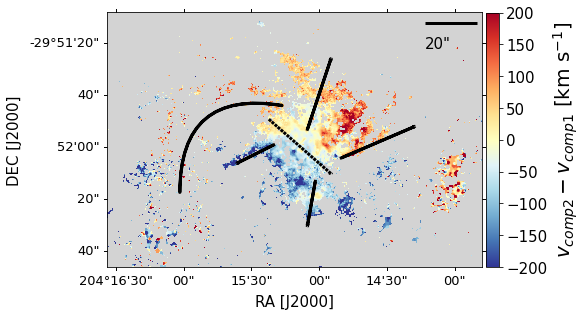}\\
\includegraphics[width=0.3\textwidth]{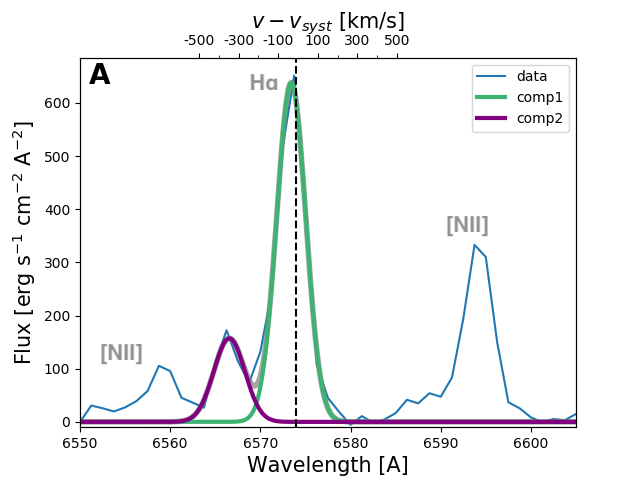}
\includegraphics[width=0.3\textwidth]{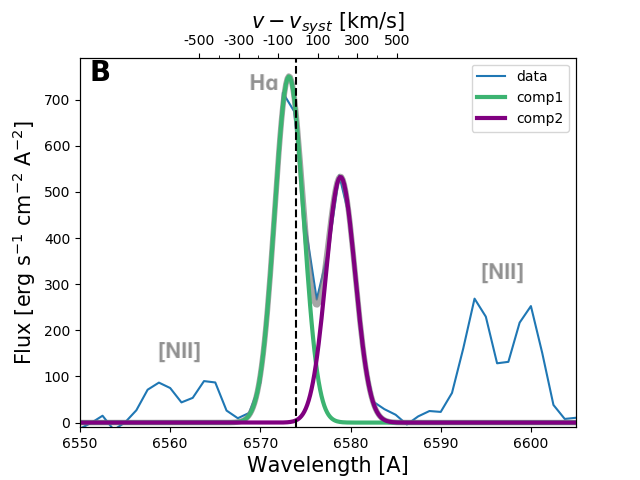}
\includegraphics[width=0.3\textwidth]{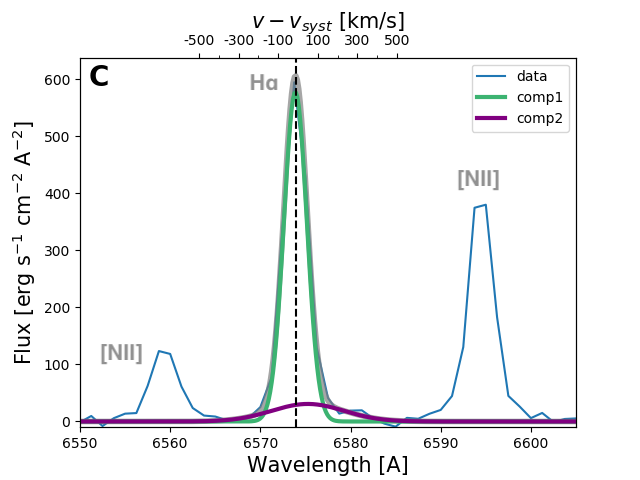}
\caption{Results from the 2-component Gaussian fit to the H$\alpha$ line. 
\textit{First and second row}: H$\alpha$ velocity (left panels) and velocity dispersion (right panels) maps, of the two components. The velocity maps have been corrected for the systemic velocity; no inclination correction has been applied. The velocity dispersion has been corrected for instrumental effects. We draw attention on the different velocity and dispersion scales spanned by the first and second component. \textit{Third row}: velocity difference between the two components (blue and red indicate, respectively, a second component that is more strongly blueshifted and redshifted with respect to the first one). Black lines mark the kinematic features discussed in Sect.~\ref{section:starburst_region}.
\textit{Bottom row}: typical spectra of features A and B and C (see labels in the figure). In panels A and B, both the H$\alpha$ line and the [\ion{N}{ii}] doublet are clearly doubly peaked, with $\Delta v \sim 300$~\kms. In panel C, both components are at similar velocity, but the second component has an extremely high $\sigma \sim$ 140~\kms.
}
\label{fig:2compfit_kinem}
\end{figure*}

In Fig.~\ref{fig:kinematics_zoomin} we show a close up view of the kinematics of the central region. In Fig.~\ref{fig:2compfit_kinem} we additionally inspect the ionised gas kinematics obtained from the 2-component Gaussian analysis. Grey shaded areas in the maps of the second component indicate regions in which the line was best fit by a single component (see description of the fitting method in Sect.~\ref{section:kinematics_ionised_gas}).

\par In the stellar kinematics (top panels in Fig.~~\ref{fig:kinematics_zoomin}) we observe a fast rotating nuclear component ($\sim$~30\arcsec{} $\simeq$ 700 pc in diameter), already reported by \citet{gadotti20}. We stress that the alignment of the rotation axis of this inner component (black dashed line) and the stellar bar is purely coincidental, and is expected to significantly vary over secular timescales. We also observe a dip in velocity dispersion ($\sigma < 60$~\kms) at the location of the starburst arc and along the dust lane west of the arc. Similar `central dispersion drops' were reported by \citet{emsellem01} and \citet{emsellem04} as being possibly due to a dynamically cold stellar component that has formed as a consequence of a bar-driven gas accretion episode.
\par The ionised gas kinematics are -- as already observed in Sect.~\ref{section:kinematics} -- more complex.
In the velocity map of the single component fit (centre left in in Fig.~\ref{fig:kinematics_zoomin}) and in the first component of the double Gaussian fit (top left in Fig.~\ref{fig:2compfit_kinem}), we observe a similar signature of a rotating circumnuclear disk, as already traced by the stellar kinematics.
On the east of the nucleus, we observe a stream of gas (labelled as feature C), extending for 50\arcsec ($\simeq$ 1250 pc) and having a velocity difference $\Delta v \simeq + 30$~\kms\ with respect to the surrounding disk rotation. The stream is surrounded by an extended region ($\sim 1000 \times 1600$ pc) having an enhanced velocity dispersion $\simeq$ 60 -- 80~\kms.
This feature was already reported in molecular gas by \citet{lundgren04} and in ionised gas by \citet{fathi08} as a potential inflowing stream of gas into the central starburst. Also \citet{piqueras_lopez12} observe a global velocity gradient in the central region that is pointing to the possible presence of an inflow.

\par Surrounding the nucleus in the map of the second velocity component (second row on the left in Fig.~\ref{fig:2compfit_kinem}), we also observe two conical features on each side of the stellar bar (labelled as features A and B). The two cones are 20\arcsec $\simeq$ 500 pc in size, have a $\Delta v \simeq \pm 100$~\kms\ on top of the disk rotation and a high velocity dispersion $\sigma_{\rm comp2} \gtrsim$ 80~\kms (second row on the right in Fig.~\ref{fig:2compfit_kinem}). Cone A appears blueshifted along our LoS, while cone B appears redshifted. In Fig.~\ref{fig:ebv_map} (bottom panels), we also observed that at the location of cone A there is a peak in the molecular gas emission, whereas the ionised gas traces low extinction. We interpreted this as the fact that the CO emitting gas is located `behind' the ionised gas along our LoS; together with the fact that cone A appears blueshifted, this could indicate that the gas is moving towards us. On the other hand, we do not see a mismatch between CO and E($B - V$) along cone B, which might indicate that the ionised gas is moving away from us.
Features A and B also clearly stand out in the H$\alpha$ rotation curve\footnote{Both on the receding side, due to their position with respect to the angular sector used to compute the curve.}, as remarked in Fig.~\ref{fig:rotcurve} (black arrows). On the third and fourth row of Fig.~\ref{fig:2compfit_kinem}, we show the velocity difference between the two Gaussian components, as well as three example spectra, corresponding to features A, B and C. We observe that at the centre of the two cones, the velocity difference between first and second component reaches 200~\kms.

\begin{figure}
\center
\includegraphics[width=8.9cm]{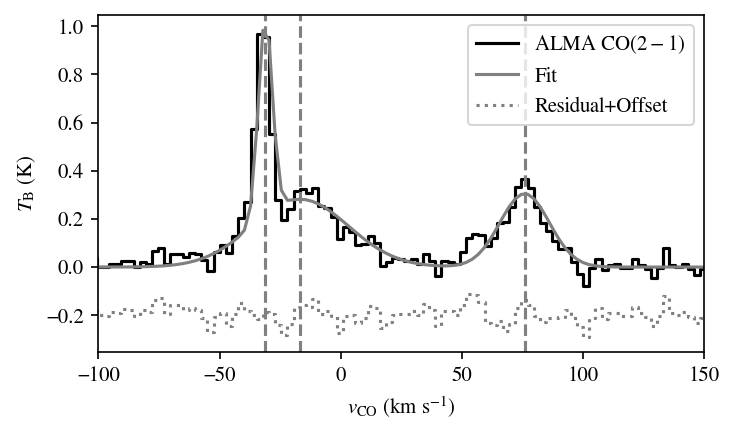}
\caption{ALMA CO(2-1) spectrum of one of the CO `blobs' along feature C in Fig.~\ref{fig:kinematics_zoomin} (one ALMA spaxel centred on RA = $204.266$, DEC = $-29.8665$). The spectrum has been corrected for the systemic velocity. We observe three separate Gaussian velocity components.
The component at $v_\mathrm{CO} \sim -30$~\kms\ is associated with the spiral arm in the galaxy, and the component at $-20$~\kms\ traces broadening around the first one, likely associated with a bar orbit.
We interpret the third component at $v_\mathrm{LSR} \sim 80$~\kms as being associated with infalling material. 
}
\label{fig:CO_spec}
\end{figure}

\par In the molecular gas (third row in Fig.~\ref{fig:kinematics_zoomin}) we observe again the signature of a circumnuclear disk. We also observe some clouds of redshifted molecular gas coincident with feature C. Along these lines of sight, we observe two or three distinct bright peaks of emission, as illustrated in Fig.~\ref{fig:CO_spec}. All but one of these components correspond to gas with a velocity compatible with the galactic rotation, whereas the remaining component is redshifted by $\sim$ 100~\kms\ with respect to the disk CO emission. 
For these complex spectra, we remark that the corresponding mom1 velocity value corresponds to an average of the features.
In fitting Gaussian models to some of these clouds, we find that the velocity dispersion of the CO emission at the velocity expected from the circular rotation field is 7 -- 20~\kms, whereas the redshifted component has a larger velocity dispersion $\simeq 25$~\kms.

The peak brightness of the redshifted CO gas blobs is 0.3 to 0.5 K, their velocity dispersions are 15 -- 25 $\mathrm{km~s}^{-1}$, and their the clouds are marginally resolved by the ALMA beam (diameters $\gtrsim 50$ pc). Using our adopted conversion factors, the redshifted blobs would have equivalent surface densities of 70 -- 200 $M_\odot~\mathrm{pc}^{-2}$ and a total mass of $4\times 10^6~M_\odot$.
At these surface densities and velocity dispersions, such gas would be typical of what is found in the central regions of barred galaxies\footnote{
We estimate however that these clouds are likely not self-gravitating -- if our assumed $\alpha_{CO}$ value is correct -- by evaluating the ratio of the kinetic energy ($K$) to the gravitational binding energy ($U_\mathrm{grav}$) under a simple spherical model \citep[e.g.][]{sun20, rosolowsky21} along a single line of sight with a radius of half the beam size ($R$ = 25 pc), $\sigma_\mathrm{CO} = 15~\mathrm{km~s}^{-1}$ and a surface density of 100~$M_\odot~\mathrm{pc}^{-2}$. The total mass in a synthesised beam is then $2 \times 10^5 M_\odot$ and $K \sim 30 U_\mathrm{grav}$. Alternatively, their opacity and corresponding conversion factors may be lower than for disk clouds, as observed in some galaxy centres \citep{sandstrom13}}.
\citep[e.g.][]{sun20}.
We remark that features A and B, on the other hand, do not stand out in the CO map. In the remainder of this section, we study features A, B and C in more detail, and in Sect.~\ref{section:discussion} we discuss their possible origin.

\subsection{BPT analysis of the starburst region}
\label{section:central_bpt}

\begin{figure}
\center
\includegraphics[width=8.5cm]{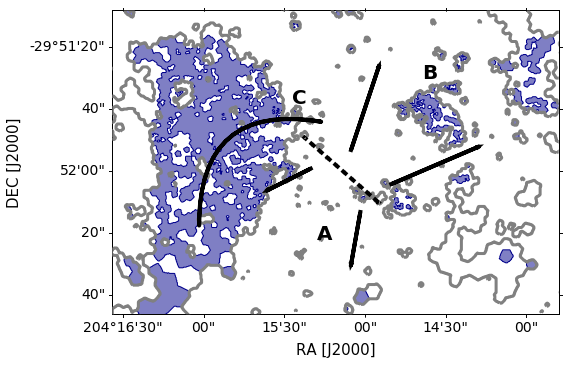}
\caption{Zoom-in into the 2D \ion{N}{ii}-BPT diagram from Fig.~\ref{fig:2D_bpt}. The grey contours correspond to $\Delta$NII = 0.
Blue shaded areas indicate regions whose emission is compatible with slow shock models only (blue grid in Fig.~\ref{fig:BPT_diagrams}).}
\label{fig:2D_bpt_zoomin}
\end{figure}

In Fig.~\ref{fig:2D_bpt_zoomin} we show a close up view of the 2D \ion{N}{ii}-BPT diagram presented in Fig.~\ref{fig:2D_bpt}. Blue shaded regions in the map correspond to areas whose emission is overlapping purely with slow shock models (area spanned exclusively by the blue grid in the top panel of Fig.~\ref{fig:BPT_diagrams}). We observe that most of the region surrounding stream C, as well as the far end of cone B (at $d \geq$ 20\arcsec{} from the galactic centre) are consistent with pure slow shocks.

\par We also performed a BPT analysis analogous to Sect.~\ref{section:BPT} for the two Gaussian components separately. We fitted the relevant emission lines in the cube binned to a S/N $\simeq$ 20 in the [\ion{O}{iii}]~$\lambda$5007 line. Given that for weaker emission lines double components might be harder to disentangle, we first performed a fit to the H$\alpha$ line and then fixed the kinematic parameters obtained from this fit for all emission lines.

\par
\begin{figure*}
\center
\includegraphics[width=7.5cm]{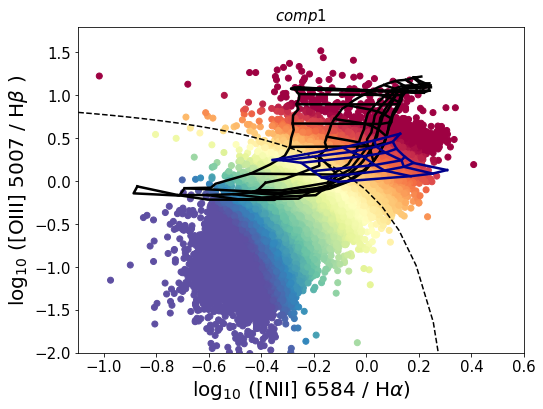}
\includegraphics[width=7.5cm]{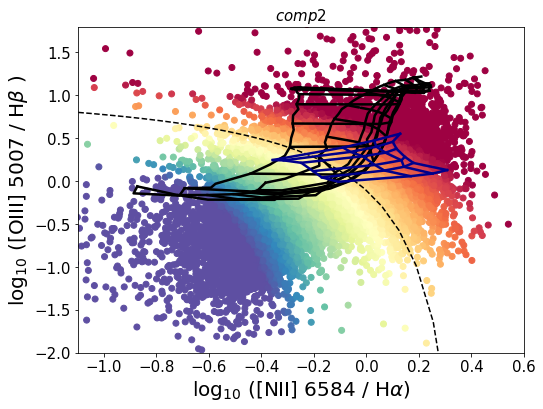}
\includegraphics[width=8.8cm]{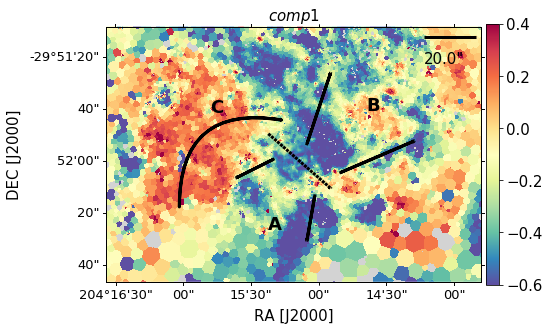}
\includegraphics[width=9.3cm]{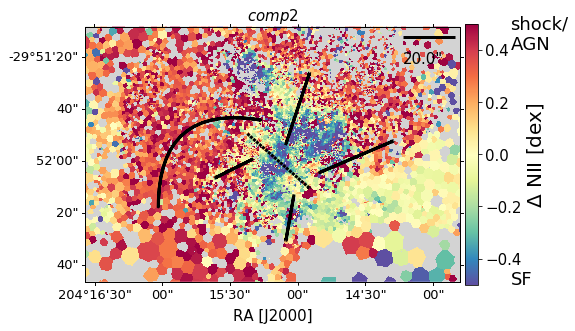}
\caption{\ion{N}{ii}-BPT diagrams obtained from a double component Gaussian fit to the emission lines. The line and model grids in the top panels are the same as in Fig.~\ref{fig:BPT_diagrams}.
Each Voronoi bin is colour coded according to its orthogonal distance from the extreme starburst line ($\Delta$~\ion{N}{ii}), ranging from blue (pure SF) to red (pure shock or AGN emission). Bins with an uncertainty $>$ 50\% on either line ratio are excluded from the top plots.
In the bottom plots, black lines mark the kinematic features discussed in Sect.~\ref{section:starburst_region}.}
\label{fig:2compfit_BPT}
\end{figure*}

 In the top panels of Fig.~\ref{fig:2compfit_BPT} we show the resulting N2-BPT diagrams. Hereby, points with an uncertainty on either [\ion{N}{ii}]/H$\alpha$ or [\ion{O}{iii}]/H$\beta$ greater than 50\% of the ratio are masked, in order to remove bad fits. We observe that both components are tracing both SF and shocks, although the second component extends to more extreme values of [\ion{N}{ii}]/H$\alpha$. This is more clearly visible in the 2D-BPT diagram shown in the bottom panels of Fig.~\ref{fig:2compfit_BPT}. We see that cones A and B are - in both components - consistent with star formation near the stellar bar, and only become shocked at a projected distance $d \geq$ 20\arcsec{} from the galactic centre, perhaps tracing an outflow originating from the starburst region that is shocking into the surrounding gas. We discuss this further in Sect.~\ref{section:discussion}.

\subsection{Shock-sensitive emission line ratios}
\label{section:o3o1maps}

\begin{figure}
\center
\includegraphics[width=9.cm]{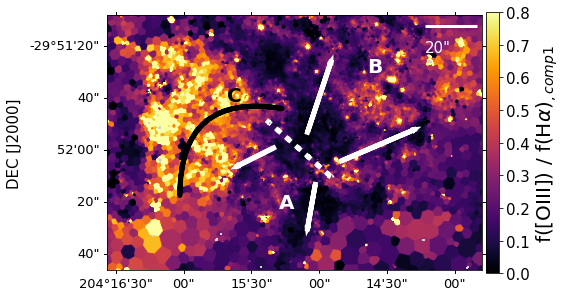}
\includegraphics[width=9.cm]{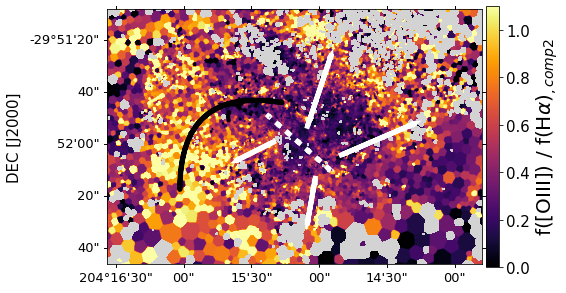}
\includegraphics[width=9.1cm]{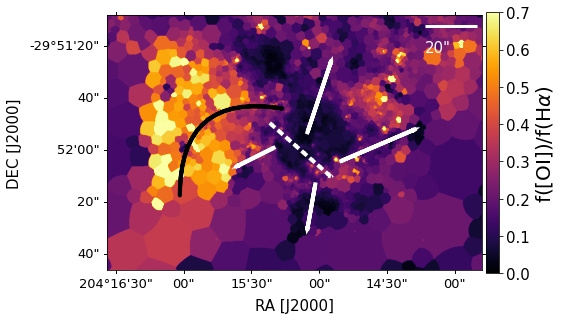}
\caption{Ratios of strong emission lines observed with MUSE. \textit{Top and centre}: Map of [\ion{O}{iii}]~$\lambda$4959,5007/H$\alpha$ obtained from a double component Gaussian fit to the emission lines.
\textit{Bottom panel}: map of [\ion{O}{i}]~$\lambda$6300/H$\alpha$ obtained from a single Gaussian component fit. 
All maps have been corrected for extinction. The labelled features are discussed in Sect.~\ref{section:o3o1maps}. Maps of the full FoV are shown in Fig.~\ref{fig:o3o1_map}.}
\label{fig:o3o1_map_zoomin}
\end{figure}

The top two panels of Fig.~\ref{fig:o3o1_map_zoomin} show a map of the [\ion{O}{iii}]~$\lambda$4959,5007/H$\alpha$ ratio for the central region (maps of the full FoV can be found in appendix~\ref{section:appendix_linemaps}). The maps were obtained by fitting the emission lines with a double Gaussian component in the gas cube spatially binned to a S/N $\simeq$ 20 in the [\ion{O}{iii}]~$\lambda$5007 line. A high [\ion{O}{iii}]/H$\alpha$ ratio is indicative of gas with a high ionisation state, where emission from doubly ionised oxygen (tracing photons with $h\nu \geq 35.1$~eV) is non-negligible with respect to ionised hydrogen \citep[e.g.][]{veilleux87}. We observe that both the region surrounding stream C and the far end of cone B (at $d \geq$ 20\arcsec{} from the galactic centre) have a high ratio both in the first ([\ion{O}{iii}]/H$\alpha~\simeq$ 0.4 -- 0.8) and in the second Gaussian component ([\ion{O}{iii}]/H$\alpha~\simeq$ 0.5 -- 1.1). Cone A, on the other hand, only shows a very locally enhanced ratio ([\ion{O}{iii}]/H$\alpha~\simeq$ 1) in the second component. This could however in part be due to the extremely high extinction at this location, as traced both by the ionised and molecular gas (bottom panel of Fig.~\ref{fig:ebv_map}).

\par In the bottom panel of Fig.~\ref{fig:o3o1_map_zoomin} we show a map of the [\ion{O}{i}]~$\lambda$6300/H$\alpha$ ratio, obtained from a single Gaussian component; we do not perform a 2-component fit to the [\ion{O}{i}] line due to its weak nature. A high ratio of [\ion{O}{i}] emission with respect to H$\alpha$ is indicative of the presence of shocks \citep{veilleux87}.
We observe that also in this map both the region surrounding stream C and the far end of cone B have an enhanced ratio ([\ion{O}{i}]/H$\alpha~\simeq$ 0.5 -- 0.7).

\section{Discussion}
\label{section:discussion}

\subsection{Kinematic features in the starburst region}
Both models and observations have shown how bars in massive disk galaxies are responsible for physical processes that result in new stellar structures such as nuclear discs or rings and inner bars \citep[][and references therein]{gadotti20}.
These processes are driven by bar-induced resonances resulting from the non-axisimmetric potential \citep[e.g.][]{binney_tremaine_87}, such as Lindblad resonances\footnote{Lindblad resonances occur for $ 2 (\Omega - \Omega_p) = \pm \kappa$, where $\kappa$ is the frequency of the radial oscillation and $\Omega_p$ and $\Omega$ are, respectively, the angular velocity of the bar and of the stars/gas (neglecting the effect of spiral arms).}.
Dynamical models for the evolution of gas in barred spiral galaxies in 1D \citep{krumholz15, krumholz17} and 2D \citep{simkin80, regan03} are able to model the creation of a nuclear ring within the inner Lindblad resonance (ILR). The main processes involved are described in detail in \citet{krumholz15}; see also \citet{renaud15}; we briefly summarise them here.
\par In a first phase, the bar exerts torques on the orbiting material. As the gas looses angular momentum, it moves inwards and by energy conservation, the gravitational potential energy is transformed into turbulent energy, resulting in an increase in velocity dispersion.
In a second phase, due to the increased velocity dispersion and the mostly flat rotation curve (low shear) within the ILR, acoustic instabilities develop in the gas. This allows for a more efficient transport of angular momentum, and leads to an inflow of gas with high turbulent pressure that is extremely gravitationally stable and has a low SFR. In a third phase, gas starts to build up on the stable ILR orbit, eventually leading in gravitational instabilities that cause fragmentation and collapse, resulting in a circumnuclear ring. The ring can form stars episodically \citep[e.g. in the 1D dynamical models of][]{krumholz15}, or having an initially steady fuelling rate before fragmenting after $\sim$ 10 Myr \citep[e.g. in the hydrodynamical simulations from][]{emsellem15}.

\par In Sect.~\ref{section:central_kinematics}, we confirmed the presence of a circumnuclear disk, both in the stellar, ionised gas and molecular gas kinematics (Fig.~\ref{fig:kinematics_zoomin}). This feature had already been recently observed with MUSE \citep{gadotti20}, and has been postulated to coincide with an ILR. Recently, \citet{callanan21} analysed high resolution ALMA data mapping the central 500 pc at scales of $\sim$ 10 pc, and put forward a model in which the gas revolves around the centre in eccentric orbits. The study proved that the gas in the ring features strong azimuthal variations in velocity dispersion and intensity which are consistent with the expectations from 1D dynamical models \citep{krumholz15,krumholz17}. Furthermore, in their simple model scenario the starburst phase resulting from the disk instability is constrained to be highly localised, both in space and in time, resulting in very efficient stellar feedback.
\par We furthermore observed that both the ionised and molecular gas are tracing a flow of gas east of the nucleus (feature C in Fig.~\ref{fig:kinematics_zoomin}). \citet{lundgren04} were the first to report this feature in CO(2-1) and (1-0). Their observations traced - on top of a regular rotating disk - 
streaming motions along the spiral arms, with the strongest deviation on the NW side of the nucleus. This feature was later confirmed by \citet{fathi08} using Fabry-Perot data of the H$\alpha$ line across the disk. More recently, the \citet{piqueras_lopez12} observed evidence of what they intepreted as a gas inflow also in high-resolution NIR IFS data mapping the central $\sim$ 200 $\times$ 200 pc. In our dataset, we observe that the stream has $\Delta v_{\rm H\alpha} \simeq +30$~\kms\ with respect to the surrounding disk rotation (Fig.~\ref{fig:kinematics_zoomin}, centre left) and multiple peaks of emission at different velocities in the molecular gas (Fig.~\ref{fig:CO_spec}), one of which is redsfhited by $\sim 100$~\kms\ with respect to the disk rotation and has an enhanced velocity dispersion ($\simeq 25$~\kms). The MUSE data furthermore trace an extended region surrounding the stream ($\sim 1000 \times 1600$ pc) featuring: (1) a high velocity dispersion ($\simeq$ 80~\kms, Fig.~\ref{fig:kinematics_zoomin}, centre right); (2) [\ion{N}{ii}]/H$\alpha$ and [\ion{O}{iii}]/H$\beta$ ratios that situate the gas clearly above the line separating SF from shocks in a BPT diagram (Figs.~\ref{fig:2D_bpt} and~\ref{fig:2compfit_BPT}), in a region consistent for most part with slow shock models only (Fig.~\ref{fig:2D_bpt_zoomin});
(3) high ratios of [\ion{O}{iii}] and [\ion{O}{i}] with respect to H$\alpha$ (Fig.~\ref{fig:o3o1_map_zoomin}), indicative of gas with a high ionisation state and of shocks; (4) the presence of bright molecular gas but weak H$\alpha$ emission (Fig.~\ref{fig:center_CO_vs_halpha}). 

\par We interpret feature C as the result of two main scenarios. 
A first possible physical picture is the superposition along the LoS of the disk and an extraplanar layer of DIG.
\citet{boettcher17} obtained a high-resolution ($\sigma_{\rm H\alpha} = 23$~\kms\ vs the MUSE resolution of 120~\kms) single slit spectrum cutting through feature C. Their data show the presence of two distinct Gaussian components: a narrow component tracing Galactic rotation and a broad component ($\sigma \simeq 95$~\kms) having a velocity lag of $\sim 25$~\kms\ (similar to what we observe in the MUSE data) and high ratios of [\ion{N}{ii}]/H$\alpha \sim 1.0$. The authors interpret this as the presence of an extraplanar layer of DIG, as broadly observed in other star-forming disks \citep[see e.g.][]{rossa03,lacerda18, levy19, rautio22}. This scenario is supported by the multiple peaks observed in the CO spectrum in Fig.~\ref{fig:CO_spec}.

\par In a second scenario, feature C could be a bar-driven inflow of gas located in the same plane as the disk; in this case the increased velocity dispersion would be tracing excess turbulence as the flow is shocked within the bar. An excess turbulence in the molecular gas could explain the lack of H$\alpha$ emission despite the bright CO emission along the stream, as the gas would be unable to collapse and form stars.

\par A final possibility could be a past interaction, possibly with the neighbouring galaxy NGC 5253 (1.8$^\circ$ in projected distance). This option has been taken into account by many authors in order explain the peculiar morphology and kinematics of the central region, where the optical nucleus is offset from the kinematic centre \citep{thatte00,mast06,diaz06,houghton08,rodrigues09, knapen10, piqueras_lopez12} and the structure of the \ion{H}{i} disk \citep{miller09, heald16}. However, given the extremely regular stellar rotation field and the general lack of global scale perturbances, we discard this hypothesis.

\par In the ionised gas kinematics resulting from the double component Gaussian fit (Fig.~\ref{fig:2compfit_kinem}), we also observed two kinematic features (labelled as cones A and B) where the H$\alpha$ line is composed of two peaks, separated by a velocity $\Delta v \leq 200$~\kms\ (third and fourth row in Fig.~\ref{fig:2compfit_kinem}) and a high velocity dispersion (up to 200~\kms, top and centre right in Fig.~\ref{fig:2compfit_kinem}). The two cones appear, respectively, blue- and redshifted along our line of sight ($v \simeq \pm 100$~\kms). 
Cone B features at its far end ($d \geq$ 20\arcsec{} from the galactic centre): (1) [\ion{N}{ii}]/H$\alpha$ and [\ion{O}{iii}]/H$\beta$ ratios above the SF limit in a N2-BPT analysis (Figs.~\ref{fig:2D_bpt} and~\ref{fig:2compfit_BPT}), in a region of the diagram that overlaps mostly with slow shock models (Fig.~\ref{fig:2D_bpt_zoomin}); (2) high ratios of [\ion{O}{iii}] and [\ion{O}{i}] with respect to H$\alpha$ (Fig.~\ref{fig:o3o1_map_zoomin}), indicative of gas with a high ionisation state and/or shocks; (3) a relatively bright CO emission in a region with little SF (bottom left in Fig.~\ref{fig:ebv_map} and left panel in Fig.~\ref{fig:halpha_map}).
Cone A stands out less clearly in these tracers, perhaps owing to the high extinction at this location (see bottom right panel of Fig.~\ref{fig:ebv_map}). Nonetheless, we observe BPT line ratios indicative of shocks in the second Gaussian component (Fig.~\ref{fig:2compfit_BPT}, for $d \geq$ 20\arcsec{}) and a locally enhanced [\ion{O}{i}]/H$\alpha$ ratio (central panel in Fig.~\ref{fig:o3o1_map_zoomin}).

\par So far, no AGN activity has been confirmed in M83, and hard X-ray observations \citep{yukita16} constrain any AGN to be either highly obscured or to have an extremely low luminosity. Given the fact that the origin of cones A and B is offset from the galaxy's centre, the lack of shock-related emission immediately surrounding the central region and the presence of slow shocks, we favour over the presence of an AGN the hypothesis of a starburst-driven outflow cone shocking into the surrounding ISM. Similar starburst-driven bi-conical outflows have been observed in M82 \citep{shopbell98, leroy15b}, in which the outflow cones also feature a double component H$\alpha$ line with $\Delta v$ = 300~\kms, low [\ion{N}{ii}]/H$\alpha$ ratios compatible with photoionisation and an [\ion{O}{iii}]/H$\alpha$ ratio that increases along the outflow, as well as in NGC 253 \citep{bolatto13} and ESO 338-IG04 \citep{bik18}.
The scenario of a starburst- (rather than AGN-) driven outflow is also in better agreement with the low [\ion{S}{ii}]~$\lambda$6716/6731 line ratio observed throughout cones A and B \citep[see e.g.][]{bik18}. The ratio is very close to the sensitivity limit ([\ion{S}{ii}]~$\lambda$6716/6731 = 1.45, corresponding to $n_e \leq$ 1 cm$^{-3}$, \citealp{draine}), and within the measurement errors is largely consistent with $n_e \leq30~\mbox{cm}^{-3}$. A starburst-driven outflow would be compatible with the scenario of a short-lived and localised starburst activity by \citet{callanan21}, resulting in very efficient stellar feedback and, potentially, in nuclear outflows.

\par An alternative explanation could be the combination of past AGN activity followed by starburst emission. In this scenario, the AGN could have carved a path for the starburst feedback, and could explain the `jet-like' morphology of the outflow and its extent.

\par Finally, what appear as redshifted and blueshifted cones along our line of sight could simply simply be the result of the superposition of bar-driven orbits, resulting from the non-axisymmetric and time varying potential. However, this effect alone would be difficult to reconcile with the large separation between the two peaks of H$\alpha$ emission (up to 200~\kms) at the centre of the cones.

\par In the near future, the increasing availability of IFS datasets comparable to the present study, covering large portions of local spiral galaxies at intermediate to high resolution, will allow for a better understanding of the signatures imprinted on the ionised gas by in- and outflows, as well as the relative importance of extraplanar gas in low inclination spirals.

\subsection{Origin of the DIG}
In Sect.~\ref{section:hii_dig} we separated the H$\alpha$ emission in our FoV into compact \ion{H}{ii} regions and diffuse ionised gas, finding a DIG fraction $f_{\rm DIG} \sim 13\%$ (20\%) based on a cut in H$\alpha$ surface brightness ([\ion{S}{ii}]/H$\alpha$ line ratio). This fraction is on the low end of the range with respect to observations in nearby spiral galaxies, where it was constrained to be between 30 and 50\% \citep{ferguson96, hoopes96,zurita00, thilker02, belfiore21}, and in some cases even up to 60\% \citep{oey07}.
This might be due to the fact that the MUSE data only cover the inner portion of the optical disk ($R \leq 1.1~R_e$). The wide field spectrograph data by \citet{poetrodjojo19} imaging most of the optical disk ($R \leq 2~R_e$ at our assumed distance and $R_e$) revealed indeed that the DIG contribution increases (up to 40\%) in the outskirts of the disk. Overall, we observe that despite the enhanced ratios of [\ion{N}{ii}], [\ion{S}{ii}] and [\ion{O}{i}] with respect to H$\alpha$, 94.9\% of the H$\alpha$ luminosity in the DIG is consistent with originating from star formation (BPT diagrams in Figs.~\ref{fig:BPT_diagrams} and~\ref{fig:2D_bpt}). This could support the hypothesis that the diffuse gas is for most part ionised by radiation escaping the SF regions.
DIG featuring stronger ratios of [\ion{N}{ii}]/H$\alpha$, [\ion{S}{ii}]/H$\alpha$ can instead be partially explained with being extraplanar, as observed by \citet{boettcher17}.
The [\ion{N}{ii}] and [\ion{S}{ii}]/H$\alpha$ ratios are indeed known to increase with distance from the galactic midplane \citep{madsen06, jones17, levy19}, due for example to shocks \citep{rand98}, turbulent mixing layers \citep{rand98, collins01} or ionisation by hot, old, low-mass evolved stars (HOLMES, \citealp{lacerda18, levy19, weber19}). We will investigate the origin of the diffuse ionised gas in more detail in a second publication (Della Bruna et al., in prep.).

\section{Conclusions}
\label{section:conclusions}
We have presented a large MUSE mosaic covering the central 3.8 kpc $\times$ 3.8 kpc of the nearby barred spiral galaxy M83 with a spatial resolution $
\sim$ 20 pc. We obtained the kinematics of the stars and the ionised gas, and compared them with molecular gas kinematics from ALMA CO(2-1). We observed that the stellar kinematics trace regular rotation along the main SF disk, plus a fast rotating circumnuclear disk of 30\arcsec{} ($\simeq$ 700 pc) in diameter, likely originating from secular processes driven by the galactic bar.

\par The gas kinematics are rich in substructures, and the ionised and molecular gas substantially match one another. At large scales, the gas maps rotation along the disk, as well as the fast rotating circumnuclear disk component. On top of the disk rotation, the gas traces a flow east of the nucleus (50\arcsec $\simeq$ 1250 pc in size), that appears redshifted on top of the disk rotation. In the ionised gas, we observe $\Delta v_{\rm H\alpha} \simeq$ +30~\kms\ with respect to the surroundings, and in the molecular gas we observe multiple peaks of emission, one of which is redshifted by $\Delta v_{\rm CO} \simeq$ +100~\kms\ with respect to the emission tracing galactic rotation. We also observe an enhanced velocity dispersion both in the ionised gas ($\sigma_{\rm H\alpha} \simeq 80$~\kms) and molecular gas ($\sigma_{\rm CO} \simeq 25$~\kms\ for the redshifted peak). The ionised gas showcases an extended region ($\sim$ 1000 $\times$ 1600 pc) surrounding the flow, whose emission lies above the upper limit for star formation in a N2-BPT emission line diagram, and is in large part consistent with models of slow shocks. The extended region also features high ratios of [\ion{O}{iii}]/H$\alpha$ ($\gtrsim 0.5$) and [\ion{O}{i}]/H$\alpha$ ($\gtrsim 0.5$), indicative of a high ionisation state and of the presence shocks. We interpreted this feature as either the superposition of non-collisional flows originating from multiple vertical layers of gas or a bar-driven inflow of shocked gas.
\par A double component Gaussian fit to the H$\alpha$ line moreover revealed the presence of two distinct cone-shaped velocity components (20\arcsec{} $\simeq$ 500 pc in size) on either side of the stellar bar, where the line features two distinct peaks that are up to 200~\kms\ apart. The two cones appear blue- and redshifted along our line of sight, with $v = \pm 100$~\kms\ and have a velocity dispersion $>$ 80~\kms\ and up to 200~\kms. At the far end of both cones, the gas emission lies above the star formation limit in a N2-BPT diagram, in an area of the diagram consistent with slow shock models, and features an enhanced [\ion{O}{iii}]/H$\alpha$ ratio ($\gtrsim$ 0.4 in one or - in the case of cone B - in both of the Gaussian components). One of the two cones also features enhanced [\ion{O}{i}]/H$\alpha$ ratios ($\gtrsim$ 0.4). We postulate that these two components are tracing a starburst-driven outflow perpendicular to the stellar bar, shocking into the surrounding ISM.

\par We estimated the gas extinction from the MUSE H$\beta$/H$\alpha$ ratio, and found that the regions more strongly affected by extinction ($E(B-V) > 0.5$) in general have a high density of molecular gas ($> 100~M_\sun$ pc$^{-2}$).
Finally, we separated the ionised gas into \ion{H}{ii} regions and DIG, using a cut in H$\alpha$ surface brightness (\ion{S}{ii}/H$\alpha$ line ratio). We obtained a DIG fraction $f_{\rm DIG} \sim$ 13\% (20\%) in our FoV, and observed that the DIG contribution varies radially between 0.8 and 46\%, peaking in the interarm region (at $R \sim 0.3~R_e$). We inspected the emission of the \ion{H}{ii} regions and DIG in BPT diagrams, finding that in \ion{H}{ii} regions 99.8\% of $L(H\alpha)$ originates from photoionisation-dominated regions, whereas the DIG has a mixed contribution from both photoionisation- (94.9\%) and shock-dominated regions (5.1\%).

\begin{acknowledgements}
We thank the anonymous referee for their very detailed and helpful feedback on this manuscript.
We thank Jeff Rich and Lisa Kewley for sharing with us the slow shock models presented in \citet{rich11}.
This work is based on observations collected at the European Southern Observatory under ESO programmes
096.B-0057(A), 0101.B-0727(A),
097.B-0899(B), 
097.B-0640(A). 
A.A. acknowledges the support of the Swedish Research Council, Vetenskapsr\aa{}det, and the Swedish National Space Agency (SNSA). FR acknowledges support from the Knut and Alice Wallenberg Foundation.
This research made use of Astropy\footnote{http://www.astropy.org}, a community-developed core Python package for Astronomy \citep{astropy:2013, astropy:2018}.
\end{acknowledgements}
\bibpunct{(}{)}{;}{a}{}{,} 
\bibliographystyle{aa}
\bibliography{paperIII}

\begin{thebibliography}{145}
\expandafter\ifx\csname natexlab\endcsname\relax\def\natexlab#1{#1}\fi

\bibitem[{{Adamo} {et~al.}(2015){Adamo}, {Kruijssen}, {Bastian}, {Silva-Villa},
  \& {Ryon}}]{adamo15}
{Adamo}, A., {Kruijssen}, J.~M.~D., {Bastian}, N., {Silva-Villa}, E., \&
  {Ryon}, J. 2015, \mnras, 452, 246

\bibitem[{{Allen} {et~al.}(2008){Allen}, {Groves}, {Dopita}, {Sutherland}, \&
  {Kewley}}]{allen08}
{Allen}, M.~G., {Groves}, B.~A., {Dopita}, M.~A., {Sutherland}, R.~S., \&
  {Kewley}, L.~J. 2008, \apjs, 178, 20

\bibitem[{{Astropy Collaboration} {et~al.}(2018){Astropy Collaboration},
  {Price-Whelan}, {Sip{\H o}cz}, {G{\"u}nther}, {Lim}, {Crawford}, {Conseil},
  {Shupe}, {Craig}, {Dencheva}, {Ginsburg}, {VanderPlas}, {Bradley},
  {P{\'e}rez-Su{\'a}rez}, {de Val-Borro}, {Aldcroft}, {Cruz}, {Robitaille},
  {Tollerud}, {Ardelean}, {Babej}, {Bach}, {Bachetti}, {Bakanov}, {Bamford},
  {Barentsen}, {Barmby}, {Baumbach}, {Berry}, {Biscani}, {Boquien}, {Bostroem},
  {Bouma}, {Brammer}, {Bray}, {Breytenbach}, {Buddelmeijer}, {Burke},
  {Calderone}, {Cano Rodr{\'{\i}}guez}, {Cara}, {Cardoso}, {Cheedella},
  {Copin}, {Corrales}, {Crichton}, {D'Avella}, {Deil}, {Depagne}, {Dietrich},
  {Donath}, {Droettboom}, {Earl}, {Erben}, {Fabbro}, {Ferreira}, {Finethy},
  {Fox}, {Garrison}, {Gibbons}, {Goldstein}, {Gommers}, {Greco}, {Greenfield},
  {Groener}, {Grollier}, {Hagen}, {Hirst}, {Homeier}, {Horton}, {Hosseinzadeh},
  {Hu}, {Hunkeler}, {Ivezi{\'c}}, {Jain}, {Jenness}, {Kanarek}, {Kendrew},
  {Kern}, {Kerzendorf}, {Khvalko}, {King}, {Kirkby}, {Kulkarni}, {Kumar},
  {Lee}, {Lenz}, {Littlefair}, {Ma}, {Macleod}, {Mastropietro}, {McCully},
  {Montagnac}, {Morris}, {Mueller}, {Mumford}, {Muna}, {Murphy}, {Nelson},
  {Nguyen}, {Ninan}, {N{\"o}the}, {Ogaz}, {Oh}, {Parejko}, {Parley}, {Pascual},
  {Patil}, {Patil}, {Plunkett}, {Prochaska}, {Rastogi}, {Reddy Janga},
  {Sabater}, {Sakurikar}, {Seifert}, {Sherbert}, {Sherwood-Taylor}, {Shih},
  {Sick}, {Silbiger}, {Singanamalla}, {Singer}, {Sladen}, {Sooley},
  {Sornarajah}, {Streicher}, {Teuben}, {Thomas}, {Tremblay}, {Turner},
  {Terr{\'o}n}, {van Kerkwijk}, {de la Vega}, {Watkins}, {Weaver}, {Whitmore},
  {Woillez}, {Zabalza}, \& {Astropy Contributors}}]{astropy:2018}
{Astropy Collaboration}, {Price-Whelan}, A.~M., {Sip{\H o}cz}, B.~M., {et~al.}
  2018, \aj, 156, 123

\bibitem[{{Astropy Collaboration} {et~al.}(2013){Astropy Collaboration},
  {Robitaille}, {Tollerud}, {Greenfield}, {Droettboom}, {Bray}, {Aldcroft},
  {Davis}, {Ginsburg}, {Price-Whelan}, {Kerzendorf}, {Conley}, {Crighton},
  {Barbary}, {Muna}, {Ferguson}, {Grollier}, {Parikh}, {Nair}, {Unther},
  {Deil}, {Woillez}, {Conseil}, {Kramer}, {Turner}, {Singer}, {Fox}, {Weaver},
  {Zabalza}, {Edwards}, {Azalee Bostroem}, {Burke}, {Casey}, {Crawford},
  {Dencheva}, {Ely}, {Jenness}, {Labrie}, {Lim}, {Pierfederici}, {Pontzen},
  {Ptak}, {Refsdal}, {Servillat}, \& {Streicher}}]{astropy:2013}
{Astropy Collaboration}, {Robitaille}, T.~P., {Tollerud}, E.~J., {et~al.} 2013,
  \aap, 558, A33

\bibitem[{{Bacon} {et~al.}(2010){Bacon}, {Accardo}, {Adjali}, {Anwand},
  {Bauer}, {Biswas}, {Blaizot}, {Boudon}, {Brau-Nogue}, {Brinchmann},
  {Caillier}, {Capoani}, {Carollo}, {Contini}, {Couderc}, {Daguis{\'e}},
  {Deiries}, {Delabre}, {Dreizler}, {Dubois}, {Dupieux}, \& {Dupuy}}]{bacon10}
{Bacon}, R., {Accardo}, M., {Adjali}, L., {et~al.} 2010, Society of
  Photo-Optical Instrumentation Engineers (SPIE) Conference Series, Vol. 7735,
  {The MUSE second-generation VLT instrument}, 773508

\bibitem[{{Baldwin} {et~al.}(1981){Baldwin}, {Phillips}, \&
  {Terlevich}}]{baldwin81}
{Baldwin}, J.~A., {Phillips}, M.~M., \& {Terlevich}, R. 1981, \pasp, 93, 5

\bibitem[{{Barrera-Ballesteros} {et~al.}(2020){Barrera-Ballesteros}, {Utomo},
  {Bolatto}, {S{\'a}nchez}, {Vogel}, {Wong}, {Levy}, {Colombo}, {Kalinova},
  {Teuben}, {Garc{\'\i}a-Benito}, {Husemann}, {Mast}, \&
  {Blitz}}]{barrera_ballesteros20}
{Barrera-Ballesteros}, J.~K., {Utomo}, D., {Bolatto}, A.~D., {et~al.} 2020,
  \mnras, 492, 2651

\bibitem[{{Belfiore} {et~al.}(2021){Belfiore}, {Santoro}, {Groves},
  {Schinnerer}, {Kreckel}, {Glover}, {Klessen}, {Emsellem}, {Blanc}, {Congiu},
  {Barnes}, {Boquien}, {Chevance}, {Dale}, {Kruijssen}, {Leroy}, {Pan},
  {Pessa}, {Schruba}, \& {Williams}}]{belfiore21}
{Belfiore}, F., {Santoro}, F., {Groves}, B., {et~al.} 2021, arXiv e-prints,
  arXiv:2111.14876

\bibitem[{{Bik} {et~al.}(2018){Bik}, {{\"O}stlin}, {Menacho}, {Adamo}, {Hayes},
  {Herenz}, \& {Melinder}}]{bik18}
{Bik}, A., {{\"O}stlin}, G., {Menacho}, V., {et~al.} 2018, \aap, 619, A131

\bibitem[{{Binney} \& {Tremaine}(1987)}]{binney_tremaine_87}
{Binney}, J. \& {Tremaine}, S. 1987, {Galactic dynamics}

\bibitem[{{Bizyaev} {et~al.}(2017){Bizyaev}, {Walterbos}, {Yoachim}, {Riffel},
  {Fern{\'a}ndez-Trincado}, {Pan}, {Diamond-Stanic}, {Jones}, {Thomas},
  {Cleary}, \& {Brinkmann}}]{bizyaev17}
{Bizyaev}, D., {Walterbos}, R.~A.~M., {Yoachim}, P., {et~al.} 2017, \apj, 839,
  87

\bibitem[{{Blair} {et~al.}(2014){Blair}, {Chandar}, {Dopita}, {Ghavamian},
  {Hammer}, {Kuntz}, {Long}, {Soria}, {Whitmore}, \& {Winkler}}]{blair14}
{Blair}, W.~P., {Chandar}, R., {Dopita}, M.~A., {et~al.} 2014, \apj, 788, 55

\bibitem[{{Boettcher} {et~al.}(2017){Boettcher}, {Gallagher}, \&
  {Zweibel}}]{boettcher17}
{Boettcher}, E., {Gallagher}, J.~S., I., \& {Zweibel}, E.~G. 2017, \apj, 845,
  155

\bibitem[{{Bohlin} {et~al.}(1978){Bohlin}, {Savage}, \& {Drake}}]{bohlin78}
{Bohlin}, R.~C., {Savage}, B.~D., \& {Drake}, J.~F. 1978, \apj, 224, 132

\bibitem[{{Bolatto} {et~al.}(2013){Bolatto}, {Wolfire}, \& {Leroy}}]{bolatto13}
{Bolatto}, A.~D., {Wolfire}, M., \& {Leroy}, A.~K. 2013, \araa, 51, 207

\bibitem[{{Bresolin} {et~al.}(2016){Bresolin}, {Kudritzki}, {Urbaneja},
  {Gieren}, {Ho}, \& {Pietrzy{\'n}ski}}]{bresolin16}
{Bresolin}, F., {Kudritzki}, R.-P., {Urbaneja}, M.~A., {et~al.} 2016, \apj,
  830, 64

\bibitem[{{Bryant} {et~al.}(2015){Bryant}, {Owers}, {Robotham}, {Croom},
  {Driver}, {Drinkwater}, {Lorente}, {Cortese}, {Scott}, {Colless}, {Schaefer},
  {Taylor}, {Konstantopoulos}, {Allen}, {Baldry}, {Barnes}, {Bauer},
  {Bland-Hawthorn}, {Bloom}, {Brooks}, {Brough}, {Cecil}, {Couch}, {Croton},
  {Davies}, {Ellis}, {Fogarty}, {Foster}, {Glazebrook}, {Goodwin}, {Green},
  {Gunawardhana}, {Hampton}, {Ho}, {Hopkins}, {Kewley}, {Lawrence},
  {Leon-Saval}, {Leslie}, {McElroy}, {Lewis}, {Liske}, {L{\'o}pez-S{\'a}nchez},
  {Mahajan}, {Medling}, {Metcalfe}, {Meyer}, {Mould}, {Obreschkow}, {O'Toole},
  {Pracy}, {Richards}, {Shanks}, {Sharp}, {Sweet}, {Thomas}, {Tonini}, \&
  {Walcher}}]{bryant15}
{Bryant}, J.~J., {Owers}, M.~S., {Robotham}, A.~S.~G., {et~al.} 2015, \mnras,
  447, 2857

\bibitem[{{Bundy} {et~al.}(2015){Bundy}, {Bershady}, {Law}, {Yan}, {Drory},
  {MacDonald}, {Wake}, {Cherinka}, {S{\'a}nchez-Gallego}, {Weijmans}, {Thomas},
  {Tremonti}, {Masters}, {Coccato}, {Diamond-Stanic}, {Arag{\'o}n-Salamanca},
  {Avila-Reese}, {Badenes}, {Falc{\'o}n-Barroso}, {Belfiore}, {Bizyaev},
  {Blanc}, {Bland-Hawthorn}, {Blanton}, {Brownstein}, {Byler}, {Cappellari},
  {Conroy}, {Dutton}, {Emsellem}, {Etherington}, {Frinchaboy}, {Fu}, {Gunn},
  {Harding}, {Johnston}, {Kauffmann}, {Kinemuchi}, {Klaene}, {Knapen},
  {Leauthaud}, {Li}, {Lin}, {Maiolino}, {Malanushenko}, {Malanushenko}, {Mao},
  {Maraston}, {McDermid}, {Merrifield}, {Nichol}, {Oravetz}, {Pan}, {Parejko},
  {Sanchez}, {Schlegel}, {Simmons}, {Steele}, {Steinmetz}, {Thanjavur},
  {Thompson}, {Tinker}, {van den Bosch}, {Westfall}, {Wilkinson}, {Wright},
  {Xiao}, \& {Zhang}}]{bundy15}
{Bundy}, K., {Bershady}, M.~A., {Law}, D.~R., {et~al.} 2015, \apj, 798, 7

\bibitem[{{Buta} \& {Crocker}(1993)}]{buta93}
{Buta}, R. \& {Crocker}, D.~A. 1993, \aj, 105, 1344

\bibitem[{{Buta} {et~al.}(2015){Buta}, {Sheth}, {Athanassoula}, {Bosma},
  {Knapen}, {Laurikainen}, {Salo}, {Elmegreen}, {Ho}, {Zaritsky}, {Courtois},
  {Hinz}, {Mu{\~n}oz-Mateos}, {Kim}, {Regan}, {Gadotti}, {Gil de Paz}, {Laine},
  {Men{\'e}ndez-Delmestre}, {Comer{\'o}n}, {Erroz Ferrer}, {Seibert},
  {Mizusawa}, {Holwerda}, \& {Madore}}]{buta15}
{Buta}, R.~J., {Sheth}, K., {Athanassoula}, E., {et~al.} 2015, \apjs, 217, 32

\bibitem[{{Callanan} {et~al.}(2021){Callanan}, {Longmore}, {Kruijssen},
  {Schruba}, {Ginsburg}, {Krumholz}, {Bastian}, {Alves}, {Henshaw}, {Knapen},
  \& {Chevance}}]{callanan21}
{Callanan}, D., {Longmore}, S.~N., {Kruijssen}, J.~M.~D., {et~al.} 2021,
  \mnras, 505, 4310

\bibitem[{{Calzetti} {et~al.}(2004){Calzetti}, {Harris}, {Gallagher}, {Smith},
  {Conselice}, {Homeier}, \& {Kewley}}]{calzetti04}
{Calzetti}, D., {Harris}, J., {Gallagher}, John~S., I., {et~al.} 2004, \aj,
  127, 1405

\bibitem[{{Cappellari}(2017)}]{cappellari17}
{Cappellari}, M. 2017, \mnras, 466, 798

\bibitem[{{Cappellari} \& {Copin}(2003)}]{cappellari03}
{Cappellari}, M. \& {Copin}, Y. 2003, \mnras, 342, 345

\bibitem[{{Cappellari} \& {Emsellem}(2004)}]{cappellari04}
{Cappellari}, M. \& {Emsellem}, E. 2004, \pasp, 116, 138

\bibitem[{{Cardelli} {et~al.}(1989){Cardelli}, {Clayton}, \&
  {Mathis}}]{cardelli89}
{Cardelli}, J.~A., {Clayton}, G.~C., \& {Mathis}, J.~S. 1989, \apj, 345, 245

\bibitem[{{Collins} \& {Rand}(2001)}]{collins01}
{Collins}, J.~A. \& {Rand}, R.~J. 2001, \apj, 551, 57

\bibitem[{{Comer{\'o}n} {et~al.}(2010){Comer{\'o}n}, {Knapen}, {Beckman},
  {Laurikainen}, {Salo}, {Mart{\'\i}nez-Valpuesta}, \& {Buta}}]{comeron10}
{Comer{\'o}n}, S., {Knapen}, J.~H., {Beckman}, J.~E., {et~al.} 2010, \mnras,
  402, 2462

\bibitem[{{Croom} {et~al.}(2012){Croom}, {Lawrence}, {Bland-Hawthorn},
  {Bryant}, {Fogarty}, {Richards}, {Goodwin}, {Farrell}, {Miziarski}, {Heald},
  {Jones}, {Lee}, {Colless}, {Brough}, {Hopkins}, {Bauer}, {Birchall}, {Ellis},
  {Horton}, {Leon-Saval}, {Lewis}, {L{\'o}pez-S{\'a}nchez}, {Min}, {Trinh}, \&
  {Trowland}}]{croom12}
{Croom}, S.~M., {Lawrence}, J.~S., {Bland-Hawthorn}, J., {et~al.} 2012, \mnras,
  421, 872

\bibitem[{{Dale}(2015)}]{dale15}
{Dale}, J.~E. 2015, \nar, 68, 1

\bibitem[{{Dale} {et~al.}(2014){Dale}, {Ngoumou}, {Ercolano}, \&
  {Bonnell}}]{dale14}
{Dale}, J.~E., {Ngoumou}, J., {Ercolano}, B., \& {Bonnell}, I.~A. 2014, \mnras,
  442, 694

\bibitem[{{de Vaucouleurs} {et~al.}(1991){de Vaucouleurs}, {de Vaucouleurs},
  {Corwin}, {Buta}, {Paturel}, \& {Fouque}}]{RC3_cat}
{de Vaucouleurs}, G., {de Vaucouleurs}, A., {Corwin}, Herold~G., J., {et~al.}
  1991, {Third Reference Catalogue of Bright Galaxies}

\bibitem[{{Della Bruna} {et~al.}(2020){Della Bruna}, {Adamo}, {Bik},
  {Fumagalli}, {Walterbos}, {{\"O}stlin}, {Bruzual}, {Calzetti}, {Charlot},
  {Grasha}, {Smith}, {Thilker}, \& {Wofford}}]{paperI}
{Della Bruna}, L., {Adamo}, A., {Bik}, A., {et~al.} 2020, \aap, 635, A134

\bibitem[{{Della Bruna} {et~al.}(2021){Della Bruna}, {Adamo}, {Lee}, {Smith},
  {Krumholz}, {Bik}, {Calzetti}, {Fox}, {Fumagalli}, {Grasha}, {Messa},
  {{\"O}stlin}, {Walterbos}, \& {Wofford}}]{paperII}
{Della Bruna}, L., {Adamo}, A., {Lee}, J.~C., {et~al.} 2021, \aap, 650, A103

\bibitem[{{den Brok} {et~al.}(2021){den Brok}, {Chatzigiannakis}, {Bigiel},
  {Puschnig}, {Barnes}, {Leroy}, {Jim{\'e}nez-Donaire}, {Usero}, {Schinnerer},
  {Rosolowsky}, {Faesi}, {Grasha}, {Hughes}, {Kruijssen}, {Liu}, {Neumann},
  {Pety}, {Querejeta}, {Saito}, {Schruba}, \& {Stuber}}]{denbrok21_alma}
{den Brok}, J.~S., {Chatzigiannakis}, D., {Bigiel}, F., {et~al.} 2021, \mnras,
  504, 3221

\bibitem[{{den Brok} {et~al.}(2020){den Brok}, {Carollo}, {Erroz-Ferrer},
  {Fagioli}, {Brinchmann}, {Emsellem}, {Krajnovi{\'c}}, {Marino}, {Onodera},
  {Tacchella}, {Weilbacher}, \& {Woo}}]{denbrok20}
{den Brok}, M., {Carollo}, C.~M., {Erroz-Ferrer}, S., {et~al.} 2020, \mnras,
  491, 4089

\bibitem[{{D{\'\i}az} {et~al.}(2006){D{\'\i}az}, {Dottori}, {Aguero},
  {Mediavilla}, {Rodrigues}, \& {Mast}}]{diaz06}
{D{\'\i}az}, R.~J., {Dottori}, H., {Aguero}, M.~P., {et~al.} 2006, \apj, 652,
  1122

\bibitem[{{Diehl} \& {Statler}(2006)}]{diehl06}
{Diehl}, S. \& {Statler}, T.~S. 2006, \mnras, 368, 497

\bibitem[{{Draine}(2011)}]{draine}
{Draine}, B.~T. 2011, {Physics of the Interstellar and Intergalactic Medium}

\bibitem[{{Drissen} {et~al.}(2019){Drissen}, {Martin}, {Rousseau-Nepton},
  {Robert}, {Martin}, {Baril}, {Prunet}, {Joncas}, {Thibault}, {Brousseau},
  {Mandar}, {Grand mont}, {Yee}, \& {Simard}}]{drissen19}
{Drissen}, L., {Martin}, T., {Rousseau-Nepton}, L., {et~al.} 2019, \mnras, 485,
  3930

\bibitem[{{Elmegreen} {et~al.}(1998){Elmegreen}, {Chromey}, \&
  {Warren}}]{elmegreen98}
{Elmegreen}, D.~M., {Chromey}, F.~R., \& {Warren}, A.~R. 1998, \aj, 116, 2834

\bibitem[{{Emsellem}(2004)}]{emsellem04}
{Emsellem}, E. 2004, in Coevolution of Black Holes and Galaxies, ed. L.~C.
  {Ho}, 11

\bibitem[{{Emsellem} {et~al.}(2001){Emsellem}, {Greusard}, {Combes}, {Friedli},
  {Leon}, {P{\'e}contal}, \& {Wozniak}}]{emsellem01}
{Emsellem}, E., {Greusard}, D., {Combes}, F., {et~al.} 2001, \aap, 368, 52

\bibitem[{{Emsellem} {et~al.}(2015){Emsellem}, {Renaud}, {Bournaud},
  {Elmegreen}, {Combes}, \& {Gabor}}]{emsellem15}
{Emsellem}, E., {Renaud}, F., {Bournaud}, F., {et~al.} 2015, \mnras, 446, 2468

\bibitem[{{Emsellem} {et~al.}(2021){Emsellem}, {Schinnerer}, {Santoro},
  {Belfiore}, {Pessa}, {McElroy}, {Blanc}, {Congiu}, {Groves}, {Ho}, {Kreckel},
  {Razza}, {Sanchez-Blazquez}, {Egorov}, {Faesi}, {Klessen}, {Leroy}, {Meidt},
  {Querejeta}, {Rosolowsky}, {Scheuermann}, {Anand}, {Barnes},
  {Be{\v{s}}li{\'c}}, {Bigiel}, {Boquien}, {Cao}, {Chevance}, {Dale},
  {Eibensteiner}, {Glover}, {Grasha}, {Henshaw}, {Hughes}, {Koch}, {Kruijssen},
  {Lee}, {Liu}, {Pan}, {Pety}, {Saito}, {Sandstrom}, {Schruba}, {Sun},
  {Thilker}, {Usero}, {Watkins}, \& {Williams}}]{emsellem21}
{Emsellem}, E., {Schinnerer}, E., {Santoro}, F., {et~al.} 2021, arXiv e-prints,
  arXiv:2110.03708

\bibitem[{{Epinat} {et~al.}(2008){Epinat}, {Amram}, {Marcelin}, {Balkowski},
  {Daigle}, {Hernandez}, {Chemin}, {Carignan}, {Gach}, \& {Balard}}]{epinat08}
{Epinat}, B., {Amram}, P., {Marcelin}, M., {et~al.} 2008, \mnras, 388, 500

\bibitem[{{Erroz-Ferrer} {et~al.}(2019){Erroz-Ferrer}, {Carollo}, {den Brok},
  {Onodera}, {Brinchmann}, {Marino}, {Monreal-Ibero}, {Schaye}, {Woo},
  {Cibinel}, {Debattista}, {Inami}, {Maseda}, {Richard}, {Tacchella}, \&
  {Wisotzki}}]{erroz-ferrer19}
{Erroz-Ferrer}, S., {Carollo}, C.~M., {den Brok}, M., {et~al.} 2019, \mnras,
  484, 5009

\bibitem[{{Farage} {et~al.}(2010){Farage}, {McGregor}, {Dopita}, \&
  {Bicknell}}]{farage10}
{Farage}, C.~L., {McGregor}, P.~J., {Dopita}, M.~A., \& {Bicknell}, G.~V. 2010,
  \apj, 724, 267

\bibitem[{{Fathi} {et~al.}(2008){Fathi}, {Beckman}, {Lundgren}, {Carignan},
  {Hernandez}, {Amram}, {Balard}, {Boulesteix}, {Gach}, {Knapen}, \&
  {Rela{\~n}o}}]{fathi08}
{Fathi}, K., {Beckman}, J.~E., {Lundgren}, A.~A., {et~al.} 2008, \apjl, 675,
  L17

\bibitem[{{Ferguson} {et~al.}(1996){Ferguson}, {Wyse}, {Gallagher}, \&
  {Hunter}}]{ferguson96}
{Ferguson}, A. M.~N., {Wyse}, R. F.~G., {Gallagher}, J.~S., I., \& {Hunter},
  D.~A. 1996, \aj, 111, 2265

\bibitem[{{Gadotti} {et~al.}(2020){Gadotti}, {Bittner}, {Falc{\'o}n-Barroso},
  {M{\'e}ndez-Abreu}, {Kim}, {Fragkoudi}, {de Lorenzo-C{\'a}ceres}, {Leaman},
  {Neumann}, {Querejeta}, {S{\'a}nchez-Bl{\'a}zquez}, {Martig},
  {Mart{\'\i}n-Navarro}, {P{\'e}rez}, {Seidel}, \& {van de Ven}}]{gadotti20}
{Gadotti}, D.~A., {Bittner}, A., {Falc{\'o}n-Barroso}, J., {et~al.} 2020, \aap,
  643, A14

\bibitem[{{Gadotti} {et~al.}(2019){Gadotti}, {S{\'a}nchez-Bl{\'a}zquez},
  {Falc{\'o}n-Barroso}, {Husemann}, {Seidel}, {P{\'e}rez}, {de
  Lorenzo-C{\'a}ceres}, {Martinez-Valpuesta}, {Fragkoudi}, {Leung}, {van de
  Ven}, {Leaman}, {Coelho}, {Martig}, {Kim}, {Neumann}, \&
  {Querejeta}}]{gadotti19}
{Gadotti}, D.~A., {S{\'a}nchez-Bl{\'a}zquez}, P., {Falc{\'o}n-Barroso}, J.,
  {et~al.} 2019, \mnras, 482, 506

\bibitem[{{Gaia Collaboration} {et~al.}(2018){Gaia Collaboration}, {Brown},
  {Vallenari}, {Prusti}, {de Bruijne}, {Babusiaux}, {Bailer-Jones}, {Biermann},
  {Evans}, {Eyer}, {Jansen}, {Jordi}, {Klioner}, {Lammers}, {Lindegren},
  {Luri}, {Mignard}, {Panem}, {Pourbaix}, {Randich}, {Sartoretti}, {Siddiqui},
  {Soubiran}, {van Leeuwen}, {Walton}, {Arenou}, {Bastian}, {Cropper},
  {Drimmel}, {Katz}, {Lattanzi}, {Bakker}, {Cacciari}, {Casta{\~n}eda},
  {Chaoul}, {Cheek}, {De Angeli}, {Fabricius}, {Guerra}, {Holl}, {Masana},
  {Messineo}, {Mowlavi}, {Nienartowicz}, {Panuzzo}, {Portell}, {Riello},
  {Seabroke}, {Tanga}, {Th{\'e}venin}, {Gracia-Abril}, {Comoretto},
  {Garcia-Reinaldos}, {Teyssier}, {Altmann}, {Andrae}, {Audard},
  {Bellas-Velidis}, {Benson}, {Berthier}, {Blomme}, {Burgess}, {Busso},
  {Carry}, {Cellino}, {Clementini}, {Clotet}, {Creevey}, {Davidson}, {De
  Ridder}, {Delchambre}, {Dell'Oro}, {Ducourant},
  {Fern{\'a}ndez-Hern{\'a}ndez}, {Fouesneau}, {Fr{\'e}mat}, {Galluccio},
  {Garc{\'\i}a-Torres}, {Gonz{\'a}lez-N{\'u}{\~n}ez}, {Gonz{\'a}lez-Vidal},
  {Gosset}, {Guy}, {Halbwachs}, {Hambly}, {Harrison}, {Hern{\'a}ndez},
  {Hestroffer}, {Hodgkin}, {Hutton}, {Jasniewicz}, {Jean-Antoine-Piccolo},
  {Jordan}, {Korn}, {Krone-Martins}, {Lanzafame}, {Lebzelter}, {L{\"o}ffler},
  {Manteiga}, {Marrese}, {Mart{\'\i}n-Fleitas}, {Moitinho}, {Mora}, {Muinonen},
  {Osinde}, {Pancino}, {Pauwels}, {Petit}, {Recio-Blanco}, {Richards},
  {Rimoldini}, {Robin}, {Sarro}, {Siopis}, {Smith}, {Sozzetti}, {S{\"u}veges},
  {Torra}, {van Reeven}, {Abbas}, {Abreu Aramburu}, {Accart}, {Aerts},
  {Altavilla}, {{\'A}lvarez}, {Alvarez}, {Alves}, {Anderson}, {Andrei},
  {Anglada Varela}, {Antiche}, {Antoja}, {Arcay}, {Astraatmadja}, {Bach},
  {Baker}, {Balaguer-N{\'u}{\~n}ez}, {Balm}, {Barache}, {Barata}, {Barbato},
  {Barblan}, {Barklem}, {Barrado}, {Barros}, {Barstow}, {Bartholom{\'e}
  Mu{\~n}oz}, {Bassilana}, {Becciani}, {Bellazzini}, {Berihuete}, {Bertone},
  {Bianchi}, {Bienaym{\'e}}, {Blanco-Cuaresma}, {Boch}, {Boeche}, {Bombrun},
  {Borrachero}, {Bossini}, {Bouquillon}, {Bourda}, {Bragaglia}, {Bramante},
  {Breddels}, {Bressan}, {Brouillet}, {Br{\"u}semeister}, {Brugaletta},
  {Bucciarelli}, {Burlacu}, {Busonero}, {Butkevich}, {Buzzi}, {Caffau},
  {Cancelliere}, {Cannizzaro}, {Cantat-Gaudin}, {Carballo}, {Carlucci},
  {Carrasco}, {Casamiquela}, {Castellani}, {Castro-Ginard}, {Charlot},
  {Chemin}, {Chiavassa}, {Cocozza}, {Costigan}, {Cowell}, {Crifo}, {Crosta},
  {Crowley}, {Cuypers}, {Dafonte}, {Damerdji}, {Dapergolas}, {David}, {David},
  {de Laverny}, {De Luise}, {De March}, {de Martino}, {de Souza}, {de Torres},
  {Debosscher}, {del Pozo}, {Delbo}, {Delgado}, {Delgado}, {Di Matteo},
  {Diakite}, {Diener}, {Distefano}, {Dolding}, {Drazinos}, {Dur{\'a}n},
  {Edvardsson}, {Enke}, {Eriksson}, {Esquej}, {Eynard Bontemps}, {Fabre},
  {Fabrizio}, {Faigler}, {Falc{\~a}o}, {Farr{\`a}s Casas}, {Federici},
  {Fedorets}, {Fernique}, {Figueras}, {Filippi}, {Findeisen}, {Fonti},
  {Fraile}, {Fraser}, {Fr{\'e}zouls}, {Gai}, {Galleti}, {Garabato},
  {Garc{\'\i}a-Sedano}, {Garofalo}, {Garralda}, {Gavel}, {Gavras}, {Gerssen},
  {Geyer}, {Giacobbe}, {Gilmore}, {Girona}, {Giuffrida}, {Glass}, {Gomes},
  {Granvik}, {Gueguen}, {Guerrier}, {Guiraud}, {Guti{\'e}rrez-S{\'a}nchez},
  {Haigron}, {Hatzidimitriou}, {Hauser}, {Haywood}, {Heiter}, {Helmi}, {Heu},
  {Hilger}, {Hobbs}, {Hofmann}, {Holland}, {Huckle}, {Hypki}, {Icardi},
  {Jan{\ss}en}, {Jevardat de Fombelle}, {Jonker}, {Juh{\'a}sz}, {Julbe},
  {Karampelas}, {Kewley}, {Klar}, {Kochoska}, {Kohley}, {Kolenberg},
  {Kontizas}, {Kontizas}, {Koposov}, {Kordopatis}, {Kostrzewa-Rutkowska},
  {Koubsky}, {Lambert}, {Lanza}, {Lasne}, {Lavigne}, {Le Fustec}, {Le
  Poncin-Lafitte}, {Lebreton}, {Leccia}, {Leclerc}, {Lecoeur-Taibi},
  {Lenhardt}, {Leroux}, {Liao}, {Licata}, {Lindstr{\o}m}, {Lister}, {Livanou},
  {Lobel}, {L{\'o}pez}, {Managau}, {Mann}, {Mantelet}, {Marchal}, {Marchant},
  {Marconi}, {Marinoni}, {Marschalk{\'o}}, {Marshall}, {Martino}, {Marton},
  {Mary}, {Massari}, {Matijevi{\v{c}}}, {Mazeh}, {McMillan}, {Messina},
  {Michalik}, {Millar}, {Molina}, {Molinaro}, {Moln{\'a}r}, {Montegriffo},
  {Mor}, {Morbidelli}, {Morel}, {Morris}, {Mulone}, {Muraveva}, {Musella},
  {Nelemans}, {Nicastro}, {Noval}, {O'Mullane}, {Ord{\'e}novic},
  {Ord{\'o}{\~n}ez-Blanco}, {Osborne}, {Pagani}, {Pagano}, {Pailler},
  {Palacin}, {Palaversa}, {Panahi}, {Pawlak}, {Piersimoni}, {Pineau}, {Plachy},
  {Plum}, {Poggio}, {Poujoulet}, {Pr{\v{s}}a}, {Pulone}, {Racero}, {Ragaini},
  {Rambaux}, {Ramos-Lerate}, {Regibo}, {Reyl{\'e}}, {Riclet}, {Ripepi}, {Riva},
  {Rivard}, {Rixon}, {Roegiers}, {Roelens}, {Romero-G{\'o}mez}, {Rowell},
  {Royer}, {Ruiz-Dern}, {Sadowski}, {Sagrist{\`a} Sell{\'e}s}, {Sahlmann},
  {Salgado}, {Salguero}, {Sanna}, {Santana-Ros}, {Sarasso}, {Savietto},
  {Schultheis}, {Sciacca}, {Segol}, {Segovia}, {S{\'e}gransan}, {Shih},
  {Siltala}, {Silva}, {Smart}, {Smith}, {Solano}, {Solitro}, {Sordo}, {Soria
  Nieto}, {Souchay}, {Spagna}, {Spoto}, {Stampa}, {Steele},
  {Steidelm{\"u}ller}, {Stephenson}, {Stoev}, {Suess}, {Surdej}, {Szabados},
  {Szegedi-Elek}, {Tapiador}, {Taris}, {Tauran}, {Taylor}, {Teixeira},
  {Terrett}, {Teyssandier}, {Thuillot}, {Titarenko}, {Torra Clotet}, {Turon},
  {Ulla}, {Utrilla}, {Uzzi}, {Vaillant}, {Valentini}, {Valette}, {van Elteren},
  {Van Hemelryck}, {van Leeuwen}, {Vaschetto}, {Vecchiato}, {Veljanoski},
  {Viala}, {Vicente}, {Vogt}, {von Essen}, {Voss}, {Votruba}, {Voutsinas},
  {Walmsley}, {Weiler}, {Wertz}, {Wevers}, {Wyrzykowski}, {Yoldas},
  {{\v{Z}}erjal}, {Ziaeepour}, {Zorec}, {Zschocke}, {Zucker}, {Zurbach}, \&
  {Zwitter}}]{gaia_dr2}
{Gaia Collaboration}, {Brown}, A.~G.~A., {Vallenari}, A., {et~al.} 2018, \aap,
  616, A1

\bibitem[{{Girardi} {et~al.}(2000){Girardi}, {Bressan}, {Bertelli}, \&
  {Chiosi}}]{girardi00}
{Girardi}, L., {Bressan}, A., {Bertelli}, G., \& {Chiosi}, C. 2000, \aaps, 141,
  371

\bibitem[{{Gu{\'e}rou} {et~al.}(2017){Gu{\'e}rou}, {Krajnovi{\'c}}, {Epinat},
  {Contini}, {Emsellem}, {Bouch{\'e}}, {Bacon}, {Michel-Dansac}, {Richard},
  {Weilbacher}, {Schaye}, {Marino}, {den Brok}, \& {Erroz- Ferrer}}]{guerou17}
{Gu{\'e}rou}, A., {Krajnovi{\'c}}, D., {Epinat}, B., {et~al.} 2017, \aap, 608,
  A5

\bibitem[{{Hadfield} {et~al.}(2005){Hadfield}, {Crowther}, {Schild}, \&
  {Schmutz}}]{hadfield05}
{Hadfield}, L.~J., {Crowther}, P.~A., {Schild}, H., \& {Schmutz}, W. 2005,
  \aap, 439, 265

\bibitem[{{Haffner} {et~al.}(2009){Haffner}, {Dettmar}, {Beckman}, {Wood},
  {Slavin}, {Giammanco}, {Madsen}, {Zurita}, \& {Reynolds}}]{haffner09}
{Haffner}, L.~M., {Dettmar}, R.~J., {Beckman}, J.~E., {et~al.} 2009, Reviews of
  Modern Physics, 81, 969

\bibitem[{{Heald} {et~al.}(2016){Heald}, {de Blok}, {Lucero}, {Carignan},
  {Jarrett}, {Elson}, {Oozeer}, {Randriamampandry}, \& {van Zee}}]{heald16}
{Heald}, G., {de Blok}, W.~J.~G., {Lucero}, D., {et~al.} 2016, \mnras, 462,
  1238

\bibitem[{{Hoopes} \& {Walterbos}(2000)}]{hoopes00}
{Hoopes}, C.~G. \& {Walterbos}, R. A.~M. 2000, \apj, 541, 597

\bibitem[{{Hoopes} \& {Walterbos}(2003)}]{hoopes03}
{Hoopes}, C.~G. \& {Walterbos}, R. A.~M. 2003, \apj, 586, 902

\bibitem[{{Hoopes} {et~al.}(1996){Hoopes}, {Walterbos}, \&
  {Greenwalt}}]{hoopes96}
{Hoopes}, C.~G., {Walterbos}, R. A.~M., \& {Greenwalt}, B.~E. 1996, \aj, 112,
  1429

\bibitem[{{Hopkins} {et~al.}(2013){Hopkins}, {Narayanan}, \&
  {Murray}}]{hopkins13}
{Hopkins}, P.~F., {Narayanan}, D., \& {Murray}, N. 2013, \mnras, 432, 2647

\bibitem[{{Hopkins} {et~al.}(2018){Hopkins}, {Wetzel}, {Kere{\v{s}}},
  {Faucher-Gigu{\`e}re}, {Quataert}, {Boylan-Kolchin}, {Murray}, {Hayward},
  {Garrison-Kimmel}, {Hummels}, {Feldmann}, {Torrey}, {Ma},
  {Angl{\'e}s-Alc{\'a}zar}, {Su}, {Orr}, {Schmitz}, {Escala}, {Sanderson},
  {Grudi{\'c}}, {Hafen}, {Kim}, {Fitts}, {Bullock}, {Wheeler}, {Chan},
  {Elbert}, \& {Narayanan}}]{hopkins18}
{Hopkins}, P.~F., {Wetzel}, A., {Kere{\v{s}}}, D., {et~al.} 2018, \mnras, 480,
  800

\bibitem[{{Houghton} \& {Thatte}(2008)}]{houghton08}
{Houghton}, R.~C.~W. \& {Thatte}, N. 2008, \mnras, 385, 1110

\bibitem[{{Howard} {et~al.}(2017){Howard}, {Pudritz}, \& {Harris}}]{howard17}
{Howard}, C.~S., {Pudritz}, R.~E., \& {Harris}, W.~E. 2017, \mnras, 470, 3346

\bibitem[{{Jacobs} {et~al.}(2009){Jacobs}, {Rizzi}, {Tully}, {Shaya},
  {Makarov}, \& {Makarova}}]{jacobs09}
{Jacobs}, B.~A., {Rizzi}, L., {Tully}, R.~B., {et~al.} 2009, \aj, 138, 332

\bibitem[{{Jones} {et~al.}(2017){Jones}, {Kauffmann}, {D'Souza}, {Bizyaev},
  {Law}, {Haffner}, {Bah{\'e}}, {Andrews}, {Bershady}, {Brownstein}, {Bundy},
  {Cherinka}, {Diamond-Stanic}, {Drory}, {Riffel}, {S{\'a}nchez}, {Thomas},
  {Wake}, {Yan}, \& {Zhang}}]{jones17}
{Jones}, A., {Kauffmann}, G., {D'Souza}, R., {et~al.} 2017, \aap, 599, A141

\bibitem[{{Kamann}(2018)}]{kamann18}
{Kamann}, S. 2018, {PampelMuse: Crowded-field 3D spectroscopy}

\bibitem[{{Kauffmann} {et~al.}(2003){Kauffmann}, {Heckman}, {Tremonti},
  {Brinchmann}, {Charlot}, {White}, {Ridgway}, {Brinkmann}, {Fukugita}, {Hall},
  {Ivezi{\'c}}, {Richards}, \& {Schneider}}]{kauffmann03}
{Kauffmann}, G., {Heckman}, T.~M., {Tremonti}, C., {et~al.} 2003, \mnras, 346,
  1055

\bibitem[{{Kewley} {et~al.}(2001){Kewley}, {Dopita}, {Sutherland}, {Heisler},
  \& {Trevena}}]{kewley01}
{Kewley}, L.~J., {Dopita}, M.~A., {Sutherland}, R.~S., {Heisler}, C.~A., \&
  {Trevena}, J. 2001, \apj, 556, 121

\bibitem[{{Kewley} {et~al.}(2019){Kewley}, {Nicholls}, \&
  {Sutherland}}]{kewley19}
{Kewley}, L.~J., {Nicholls}, D.~C., \& {Sutherland}, R.~S. 2019, \araa, 57, 511

\bibitem[{{Knapen} {et~al.}(2010){Knapen}, {Sharp}, {Ryder},
  {Falc{\'o}n-Barroso}, {Fathi}, \& {Guti{\'e}rrez}}]{knapen10}
{Knapen}, J.~H., {Sharp}, R.~G., {Ryder}, S.~D., {et~al.} 2010, \mnras, 408,
  797

\bibitem[{{Koch} {et~al.}(2021){Koch}, {Rosolowsky}, {Leroy}, {Chastenet},
  {Chiang (江宜達)}, {Dalcanton}, {Kepley}, {Sandstrom}, {Schruba},
  {Stanimirovi{\'c}}, {Utomo}, \& {Williams}}]{koch21}
{Koch}, E.~W., {Rosolowsky}, E.~W., {Leroy}, A.~K., {et~al.} 2021, \mnras, 504,
  1801

\bibitem[{{Kreckel} {et~al.}(2016){Kreckel}, {Blanc}, {Schinnerer}, {Groves},
  {Adamo}, {Hughes}, \& {Meidt}}]{kreckel16}
{Kreckel}, K., {Blanc}, G.~A., {Schinnerer}, E., {et~al.} 2016, \apj, 827, 103

\bibitem[{{Kreckel} {et~al.}(2019){Kreckel}, {Ho}, {Blanc}, {Groves},
  {Santoro}, {Schinnerer}, {Bigiel}, {Chevance}, {Congiu}, {Emsellem}, {Faesi},
  {Glover}, {Grasha}, {Kruijssen}, {Lang}, {Leroy}, {Meidt}, {McElroy}, {Pety},
  {Rosolowsky}, {Saito}, {Sandstrom}, {Sanchez-Blazquez}, \&
  {Schruba}}]{kreckel19}
{Kreckel}, K., {Ho}, I.~T., {Blanc}, G.~A., {et~al.} 2019, \apj, 887, 80

\bibitem[{{Krumholz} {et~al.}(2014){Krumholz}, {Bate}, {Arce}, {Dale},
  {Gutermuth}, {Klein}, {Li}, {Nakamura}, \& {Zhang}}]{krumholz14_review}
{Krumholz}, M.~R., {Bate}, M.~R., {Arce}, H.~G., {et~al.} 2014, in Protostars
  and Planets VI, ed. H.~{Beuther}, R.~S. {Klessen}, C.~P. {Dullemond}, \&
  T.~{Henning}, 243

\bibitem[{{Krumholz} \& {Kruijssen}(2015)}]{krumholz15}
{Krumholz}, M.~R. \& {Kruijssen}, J.~M.~D. 2015, \mnras, 453, 739

\bibitem[{{Krumholz} {et~al.}(2017){Krumholz}, {Kruijssen}, \&
  {Crocker}}]{krumholz17}
{Krumholz}, M.~R., {Kruijssen}, J.~M.~D., \& {Crocker}, R.~M. 2017, \mnras,
  466, 1213

\bibitem[{{Lacerda} {et~al.}(2018){Lacerda}, {Cid Fernandes}, {Couto},
  {Stasi{\'n}ska}, {Garc{\'\i}a-Benito}, {Vale Asari}, {P{\'e}rez},
  {Gonz{\'a}lez Delgado}, {S{\'a}nchez}, \& {de Amorim}}]{lacerda18}
{Lacerda}, E.~A.~D., {Cid Fernandes}, R., {Couto}, G.~S., {et~al.} 2018,
  \mnras, 474, 3727

\bibitem[{{Leroy} {et~al.}(2015{\natexlab{a}}){Leroy}, {Bolatto}, {Ostriker},
  {Rosolowsky}, {Walter}, {Warren}, {Donovan Meyer}, {Hodge}, {Meier}, {Ott},
  {Sandstrom}, {Schruba}, {Veilleux}, \& {Zwaan}}]{leroy15a}
{Leroy}, A.~K., {Bolatto}, A.~D., {Ostriker}, E.~C., {et~al.}
  2015{\natexlab{a}}, \apj, 801, 25

\bibitem[{{Leroy} {et~al.}(2021{\natexlab{a}}){Leroy}, {Hughes}, {Liu}, {Pety},
  {Rosolowsky}, {Saito}, {Schinnerer}, {Schruba}, {Usero}, {Faesi}, {Herrera},
  {Chevance}, {Hygate}, {Kepley}, {Koch}, {Querejeta}, {Sliwa}, {Will},
  {Wilson}, {Anand}, {Barnes}, {Belfiore}, {Be{\v{s}}li{\'c}}, {Bigiel},
  {Blanc}, {Bolatto}, {Boquien}, {Cao}, {Chandar}, {Chastenet}, {Chiang},
  {Congiu}, {Dale}, {Deger}, {den Brok}, {Eibensteiner}, {Emsellem},
  {Garc{\'\i}a-Rodr{\'\i}guez}, {Glover}, {Grasha}, {Groves}, {Henshaw},
  {Jim{\'e}nez Donaire}, {Kim}, {Klessen}, {Kreckel}, {Kruijssen}, {Larson},
  {Lee}, {Mayker}, {McElroy}, {Meidt}, {Mok}, {Pan}, {Puschnig}, {Razza},
  {S{\'a}nchez-Bl'azquez}, {Sandstrom}, {Santoro}, {Sardone}, {Scheuermann},
  {Sun}, {Thilker}, {Turner}, {Ubeda}, {Utomo}, {Watkins}, \&
  {Williams}}]{leroy21_pipeline}
{Leroy}, A.~K., {Hughes}, A., {Liu}, D., {et~al.} 2021{\natexlab{a}}, \apjs,
  255, 19

\bibitem[{{Leroy} {et~al.}(2021{\natexlab{b}}){Leroy}, {Rosolowsky}, {Usero},
  {Sandstrom}, {Schinnerer}, {Schruba}, {Bolatto}, {Sun}, {Barnes}, {Belfiore},
  {Bigiel}, {den Brok}, {Cao}, {Chiang}, {Chevance}, {Dale}, {Eibensteiner},
  {Faesi}, {Glover}, {Hughes}, {Jim{\'e}nez Donaire}, {Klessen}, {Koch},
  {Kruijssen}, {Liu}, {Meidt}, {Pan}, {Pety}, {Puschnig}, {Querejeta}, {Saito},
  {Sardone}, {Watkins}, {Weiss}, \& {Williams}}]{leroy21_subm}
{Leroy}, A.~K., {Rosolowsky}, E., {Usero}, A., {et~al.} 2021{\natexlab{b}},
  arXiv e-prints, arXiv:2109.11583

\bibitem[{{Leroy} {et~al.}(2021{\natexlab{c}}){Leroy}, {Schinnerer}, {Hughes},
  {Rosolowsky}, {Pety}, {Schruba}, {Usero}, {Blanc}, {Chevance}, {Emsellem},
  {Faesi}, {Herrera}, {Liu}, {Meidt}, {Querejeta}, {Saito}, {Sandstrom}, {Sun},
  {Williams}, {Anand}, {Barnes}, {Behrens}, {Belfiore}, {Benincasa},
  {Be{\v{s}}li{\'c}}, {Bigiel}, {Bolatto}, {den Brok}, {Cao}, {Chandar},
  {Chastenet}, {Chiang}, {Congiu}, {Dale}, {Deger}, {Eibensteiner}, {Egorov},
  {Garc{\'\i}a-Rodr{\'\i}guez}, {Glover}, {Grasha}, {Henshaw}, {Ho}, {Kepley},
  {Kim}, {Klessen}, {Kreckel}, {Koch}, {Kruijssen}, {Larson}, {Lee}, {Lopez},
  {Machado}, {Mayker}, {McElroy}, {Murphy}, {Ostriker}, {Pan}, {Pessa},
  {Puschnig}, {Razza}, {S{\'a}nchez-Bl{\'a}zquez}, {Santoro}, {Sardone},
  {Scheuermann}, {Sliwa}, {Sormani}, {Stuber}, {Thilker}, {Turner}, {Utomo},
  {Watkins}, \& {Whitmore}}]{leroy21}
{Leroy}, A.~K., {Schinnerer}, E., {Hughes}, A., {et~al.} 2021{\natexlab{c}},
  \apjs, 257, 43

\bibitem[{{Leroy} {et~al.}(2015{\natexlab{b}}){Leroy}, {Walter}, {Martini},
  {Roussel}, {Sandstrom}, {Ott}, {Weiss}, {Bolatto}, {Schuster}, \&
  {Dessauges-Zavadsky}}]{leroy15b}
{Leroy}, A.~K., {Walter}, F., {Martini}, P., {et~al.} 2015{\natexlab{b}}, \apj,
  814, 83

\bibitem[{{Levy} {et~al.}(2019){Levy}, {Bolatto}, {S{\'a}nchez}, {Blitz},
  {Colombo}, {Kalinova}, {L{\'o}pez-Cob{\'a}}, {Ostriker}, {Teuben}, {Utomo},
  {Vogel}, \& {Wong}}]{levy19}
{Levy}, R.~C., {Bolatto}, A.~D., {S{\'a}nchez}, S.~F., {et~al.} 2019, \apj,
  882, 84

\bibitem[{{Lundgren} {et~al.}(2004){Lundgren}, {Olofsson}, {Wiklind}, \&
  {Rydbeck}}]{lundgren04}
{Lundgren}, A.~A., {Olofsson}, H., {Wiklind}, T., \& {Rydbeck}, G. 2004, \aap,
  422, 865

\bibitem[{{Luridiana} {et~al.}(2015){Luridiana}, {Morisset}, \&
  {Shaw}}]{Luridiana15}
{Luridiana}, V., {Morisset}, C., \& {Shaw}, R.~A. 2015, \aap, 573, A42

\bibitem[{{Madsen} {et~al.}(2006){Madsen}, {Reynolds}, \& {Haffner}}]{madsen06}
{Madsen}, G.~J., {Reynolds}, R.~J., \& {Haffner}, L.~M. 2006, \apj, 652, 401

\bibitem[{{Maiolino} \& {Mannucci}(2019)}]{maiolino19}
{Maiolino}, R. \& {Mannucci}, F. 2019, \aapr, 27, 3

\bibitem[{{Makarov} {et~al.}(2014){Makarov}, {Prugniel}, {Terekhova},
  {Courtois}, \& {Vauglin}}]{makarov14}
{Makarov}, D., {Prugniel}, P., {Terekhova}, N., {Courtois}, H., \& {Vauglin},
  I. 2014, \aap, 570, A13

\bibitem[{{Martins} {et~al.}(2005){Martins}, {Schaerer}, \&
  {Hillier}}]{martins05}
{Martins}, F., {Schaerer}, D., \& {Hillier}, D.~J. 2005, \aap, 436, 1049

\bibitem[{{Mast} {et~al.}(2006){Mast}, {D{\'\i}az}, \& {Ag{\"u}ero}}]{mast06}
{Mast}, D., {D{\'\i}az}, R.~J., \& {Ag{\"u}ero}, M.~P. 2006, \aj, 131, 1394

\bibitem[{{McLeod} {et~al.}(2021){McLeod}, {Ali}, {Chevance}, {Della Bruna},
  {Schruba}, {Stevance}, {Adamo}, {Kruijssen}, {Longmore}, {Weisz}, \&
  {Zeidler}}]{mcleod21}
{McLeod}, A.~F., {Ali}, A.~A., {Chevance}, M., {et~al.} 2021, \mnras, 508, 5425

\bibitem[{{McLeod} {et~al.}(2019){McLeod}, {Dale}, {Evans}, {Ginsburg},
  {Kruijssen}, {Pellegrini}, {Ramsay}, \& {Testi}}]{mcleod19_lmc}
{McLeod}, A.~F., {Dale}, J.~E., {Evans}, C.~J., {et~al.} 2019, \mnras, 486,
  5263

\bibitem[{{McLeod} {et~al.}(2020){McLeod}, {Kruijssen}, {Weisz}, {Zeidler},
  {Schruba}, {Dalcanton}, {Longmore}, {Chevance}, {Faesi}, \&
  {Byler}}]{mcleod20}
{McLeod}, A.~F., {Kruijssen}, J.~M.~D., {Weisz}, D.~R., {et~al.} 2020, \apj,
  891, 25

\bibitem[{{Miller} {et~al.}(2009){Miller}, {Bregman}, \& {Wakker}}]{miller09}
{Miller}, E.~D., {Bregman}, J.~N., \& {Wakker}, B.~P. 2009, \apj, 692, 470

\bibitem[{{Morisset} {et~al.}(2016){Morisset}, {Delgado-Inglada},
  {S{\'a}nchez}, {Galbany}, {Garc{\'\i}a-Benito}, {Husemann}, {Marino}, {Mast},
  \& {Roth}}]{morisset16}
{Morisset}, C., {Delgado-Inglada}, G., {S{\'a}nchez}, S.~F., {et~al.} 2016,
  \aap, 594, A37

\bibitem[{{Oey} {et~al.}(2007){Oey}, {Meurer}, {Yelda}, {Furst},
  {Caballero-Nieves}, {Hanish}, {Levesque}, {Thilker}, {Walth},
  {Bland-Hawthorn}, {Dopita}, {Ferguson}, {Heckman}, {Doyle}, {Drinkwater},
  {Freeman}, {Kennicutt}, {Kilborn}, {Knezek}, {Koribalski}, {Meyer}, {Putman},
  {Ryan-Weber}, {Smith}, {Staveley-Smith}, {Webster}, {Werk}, \&
  {Zwaan}}]{oey07}
{Oey}, M.~S., {Meurer}, G.~R., {Yelda}, S., {et~al.} 2007, \apj, 661, 801

\bibitem[{{Osterbrock} \& {Ferland}(2006)}]{osterbrock}
{Osterbrock}, D.~E. \& {Ferland}, G.~J. 2006, {Astrophysics of gaseous nebulae
  and active galactic nuclei}

\bibitem[{{Paturel} {et~al.}(2003){Paturel}, {Petit}, {Prugniel}, {Theureau},
  {Rousseau}, {Brouty}, {Dubois}, \& {Cambr{\'e}sy}}]{paturel03}
{Paturel}, G., {Petit}, C., {Prugniel}, P., {et~al.} 2003, \aap, 412, 45

\bibitem[{{Piqueras L{\'o}pez} {et~al.}(2012){Piqueras L{\'o}pez}, {Davies},
  {Colina}, \& {Orban de Xivry}}]{piqueras_lopez12}
{Piqueras L{\'o}pez}, J., {Davies}, R., {Colina}, L., \& {Orban de Xivry}, G.
  2012, \apj, 752, 47

\bibitem[{{Plummer}(1911)}]{plummer1911}
{Plummer}, H.~C. 1911, \mnras, 71, 460

\bibitem[{{Poetrodjojo} {et~al.}(2019){Poetrodjojo}, {D'Agostino}, {Groves},
  {Kewley}, {Ho}, {Rich}, {Madore}, \& {Seibert}}]{poetrodjojo19}
{Poetrodjojo}, H., {D'Agostino}, J.~J., {Groves}, B., {et~al.} 2019, \mnras,
  487, 79

\bibitem[{{Popesso} {et~al.}(2019){Popesso}, {Concas}, {Morselli}, {Schreiber},
  {Rodighiero}, {Cresci}, {Belli}, {Erfanianfar}, {Mancini}, {Inami},
  {Dickinson}, {Ilbert}, {Pannella}, \& {Elbaz}}]{popesso19}
{Popesso}, P., {Concas}, A., {Morselli}, L., {et~al.} 2019, \mnras, 483, 3213

\bibitem[{{Rand}(1998)}]{rand98}
{Rand}, R.~J. 1998, \apj, 501, 137

\bibitem[{{Rautio} {et~al.}(2022){Rautio}, {Watkins}, {Comer{\'o}n}, {Salo},
  {D{\'\i}az-Garc{\'\i}a}, \& {Janz}}]{rautio22}
{Rautio}, R.~P.~V., {Watkins}, A.~E., {Comer{\'o}n}, S., {et~al.} 2022, arXiv
  e-prints, arXiv:2201.00566

\bibitem[{{Regan} \& {Teuben}(2003)}]{regan03}
{Regan}, M.~W. \& {Teuben}, P. 2003, \apj, 582, 723

\bibitem[{{Renaud} {et~al.}(2015){Renaud}, {Bournaud}, {Emsellem}, {Agertz},
  {Athanassoula}, {Combes}, {Elmegreen}, {Kraljic}, {Motte}, \&
  {Teyssier}}]{renaud15}
{Renaud}, F., {Bournaud}, F., {Emsellem}, E., {et~al.} 2015, \mnras, 454, 3299

\bibitem[{{Rich} {et~al.}(2011){Rich}, {Kewley}, \& {Dopita}}]{rich11}
{Rich}, J.~A., {Kewley}, L.~J., \& {Dopita}, M.~A. 2011, \apj, 734, 87

\bibitem[{{Rodrigues} {et~al.}(2009){Rodrigues}, {Dottori}, {D{\'\i}az},
  {Ag{\"u}ero}, \& {Mast}}]{rodrigues09}
{Rodrigues}, I., {Dottori}, H., {D{\'\i}az}, R.~J., {Ag{\"u}ero}, M.~P., \&
  {Mast}, D. 2009, \aj, 137, 4083

\bibitem[{{Rosolowsky} {et~al.}(2021){Rosolowsky}, {Hughes}, {Leroy}, {Sun},
  {Querejeta}, {Schruba}, {Usero}, {Herrera}, {Liu}, {Pety}, {Saito},
  {Be{\v{s}}li{\'c}}, {Bigiel}, {Blanc}, {Chevance}, {Dale}, {Deger}, {Faesi},
  {Glover}, {Henshaw}, {Klessen}, {Kruijssen}, {Larson}, {Lee}, {Meidt}, {Mok},
  {Schinnerer}, {Thilker}, \& {Williams}}]{rosolowsky21}
{Rosolowsky}, E., {Hughes}, A., {Leroy}, A.~K., {et~al.} 2021, \mnras, 502,
  1218

\bibitem[{{Rossa} \& {Dettmar}(2003)}]{rossa03}
{Rossa}, J. \& {Dettmar}, R.~J. 2003, \aap, 406, 493

\bibitem[{{Rousseau-Nepton} {et~al.}(2019){Rousseau-Nepton}, {Martin},
  {Robert}, {Drissen}, {Amram}, {Prunet}, {Martin}, {Moumen}, {Adamo},
  {Alarie}, {Barmby}, {Boselli}, {Bresolin}, {Bureau}, {Chemin}, {Fernandes},
  {Combes}, {Crowder}, {Della Bruna}, {Duarte Puertas}, {Egusa}, {Epinat},
  {Ksoll}, {Girard}, {G{\'o}mez Llanos}, {Gouliermis}, {Grasha}, {Higgs},
  {Hlavacek-Larrondo}, {Ho}, {Iglesias-P{\'a}ramo}, {Joncas}, {Kam}, {Karera},
  {Kennicutt}, {Klessen}, {Lianou}, {Liu}, {Liu}, {de Amorim}, {Lyman},
  {Martel}, {Mazzilli-Ciraulo}, {McLeod}, {Melchior}, {Millan}, {Moll{\'a}},
  {Momose}, {Morisset}, {Pan}, {Pati}, {Pellerin}, {Pellegrini}, {P{\'e}rez},
  {Petric}, {Plana}, {Rahner}, {Ruiz Lara}, {S{\'a}nchez-Menguiano},
  {Spekkens}, {Stasi{\'n}ska}, {Takamiya}, {Vale Asari}, \&
  {V{\'\i}lchez}}]{rousseau19}
{Rousseau-Nepton}, L., {Martin}, R.~P., {Robert}, C., {et~al.} 2019, \mnras,
  489, 5530

\bibitem[{{Russell} {et~al.}(2020){Russell}, {White}, {Long}, {Blair}, {Soria},
  \& {Winkler}}]{russell20}
{Russell}, T.~D., {White}, R.~L., {Long}, K.~S., {et~al.} 2020, \mnras, 495,
  479

\bibitem[{{Salo} {et~al.}(2015){Salo}, {Laurikainen}, {Laine}, {Comer{\'o}n},
  {Gadotti}, {Buta}, {Sheth}, {Zaritsky}, {Ho}, {Knapen}, {Athanassoula},
  {Bosma}, {Laine}, {Cisternas}, {Kim}, {Mu{\~n}oz-Mateos}, {Regan}, {Hinz},
  {Gil de Paz}, {Menendez-Delmestre}, {Mizusawa}, {Erroz-Ferrer}, {Meidt}, \&
  {Querejeta}}]{salo15}
{Salo}, H., {Laurikainen}, E., {Laine}, J., {et~al.} 2015, \apjs, 219, 4

\bibitem[{{S{\'a}nchez} {et~al.}(2012){S{\'a}nchez}, {Kennicutt}, {Gil de Paz},
  {van de Ven}, {V{\'\i}lchez}, {Wisotzki}, {Walcher}, {Mast}, {Aguerri},
  {Albiol-P{\'e}rez}, {Alonso-Herrero}, {Alves}, {Bakos}, {Bart{\'a}kov{\'a}},
  {Bland-Hawthorn}, {Boselli}, {Bomans}, {Castillo-Morales}, {Cortijo-Ferrero},
  {de Lorenzo-C{\'a}ceres}, {Del Olmo}, {Dettmar}, {D{\'\i}az}, {Ellis},
  {Falc{\'o}n-Barroso}, {Flores}, {Gallazzi}, {Garc{\'\i}a-Lorenzo},
  {Gonz{\'a}lez Delgado}, {Gruel}, {Haines}, {Hao}, {Husemann},
  {Igl{\'e}sias-P{\'a}ramo}, {Jahnke}, {Johnson}, {Jungwiert}, {Kalinova},
  {Kehrig}, {Kupko}, {L{\'o}pez-S{\'a}nchez}, {Lyubenova}, {Marino},
  {M{\'a}rmol-Queralt{\'o}}, {M{\'a}rquez}, {Masegosa}, {Meidt},
  {Mendez-Abreu}, {Monreal-Ibero}, {Montijo}, {Mour{\~a}o}, {Palacios-Navarro},
  {Papaderos}, {Pasquali}, {Peletier}, {P{\'e}rez}, {P{\'e}rez}, {Quirrenbach},
  {Rela{\~n}o}, {Rosales-Ortega}, {Roth}, {Ruiz-Lara},
  {S{\'a}nchez-Bl{\'a}zquez}, {Sengupta}, {Singh}, {Stanishev}, {Trager},
  {Vazdekis}, {Viironen}, {Wild}, {Zibetti}, \& {Ziegler}}]{sanchez12}
{S{\'a}nchez}, S.~F., {Kennicutt}, R.~C., {Gil de Paz}, A., {et~al.} 2012,
  \aap, 538, A8

\bibitem[{{S{\'a}nchez} {et~al.}(2015){S{\'a}nchez}, {P{\'e}rez},
  {Rosales-Ortega}, {Miralles-Caballero}, {L{\'o}pez-S{\'a}nchez},
  {Iglesias-P{\'a}ramo}, {Marino}, {S{\'a}nchez-Menguiano},
  {Garc{\'\i}a-Benito}, {Mast}, {Mendoza}, {Papaderos}, {Ellis}, {Galbany},
  {Kehrig}, {Monreal-Ibero}, {Gonz{\'a}lez Delgado}, {Moll{\'a}}, {Ziegler},
  {de Lorenzo-C{\'a}ceres}, {Mendez-Abreu}, {Bland -Hawthorn},
  {Bekerait{\.{e}}}, {Roth}, {Pasquali}, {D{\'\i}az}, {Bomans}, {van de Ven},
  \& {Wisotzki}}]{sanchez15}
{S{\'a}nchez}, S.~F., {P{\'e}rez}, E., {Rosales-Ortega}, F.~F., {et~al.} 2015,
  \aap, 574, A47

\bibitem[{{Sandstrom} {et~al.}(2013){Sandstrom}, {Leroy}, {Walter}, {Bolatto},
  {Croxall}, {Draine}, {Wilson}, {Wolfire}, {Calzetti}, {Kennicutt}, {Aniano},
  {Donovan Meyer}, {Usero}, {Bigiel}, {Brinks}, {de Blok}, {Crocker}, {Dale},
  {Engelbracht}, {Galametz}, {Groves}, {Hunt}, {Koda}, {Kreckel}, {Linz},
  {Meidt}, {Pellegrini}, {Rix}, {Roussel}, {Schinnerer}, {Schruba}, {Schuster},
  {Skibba}, {van der Laan}, {Appleton}, {Armus}, {Brandl}, {Gordon}, {Hinz},
  {Krause}, {Montiel}, {Sauvage}, {Schmiedeke}, {Smith}, \&
  {Vigroux}}]{sandstrom13}
{Sandstrom}, K.~M., {Leroy}, A.~K., {Walter}, F., {et~al.} 2013, \apj, 777, 5

\bibitem[{{Scannapieco} {et~al.}(2012){Scannapieco}, {Wadepuhl}, {Parry},
  {Navarro}, {Jenkins}, {Springel}, {Teyssier}, {Carlson}, {Couchman}, {Crain},
  {Dalla Vecchia}, {Frenk}, {Kobayashi}, {Monaco}, {Murante}, {Okamoto},
  {Quinn}, {Schaye}, {Stinson}, {Theuns}, {Wadsley}, {White}, \&
  {Woods}}]{scannapieco12}
{Scannapieco}, C., {Wadepuhl}, M., {Parry}, O.~H., {et~al.} 2012, \mnras, 423,
  1726

\bibitem[{{Schaye} {et~al.}(2015){Schaye}, {Crain}, {Bower}, {Furlong},
  {Schaller}, {Theuns}, {Dalla Vecchia}, {Frenk}, {McCarthy}, {Helly},
  {Jenkins}, {Rosas-Guevara}, {White}, {Baes}, {Booth}, {Camps}, {Navarro},
  {Qu}, {Rahmati}, {Sawala}, {Thomas}, \& {Trayford}}]{schaye15}
{Schaye}, J., {Crain}, R.~A., {Bower}, R.~G., {et~al.} 2015, \mnras, 446, 521

\bibitem[{{Schwarz}(1978)}]{schwarz78}
{Schwarz}, G. 1978, Annals of Statistics, 6, 461

\bibitem[{{Seon} \& {Witt}(2012)}]{seon12}
{Seon}, K.-I. \& {Witt}, A.~N. 2012, \apj, 758, 109

\bibitem[{{S{\'e}rsic} \& {Pastoriza}(1965)}]{sersic65}
{S{\'e}rsic}, J.~L. \& {Pastoriza}, M. 1965, \pasp, 77, 287

\bibitem[{{Sheth} {et~al.}(2010){Sheth}, {Regan}, {Hinz}, {Gil de Paz},
  {Men{\'e}ndez-Delmestre}, {Mu{\~n}oz-Mateos}, {Seibert}, {Kim},
  {Laurikainen}, {Salo}, {Gadotti}, {Laine}, {Mizusawa}, {Armus},
  {Athanassoula}, {Bosma}, {Buta}, {Capak}, {Jarrett}, {Elmegreen},
  {Elmegreen}, {Knapen}, {Koda}, {Helou}, {Ho}, {Madore}, {Masters},
  {Mobasher}, {Ogle}, {Peng}, {Schinnerer}, {Surace}, {Zaritsky},
  {Comer{\'o}n}, {de Swardt}, {Meidt}, {Kasliwal}, \& {Aravena}}]{sheth10}
{Sheth}, K., {Regan}, M., {Hinz}, J.~L., {et~al.} 2010, \pasp, 122, 1397

\bibitem[{{Shopbell} \& {Bland-Hawthorn}(1998)}]{shopbell98}
{Shopbell}, P.~L. \& {Bland-Hawthorn}, J. 1998, \apj, 493, 129

\bibitem[{{Silva-Villa} {et~al.}(2014){Silva-Villa}, {Adamo}, {Bastian},
  {Fouesneau}, \& {Zackrisson}}]{silva-villa14}
{Silva-Villa}, E., {Adamo}, A., {Bastian}, N., {Fouesneau}, M., \&
  {Zackrisson}, E. 2014, \mnras, 440, L116

\bibitem[{{Simkin} {et~al.}(1980){Simkin}, {Su}, \& {Schwarz}}]{simkin80}
{Simkin}, S.~M., {Su}, H.~J., \& {Schwarz}, M.~P. 1980, \apj, 237, 404

\bibitem[{{Sun} {et~al.}(2020){Sun}, {Leroy}, {Schinnerer}, {Hughes},
  {Rosolowsky}, {Querejeta}, {Schruba}, {Liu}, {Saito}, {Herrera}, {Faesi},
  {Usero}, {Pety}, {Kruijssen}, {Ostriker}, {Bigiel}, {Blanc}, {Bolatto},
  {Boquien}, {Chevance}, {Dale}, {Deger}, {Emsellem}, {Glover}, {Grasha},
  {Groves}, {Henshaw}, {Jimenez-Donaire}, {Kim}, {Klessen}, {Kreckel}, {Lee},
  {Meidt}, {Sandstrom}, {Sardone}, {Utomo}, \& {Williams}}]{sun20}
{Sun}, J., {Leroy}, A.~K., {Schinnerer}, E., {et~al.} 2020, \apjl, 901, L8

\bibitem[{{Thatte} {et~al.}(2000){Thatte}, {Tecza}, \& {Genzel}}]{thatte00}
{Thatte}, N., {Tecza}, M., \& {Genzel}, R. 2000, \aap, 364, L47

\bibitem[{{Thilker} {et~al.}(2002){Thilker}, {Walterbos}, {Braun}, \&
  {Hoopes}}]{thilker02}
{Thilker}, D.~A., {Walterbos}, R. A.~M., {Braun}, R., \& {Hoopes}, C.~G. 2002,
  \aj, 124, 3118

\bibitem[{{Vandenbroucke} {et~al.}(2018){Vandenbroucke}, {Wood}, {Girichidis},
  {Hill}, \& {Peters}}]{vandenbroucke18}
{Vandenbroucke}, B., {Wood}, K., {Girichidis}, P., {Hill}, A.~S., \& {Peters},
  T. 2018, \mnras, 476, 4032

\bibitem[{{Vazdekis} {et~al.}(2016){Vazdekis}, {Koleva}, {Ricciardelli},
  {R{\"o}ck}, \& {Falc{\'o}n-Barroso}}]{vazdekis16}
{Vazdekis}, A., {Koleva}, M., {Ricciardelli}, E., {R{\"o}ck}, B., \&
  {Falc{\'o}n-Barroso}, J. 2016, \mnras, 463, 3409

\bibitem[{{Veilleux} \& {Osterbrock}(1987)}]{veilleux87}
{Veilleux}, S. \& {Osterbrock}, D.~E. 1987, \apjs, 63, 295

\bibitem[{{Warner} {et~al.}(1973){Warner}, {Wright}, \& {Baldwin}}]{warner73}
{Warner}, P.~J., {Wright}, M.~C.~H., \& {Baldwin}, J.~E. 1973, \mnras, 163, 163

\bibitem[{{Weber} {et~al.}(2019){Weber}, {Pauldrach}, \& {Hoffmann}}]{weber19}
{Weber}, J.~A., {Pauldrach}, A.~W.~A., \& {Hoffmann}, T.~L. 2019, \aap, 622,
  A115

\bibitem[{{Weilbacher} {et~al.}(2018){Weilbacher}, {Monreal-Ibero}, {Verhamme},
  {Sandin}, {Steinmetz}, {Kollatschny}, {Krajnovi{\'c}}, {Kamann}, {Roth},
  {Erroz-Ferrer}, {Marino}, {Maseda}, {Wendt}, {Bacon}, {Dreizler}, {Richard},
  \& {Wisotzki}}]{weilbacher18}
{Weilbacher}, P.~M., {Monreal-Ibero}, A., {Verhamme}, A., {et~al.} 2018, \aap,
  611, A95

\bibitem[{{Weilbacher} {et~al.}(2020){Weilbacher}, {Palsa}, {Streicher},
  {Bacon}, {Urrutia}, {Wisotzki}, {Conseil}, {Husemann}, {Jarno}, {Kelz},
  {P{\'e}contal-Rousset}, {Richard}, {Roth}, {Selman}, \&
  {Vernet}}]{weilbacher20_pipeline}
{Weilbacher}, P.~M., {Palsa}, R., {Streicher}, O., {et~al.} 2020, \aap, 641,
  A28

\bibitem[{{Weilbacher} {et~al.}(2014){Weilbacher}, {Streicher}, {Urrutia},
  {P{\'e }contal-Rousset}, {Jarno}, \& {Bacon}}]{weilbacher14}
{Weilbacher}, P.~M., {Streicher}, O., {Urrutia}, T., {et~al.} 2014, in
  Astronomical Society of the Pacific Conference Series, Vol. 485, Astronomical
  Data Analysis Software and Systems XXIII, ed. N.~{Manset} \& P.~{Forshay},
  451

\bibitem[{{Williams} {et~al.}(2019){Williams}, {Hillis}, {Blair}, {Long},
  {Murphy}, {Dolphin}, {Khan}, \& {Dalcanton}}]{williams19}
{Williams}, B.~F., {Hillis}, T.~J., {Blair}, W.~P., {et~al.} 2019, \apj, 881,
  54

\bibitem[{{Winkler} {et~al.}(2017){Winkler}, {Blair}, \& {Long}}]{winkler17}
{Winkler}, P.~F., {Blair}, W.~P., \& {Long}, K.~S. 2017, \apj, 839, 83

\bibitem[{{Wood} \& {Mathis}(2004)}]{wood04}
{Wood}, K. \& {Mathis}, J.~S. 2004, \mnras, 353, 1126

\bibitem[{{Yukita} {et~al.}(2016){Yukita}, {Hornschemeier}, {Lehmer}, {Ptak},
  {Wik}, {Zezas}, {Antoniou}, {Maccarone}, {Replicon}, {Tyler}, {Venters},
  {Argo}, {Bechtol}, {Boggs}, {Christensen}, {Craig}, {Hailey}, {Harrison},
  {Krivonos}, {Kuntz}, {Stern}, \& {Zhang}}]{yukita16}
{Yukita}, M., {Hornschemeier}, A.~E., {Lehmer}, B.~D., {et~al.} 2016, \apj,
  824, 107

\bibitem[{{Zhang} {et~al.}(2017){Zhang}, {Yan}, {Bundy}, {Bershady}, {Haffner},
  {Walterbos}, {Maiolino}, {Tremonti}, {Thomas}, {Drory}, {Jones}, {Belfiore},
  {S{\'a}nchez}, {Diamond-Stanic}, {Bizyaev}, {Nitschelm}, {Andrews},
  {Brinkmann}, {Brownstein}, {Cheung}, {Li}, {Law}, {Roman Lopes}, {Oravetz},
  {Pan}, {Storchi Bergmann}, \& {Simmons}}]{zhang17}
{Zhang}, K., {Yan}, R., {Bundy}, K., {et~al.} 2017, \mnras, 466, 3217

\bibitem[{{Zurita} {et~al.}(2002){Zurita}, {Beckman}, {Rozas}, \&
  {Ryder}}]{zurita02}
{Zurita}, A., {Beckman}, J.~E., {Rozas}, M., \& {Ryder}, S. 2002, \aap, 386,
  801

\bibitem[{{Zurita} {et~al.}(2000){Zurita}, {Rozas}, \& {Beckman}}]{zurita00}
{Zurita}, A., {Rozas}, M., \& {Beckman}, J.~E. 2000, \aap, 363, 9

\end{thebibliography}

\begin{appendix}
\onecolumn

\section{Additional material on the kinematics fitting analysis}
\label{section:appendix_kinem_fitting}

\subsection{Rotation curves and 2D velocity fields}
\begin{figure}
\center
\includegraphics[width=9.1cm]{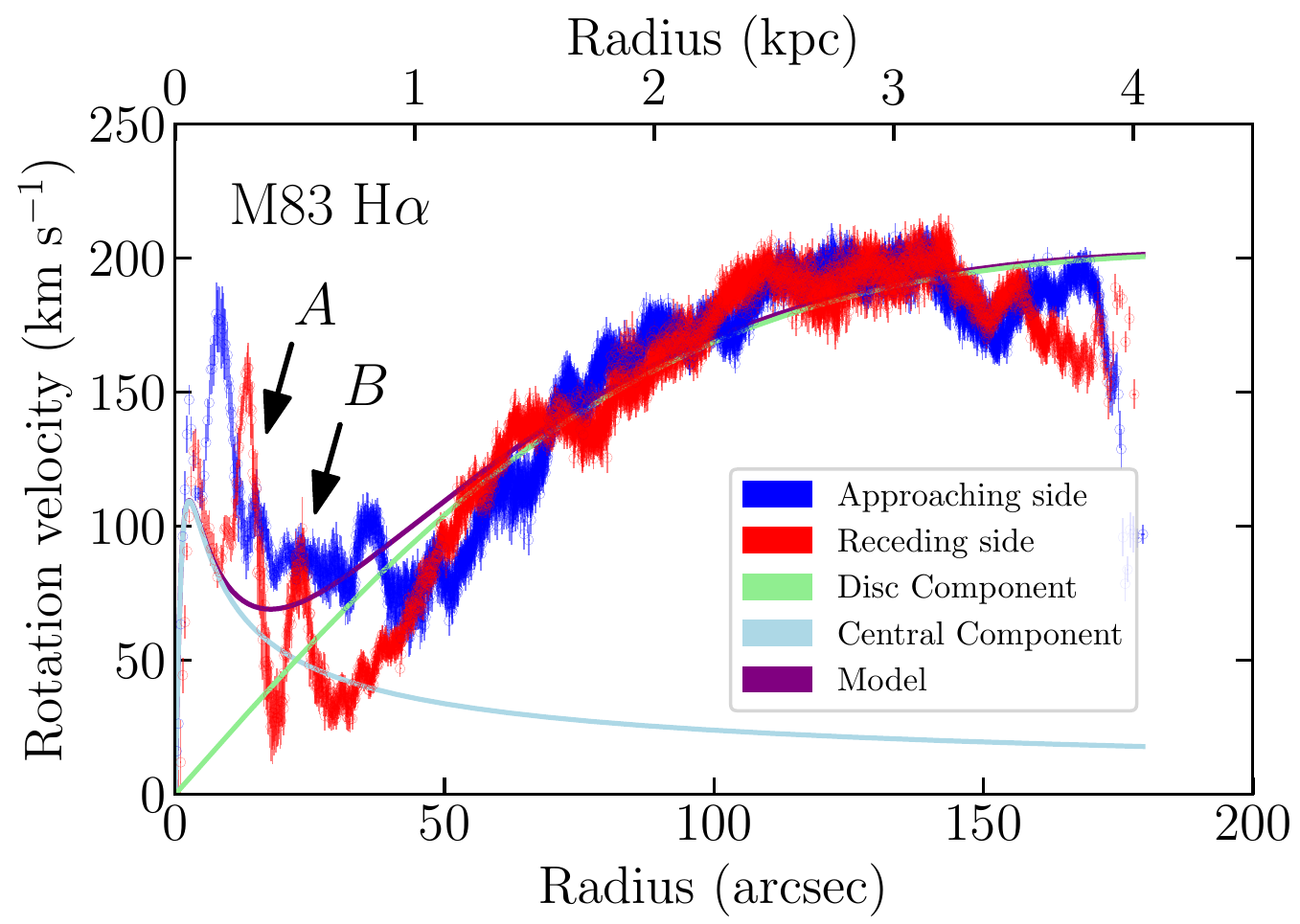}
\caption{Same as the middle panel of Fig.~\ref{fig:rotcurve}, but for model $(e)$ in Table~\ref{table:kinematic_models}, which enforces a symmetrical curve.}
\label{fig:rotcurve_extra}
\end{figure}

\begin{figure*}
\center
\includegraphics[width=9.1cm]{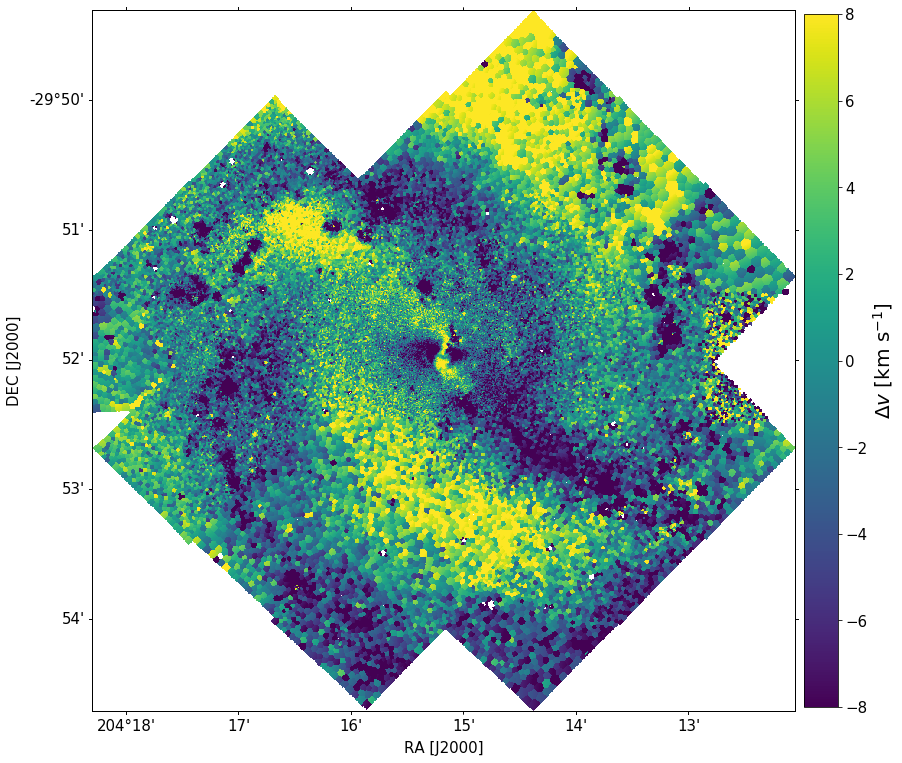}
\includegraphics[width=9.1cm]{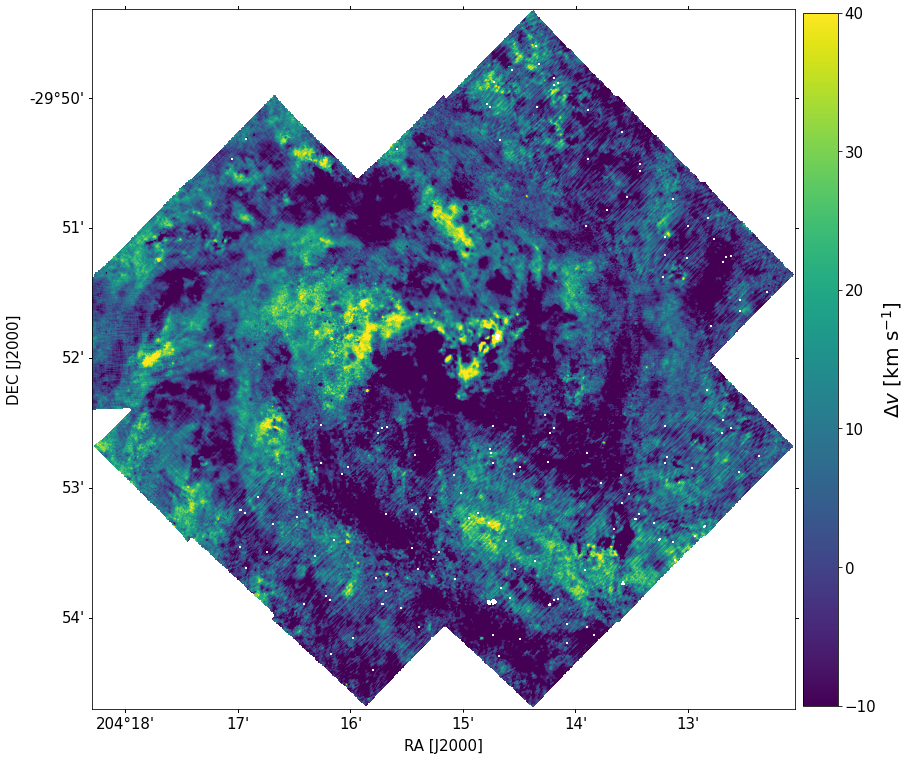}
\caption{2D residual velocity for model $(A)$ (left) and $(d)$ (right)
, as described in Table~\ref{table:kinematic_models}.}
\label{fig:residual_vel}
\end{figure*}

In Fig.~\ref{fig:rotcurve_extra} we present the rotation curve for model $(e)$ (same as in Fig.~\ref{fig:rotcurve} but enforcing $\Delta V_{\rm rot} = 0$).
In Fig.~\ref{fig:residual_vel} we show two-dimensional (2D) residual velocity fields for the best-fit models, $(A)$ and $(a)$, respectively. 2D--residual fields are used to optimise the free parameter determination of the rotation curves that are computed from 2D--velocity fields. The residual velocity fields ($V_{\rm res}$) shown in the figure are obtained after subtracting the best fit models (models $(A)$ and $(a)$ reported in Table~\ref{table:kinematic_models}) from the observed velocity field. Because rotation curves are based on axisymmetric models, specific patterns observed on the residual velocity field would be due to an incorrect parameter determination \citep{warner73}. As a consequence, when the free parameters of the rotation curve are optimised, the residual velocity field is typical-pattern-free and the amplitude in the 2D residual velocity is minimum. No artefacts are observed in the residual velocity fields, indicating that the models are optimal. The main residual patterns that remain are due to non-axisymmetirc features (e.g. the large-scale spiral structure and the central component which features an asymmetric X-shape pattern).

\subsection{Uncertainty estimation}
\begin{figure*}
\center
\includegraphics[width=1\linewidth]{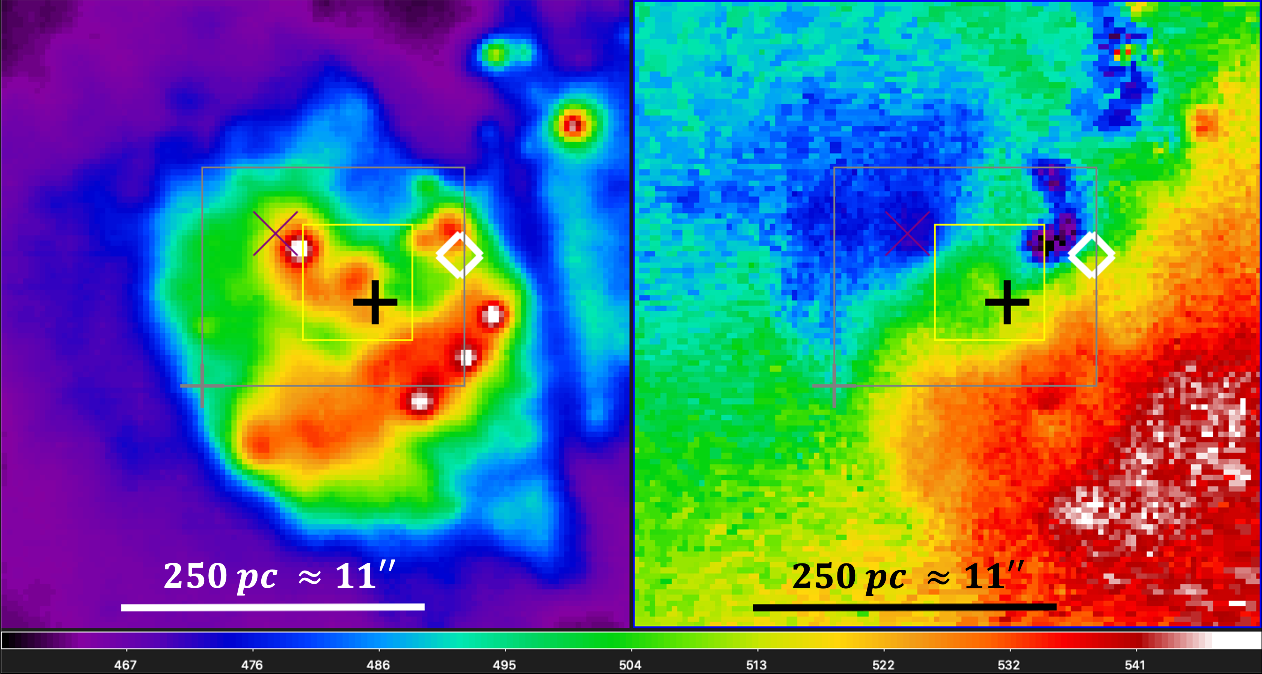}
\caption{Different centre positions from this work and the literature. Boxes indicate uncertainty determinations from centres as calculated by other authors.
The background images show: continuum emission around the H$\alpha$ line, rescaled in intensity (left panel, arbitrary units) and the observed stellar velocity field (right panel and horizontal colour bar, not corrected for the systemic velocity). The coordinates of the symbols and the box centres are given in Table~\ref{table:centers_literature}.
Yellow box  (Fabry-Perot kinematic centre, \citealp{fathi08}, \citealp{knapen10}, size of the box: 4.0$\times$4.2 arcsec$^2$);
grey box (Pa $\beta$ kinematic centre, \citealp{knapen10}, size of the box: 9.6$\times$8.0 arcsec$^2$);
grey '{\Large{+}}' (corner of the Pa $\beta$ uncertainty box that is the furthest away from the centre determined using the MUSE stellar velocity map);
purple '{\Large{$\times$}}' (optical nucleus, \citealp{diaz06});
white '{\Large{$\Diamond$}}' (MUSE stellar kinematic centre from model $(A)$, this work);
black '{\Large{+}}' (CO kinematic centre from model $(\beta)$, this work).}
\label{fig:kinem_center}
\end{figure*}

The raw stellar LoS velocity field consists of 1,678,471 velocity measurements. 
To remove spaxels possibly containing spurious information, the raw stellar field was masked accounting for spaxels with low S/N and/or large errors in the velocity or velocity dispersion determination,
which resulted in masking of 0.16\% of the spaxels.
Despite the nine free parameters, due to the large number of spaxels used in the fit, the associated statistical uncertainties are very low. Therefore, in order to study the robustness of the fit we inspect three additional models for which we set centre locations estimated in the literature. Fig.~\ref{fig:kinem_center} shows these locations, all found within a box of $\sim$3$\times$3 arcsec$^2$, close to the brightest central clumps of M83. For models  $(B)$, $(C)$ and $(D)$ we fix the centre to the coordinates computed in two different studies for which uncertainties have been computed.  In model $(B)$, we utilise the kinematic centre computed by \citet{fathi08} that used the H$\alpha$ line and 2D-spectroimagery (Fabry-Perot); for model $(C)$ we use the one determined by \citet{knapen10} from kinematic Pa $\beta$ data. Finally, to estimate our own uncertainty on the centre, for model $(D)$ we consider 
the corner of the uncertainty box from \citet{knapen10} (grey box in Fig.~\ref{fig:kinem_center}) that is the furthest away from the centre determined in model $(A)$.
The fits lead to very similar parameters (see table \ref{table:kinematic_models}).  Even if model $(A)$ provides the smallest $\chi^2_r$, model $(B)$ cannot be discarded just on the relevance of its $\chi^2_r$ which is only less than 2\% higher than those of model $(A)$.  Model $(C)$ and $(D)$ lead to $\chi^2_r$ respectively 5 and 12\% larger than model $(A)$. In conclusion we can estimate that our centre uncertainty within a $\chi^2_r$ variation of 2\% is $\Delta \alpha\pm$4 arcsec and $\Delta \delta\pm$1 arcsec. We observe that the disagreement in the rotation curves between the approaching and receding sides within the first kpc - and especially within the first 100 pc - becomes increasingly large with models $(B)$, $(C)$ and $(D)$, confirming that model $(A)$ is the best fit.

\begin{table*}
\caption{Centre positions from this work and the literature shown in Fig.~\ref{fig:kinem_center}.}
\begin{tabular}{lcccc}
\hline \hline
Centre & RA (J2000) & DEC (J2000) & Symbol & Ref \\ \hline
MUSE stellar kinematic centre (model $(A$)) & 13:37:00.43 & -29:51:56.2 & white '{\Large{$\Diamond$}}' & This work \\
CO kinematic centre (model $(\beta$)) & 13:37:01.14 & -29:51:52.5 &  black '{\Large{+}}' & This work \\
Fabry-Perot kinematic centre & 13:37:00.75 & -29:51:57.3 & yellow box &  (1), (2)\\
Pa$\beta$ kinematic centre & 13:37:00.81 & -29:51:57.1 & grey box & (2) \\
Outermost Pa$\beta$ centre\tablefootmark{(a)} & 13:37:01.13 & -29:51:59.5 & grey '{\Large{+}}' & (3) \\
Optical nucleus & 13:37:00.95 & -29:51:55.5 & purple '{\Large{$\times$}}' & (3), (2) \\
\hline
\end{tabular}
\tablefoot{
\tablefoottext{a}{Maximum distance to the MUSE stellar kinematic centre within the uncertainty on the Pa $\beta$ centre.}
}
\tablebib{(1) \citet{fathi08}; (2) \citet{knapen10}; (3) \citet{diaz06}.}
\label{table:centers_literature}
\end{table*}

\clearpage

\section{Supplementary linemaps}
\label{section:appendix_linemaps}

\begin{figure*}[!htb]
\center
\includegraphics[width=0.49\textwidth]{alma_mom0_zoomin.pdf}
\includegraphics[width=0.49\textwidth]{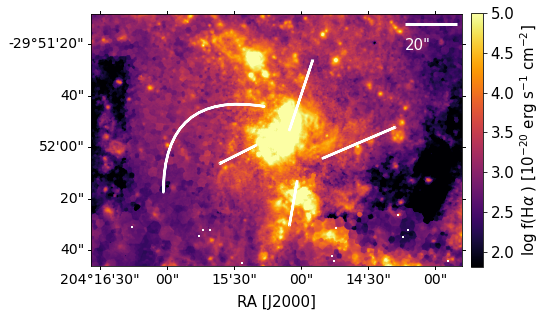}
\caption{Central starburst region observed in ALMA CO(2-1) (left panel) and in MUSE H$\alpha$ (reddening corrected, right panel). The white lines mark the position of the features discussed in Sect.~\ref{section:starburst_region}.}
\label{fig:center_CO_vs_halpha}
\end{figure*}

\begin{figure*}[!htb]
\center
\includegraphics[width=9.1cm]{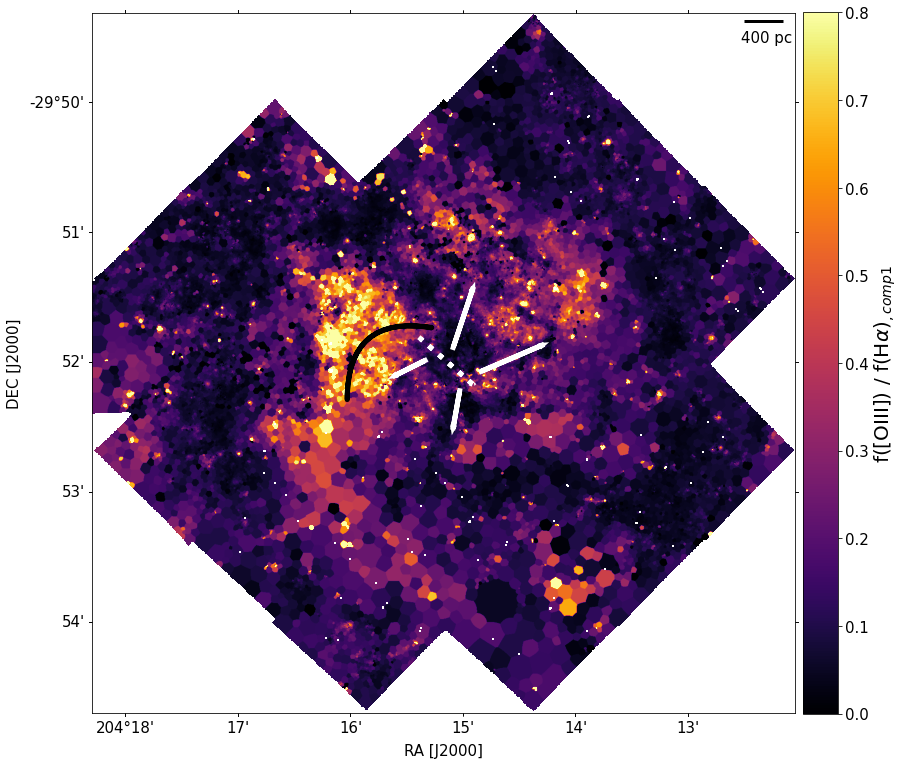}
\includegraphics[width=8.9cm]{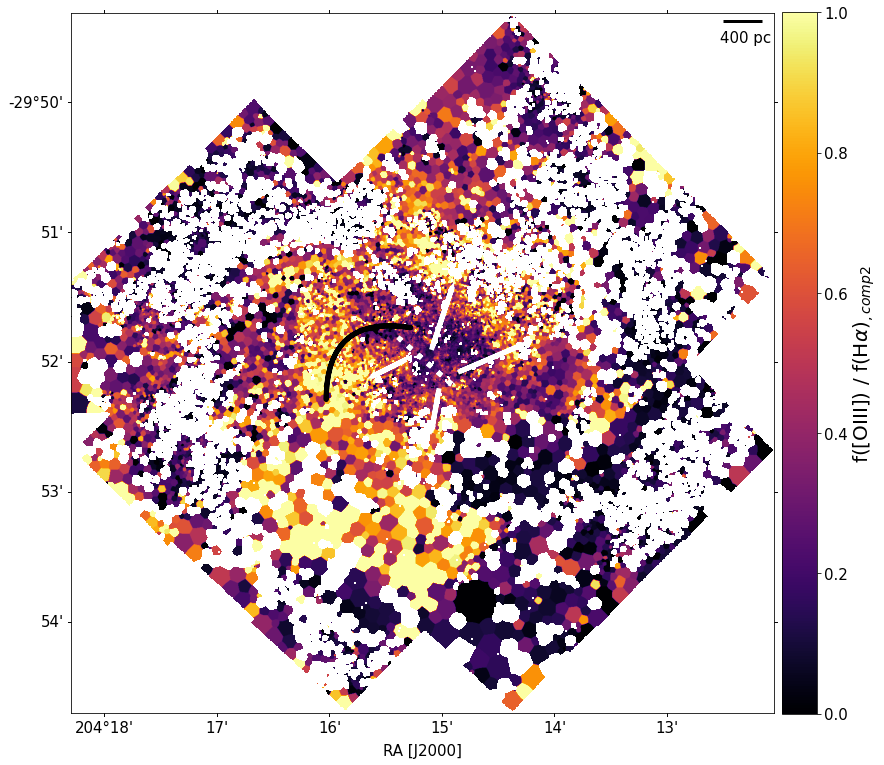}
\includegraphics[width=9.1cm]{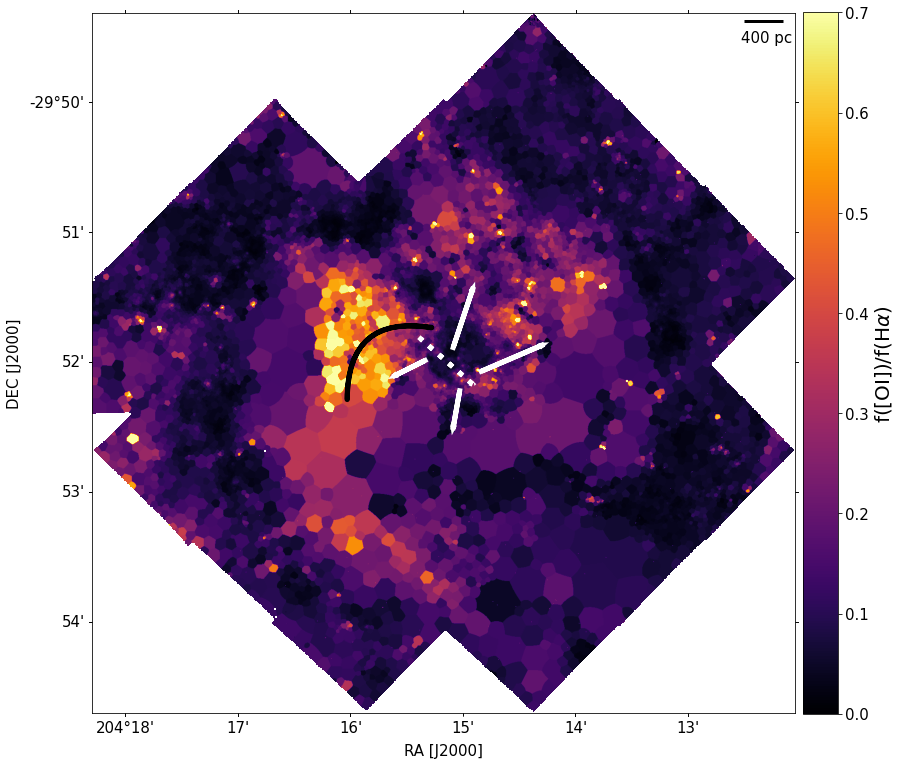}
\caption{Same as Fig.~\ref{fig:o3o1_map_zoomin}, but for the full FoV.
}
\label{fig:o3o1_map}
\end{figure*}

\clearpage

\section{MUSE dataset coverage and properties}
\label{section:appendix_muse_data_coverage}

\begin{table*}[!htb]
\caption{Central coordinates, exposure time, and seeing of the MUSE mosaic tiles. The seeing has been measured at 7000~\AA{}, as described in Sect.~\ref{section:data_description}.}
\begin{tabular}{llllll}
\hline \hline
ID & Programme ID & RA (J2000) & DEC (J2000) & $t_{\rm exp,tot}$ [s] & Seeing [arcsec] \\ \hline
Pnt1 & 096.B-0057(A) & 13:37:00.4900 & -29:52:00.300 & 2750 & 0.72 \\
Pnt2 & 096.B-0057(A) & 13:37:03.6200 & -29:51:21.700 & 2200 & 0.77 \\
Pnt3 & 096.B-0057(A) & 13:37:00.5000 & -29:50:41.100 & 2200 & 0.80 \\
Pnt4 & 096.B-0057(A) & 13:36:57.4500 & -29:51:20.600 & 2750 & 0.80 \\
Pnt5 & 096.B-0057(A), 0101.B-0727(A) & 13:37:06.6600 & -29:52:00.300 & 2200 & 0.81 \\
Pnt6 & 096.B-0057(A), 0101.B-0727(A) & 13:37:09.7000 & -29:52:39.900 & 2200 & 0.84 \\
Pnt7 & 096.B-0057(A), 0101.B-0727(A) & 13:37:09.7100 & -29:51:20.700 & 2200 & 0.68 \\
Pnt8 & 096.B-0057(A), 0101.B-0727(A) & 13:37:06.6700 & -29:50:41.100 & 2200 & 0.73 \\
Pnt9 & 096.B-0057(A) & 13:37:03.6200 & -29:52:39.900  & 550 & 0.63 \\
Pnt10 & 096.B-0057(A) & 13:37:06.6600 & -29:53:19.500 & 550 & 0.91 \\
Pnt11 & 096.B-0057(A) & 13:37:03.6100 & -29:53:59.100 & 550 & 0.73 \\
Pnt12 & 096.B-0057(A) & 13:37:00.5700 & -29:53:19.400 & 550 & 0.67 \\
Pnt13 & 096.B-0057(A) & 13:36:57.5300 & -29:52:39.900 & 550 & 0.73 \\
Pnt14 & 096.B-0057(A) & 13:36:57.5200 & -29:53:59.100 & 550 & 0.57 \\
Pnt15 & 096.B-0057(A) & 13:36:54.4800 & -29:53:19.400 & 550 & 0.73 \\
Pnt16 & 096.B-0057(A) & 13:36:51.4400 & -29:52:39.800 & 550 & 0.66\\
Pnt17 & 096.B-0057(A) & 13:36:54.4800 & -29:52:00.300 & 550 & 0.62 \\
Pnt18 & 096.B-0057(A) & 13:36:51.4400 & -29:51:20.600 & 550 & 0.77 \\
Pnt19 & 096.B-0057(A) & 13:36:54.4900 & -29:50:41.100 & 550 & 0.67 \\
Pnt20 & 096.B-0057(A) & 13:36:57.5300 & -29:50:01.500 & 550 & 0.71 \\
A1 & 097.B-0899(B) & 13:36:58.2800 & -29:51:55.900 & 1800 & 0.86 \\
A2 & 097.B-0899(B) & 13:36:53.7300 & -29:51:57.100 & 1800 & 0.80 \\
A3 & 097.B-0899(B) & 13:37:02.7000 & -29:51:55.500 & 1800 & 0.87 \\
A4 & 097.B-0899(B) & 13:37:07.2100 & -29:51:54.600 & 1800 & 0.71 \\
A7 & 097.B-0899(B) & 13:37:11.7100 & -29:51:53.700 & 1800 & 1.10\\
B1 & 097.B-0640(A) & 13:37:00.8300 & -29:51:55.500 & 1920 & 0.80 \\
\hline
\end{tabular}
\label{table:pointings_info}
\end{table*}

\begin{figure*}
\center
\includegraphics[width=12cm]{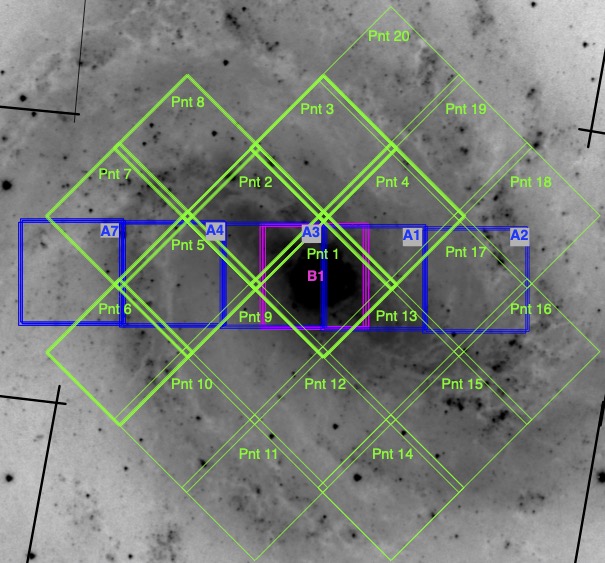}\\
\includegraphics[width=13cm]{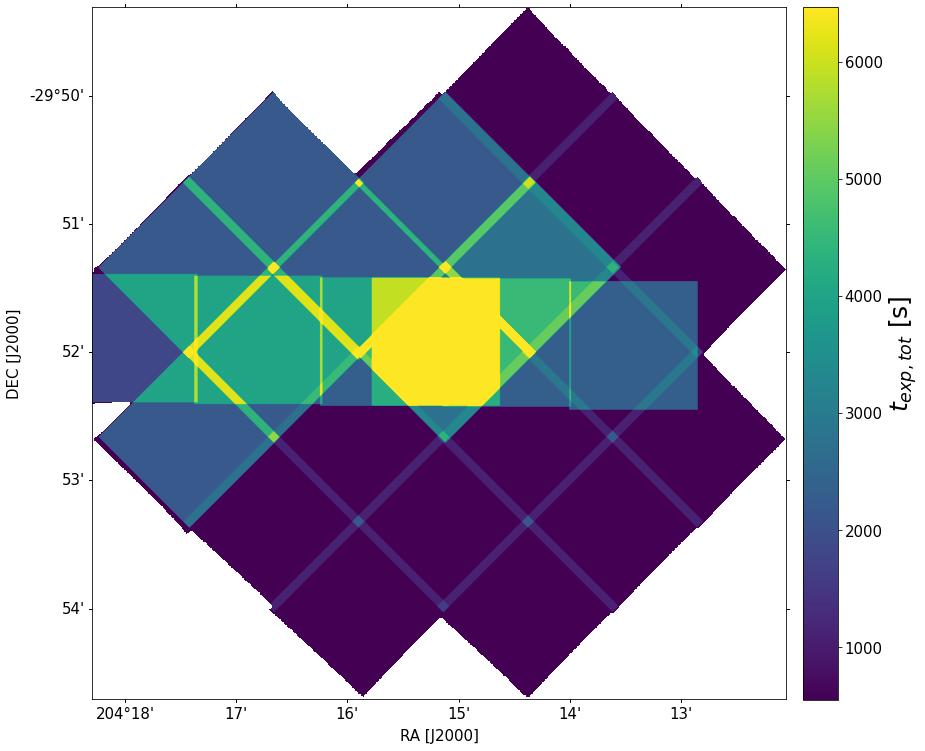}
\caption{Layout of the MUSE mosaic. \textit{Top panel}: location of the MUSE mosaic tiles. In green we indicate the data acquired as part of this project (PI Adamo); in blue and purple archival data that we have included in the final mosaic (PI Ibar and Gadotti). \textit{Bottom panel}: total exposure time of each mosaic tile.}
\label{fig:texp_coverage}
\end{figure*}

\end{appendix}

\end{document}